%% file: ms.tex
\documentclass[11pt]{article}
\input{header.tex}
\DontPrintSemicolon

\title{Finding Subgroups with Significant Treatment Effects}

\author{
	Jann Spiess
	\\ Stanford University
	\and
	Vasilis Syrgkanis\\
	Stanford University
	\and 
	Victor Yaneng Wang \\
	University of Oxford}

\date{
	First public version: March 2021 \\
	This version: December 2023 \\
	\bigskip
	\textit{Comments welcome!}
}

\begin{document}
	
	\maketitle
	
	\begin{abstract}%
		
		Researchers often run resource-intensive randomized controlled trials (RCTs) to estimate the causal effects of interventions on outcomes of interest. Yet these outcomes are often noisy, and estimated overall effects can be small or imprecise.
		Nevertheless, we may still be able to produce reliable evidence of the efficacy of an intervention by finding subgroups with significant effects.
		In this paper, we propose a machine-learning method that is specifically optimized for finding such subgroups in noisy data.
		Unlike available methods for personalized treatment assignment, our tool is fundamentally designed to take significance testing into account: it produces a subgroup that is chosen to maximize the probability of obtaining a statistically significant positive treatment effect.
		We provide a computationally efficient implementation using decision trees and demonstrate its gain over selecting subgroups based on positive (estimated) treatment effects. Compared to standard tree-based regression and classification tools, this approach tends to yield higher power in detecting subgroups affected by the treatment.
	\end{abstract}
	
	\blfootnote{
		Jann Spiess, Graduate School of Business, Stanford University, \href{mailto:jspiess@stanford.edu}{jspiess@stanford.edu};
		Vasilis Syrgkanis, Department of Management Science \& Engineering, Stanford University, \href{mailto:vsyrgk@stanford.edu}{vsyrgk@stanford.edu};
		Victor Yaneng Wang, Global Priorities Institute, University of Oxford, \href{mailto:victor.wang@philosophy.ox.ac.uk}{victor.wang@philosophy.ox.ac.uk}.
		We thank Saarthak Gupta for expert research assistance and audiences at Wharton, UT Austin, and CLeaR for helpful comments. This manuscript supersedes an earlier version that was selected for oral presentation at the 1st Conference on Causal Learning and Reasoning (CLeaR 2022) under the title ``Evidence-Based Policy Learning'' (Jann Spiess and Vasilis Syrgkanis).
	}
	
	\bigskip
	
	
	\clearpage
	
	\section{Introduction}
	
	Randomized controlled trials (RCTs) are a standard way of inferring the causal effect of an intervention on outcomes of interest. However, such outcomes are often noisy, and estimated overall effects in RCTs can be small or imprecise. One could attempt to overcome these issues by collecting more data to increase sample sizes, yet in practice, this may be infeasible or too costly.
	We present an alternative solution: we develop tools that provide reliable evidence of the efficacy of the intervention by finding \emph{subgroups with statistically significant treatment effects}, even if overall effects are noisy or small.
	
	Our tool for finding subgroups with statistically significant effects builds upon a fast-growing literature on causal machine-learning methods, including estimators for heterogeneous treatment effects and personalized treatment-assignment policies. These methods allow us not only to estimate \emph{whether} an intervention works on average but also to analyze \emph{for whom} the intervention is particularly effective. However, available tools are not typically optimized for finding evidence of treatment effects in very noisy data.
	In particular, our article addresses two related shortcomings of existing approaches.
	First, existing machine-learning methods for the estimation of heterogeneous treatment effects and personalized treatment assignments tend to perform poorly in data with small and noisy effects.
	Second, unlike existing methods, we fundamentally incorporate significance testing into our procedure. We show that this motivates a particular objective function, which in turn requires new tools to implement.
	
	We formalize the task of finding a subgroup with significant treatment effects in \autoref{sect:setup}, and show in \autoref{sect:objective} that it translates into an objective that differs from the standard goal of finding those observations who have high treatment effects.
	We consider procedures that learn assignment rules based on training data from a randomized experiment.
	These assignment rules map observable characteristics of observations in the experiment to a binary indicator of whether an observation should be included in a subsequent hypothesis test.
	The chosen assignment rule is then deployed on held-out data, where we test whether there is a significant treatment effect on the chosen subgroup.
	We optimize for maximizing the probability of rejecting the null hypothesis of a zero effect, corresponding to finding reliable evidence of treatment effects.
	Under a Normal large-sample approximation, we show that this goal is equivalent to maximizing the ratio of mean to standard error of treatment effect estimates in the hold-out data.
	When conditional variances differ across covariates, then the solution to this goal can differ markedly from subgroups formed solely by those observations that have a predictably positive or large treatment effect.
	
	Having motivated an objective function for finding subgroups with statistically significant effects, in \autoref{sect:significance_trees} we discuss an implementation based on a ``significance tree'' algorithm that performs greedy recursive partitioning.
	In \autoref{sect:submodular}, we then consider approximate solutions based on submodular optimization.
	We show that empirical optimization of our target goal approximates an optimal assignment and reduces to a submodular minimization problem, for which we present approximation bounds for finite support under a sparsity assumption on conditional treatment effects and conditional variances.
	
	In the simulation study in \autoref{sect:simulation}, we show the gain from using our purpose-built significance tree for finding subgroups with statistically significant treatment effects, as well as the value of submodular (re-)optimization.
	In the simulation study, we compare the performance of our algorithm to tree-based algorithms that form subgroups based on positive estimated treatment effects alone.
	We demonstrate the improvement in test performance and highlight cases in which our assignment tests better than a simple overall test in the full training and hold-out sample combined.
	
\begin{figure}[h!]
		\centering
		\includegraphics[width=.6\textwidth]{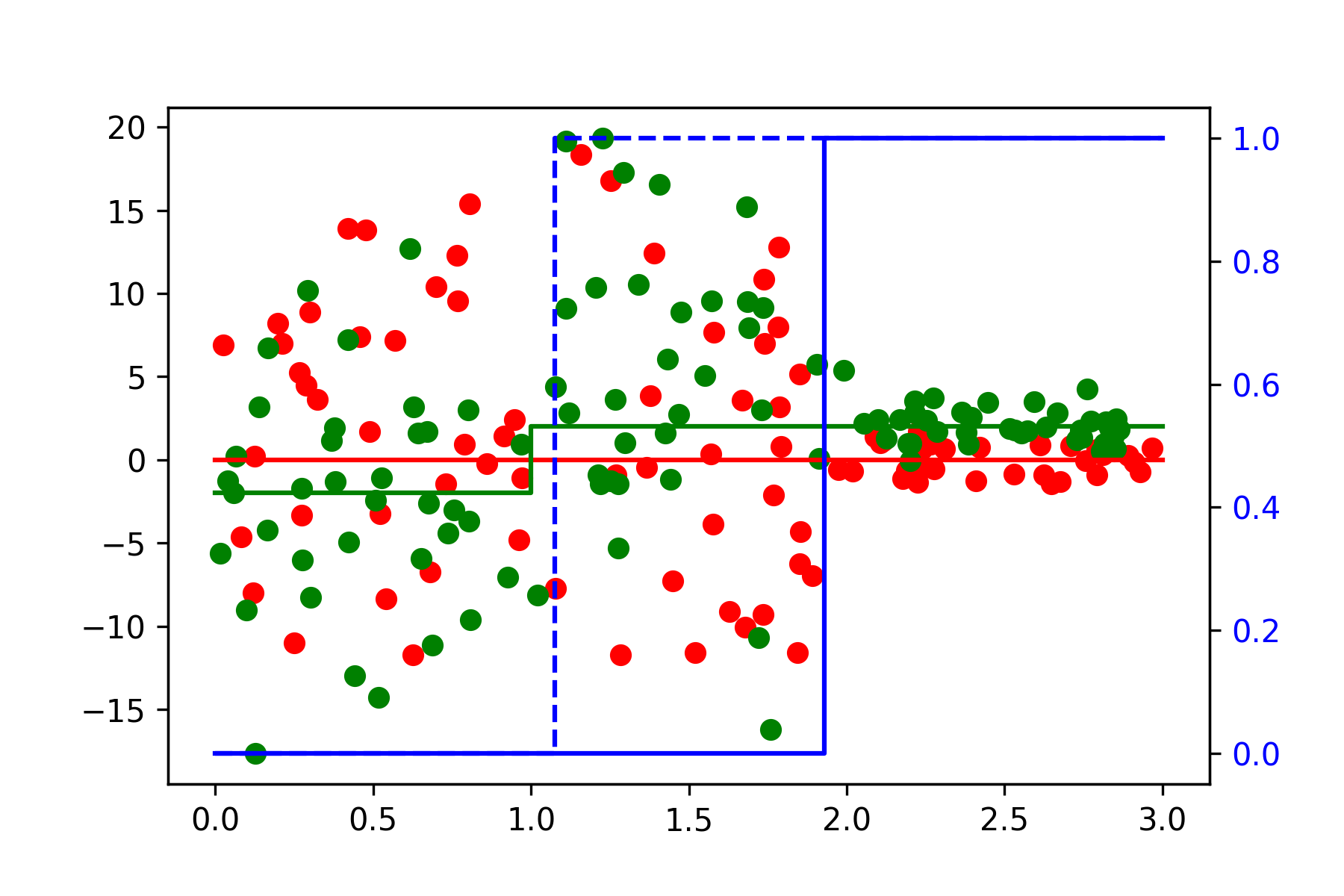}
		\vspace{-.5em}
		\caption{Example of a single tree with $n=100$ for test-based (solid) and simple classification-based (dashed) assignment, where on the blue axis $Y=1.0$ corresponds to parts of the $x$-axis assigned to treatment, and $Y=0.0$ to parts assigned to control. Green dots are treatment observations and red dots are control observations, while green and red lines correspond to the true conditional averages of the treatment and control groups, respectively. The simulation follows the setup from \autoref{sect:simple-synth} with treatment effects $\tau = (2,2,0)$.}
		\label{fig:illustration}
	\end{figure}

	We can illustrate the main idea behind our approach in a simple example with just one covariate, which we expand on in our simulation study. 
	This example shows the main intuition for why optimizing for significant treatment effects can improve power, compared to testing effects using the full sample or selecting those observations with positive or high treatment effects.
	\autoref{fig:illustration} plots a draw of treatment and control outcomes on the $y$-axis against their univariate covariate on the $x$-axis.
	In this simulated example, there are three parts of the covariate space: the left ($x \in [0,1]$) has negative treatment effects and high variance, the middle ($x \in [1,2]$) has positive average treatment effect and high noise, and the right ($x \in [2,3]$) has positive treatment effects with small noise.
	
	In our example, we now compare different methods of selecting observations for testing. If we used the full sample for testing, we would include the left area with negative treatment effects, which would go against our goal of finding evidence for a positive treatment effect.
	We therefore hope to do better by using data to select a subset on which we can achieve higher power.
	A standard machine-learning method that aims to find observations with positive treatment effects would often separate the leftmost from the remaining parts (dashed blue line).
	This would improve the power of the test, but still include the middle part,
	yielding a test with low power on a hold-out dataset because of the noisy outcomes in this region.
	In this example, our significance tree (blue solid line) selects instead only the right-most segment into the subgroup that gets tested, yielding better evidence of a positive effect.
	This is because the right-most part presents the clearest evidence of the efficacy of the treatment, since it combines a positive effect with small noise.
	
	Building upon our baseline algorithm, \autoref{sect:modifications} provides two modifications that can further improve the power of finding evidence for positive treatment effects.
	First, we consider a continuous relaxation that tests for significant effects based on a weighted average of observations, rather than a simple subgroup. 
	Second, we aggregate individual significance trees into a forest that can capture more flexible subgroups. We demonstrate the corresponding performance gains in a simulation.
	
	Finally, \autoref{sect:extensions} describes additional extensions of our setup and algorithm.
	For instance, we could instead address the problem of finding significant heterogeneity in treatment effects.
	Alternatively, we could consider a mixed criterion that incorporates both the probability of passing a hypothesis test on the hold-out sample \emph{and} the value of assigning the chosen observations to treatment conditional on passing that test.
	In addition, we consider alternative outcomes and reference assignments, estimated propensity scores, and cross-evaluated tests.
	
	\paragraph{Related work.} 
	
	We build upon an extensive machine-learning literature on learning personalized policies from experimental or observational data \citep{dudik2011doubly,swaminathan2015counterfactual,nie2017quasi,kallus2018policy,kitagawa2018should}. A large volume of works uses ideas from the doubly robust literature in semi-parametric inference \citep{rubin2005general,rubin2007doubly,diaz2013targeted,van2014targeted,kennedy2017non,kennedy2019robust} to construct objectives to be optimized over a space of policies. However, the majority of these works optimize the mean policy value and do not focus on the objective of succeeding in a subsequent hypothesis test on a fresh sample.
	Closest in spirit to our approach is \cite{kallus2021more}, which improves the efficiency of policy-learning estimators in data with limited overlap by a retargeting procedure that assigns a higher weight to regions with good overlap. Somewhat analogously, our paper shows that for data where we struggle to obtain statistically significant overall effects, we can extract more reliable evidence by focusing on subgroups in which effects are large and measured more precisely. However, our approach aims to maximize performance on a hold-out hypothesis test, whereas \cite{kallus2021more} instead aims to improve the efficiency of a policy-learning estimator.
	
	Our work also builds upon work on recursive partitioning and treatment effect estimation in the causal machine-learning literature \citep{kunzel2017meta,athey2016recursive,athey2016generalized,nie2017quasi,athey2017efficient,zhou2018offline,oprescu2018orthogonal,friedberg2018,chernozhukov2017orthogonal,chernozhukov2018plugin,foster2019orthogonal,Bertsimas2019-ms}. Unlike this treatment-effect literature, we do not care about learning an accurate model of how the effect varies, but rather only learn a model of which subgroup to select for further testing. 
	
	Moreover, the majority of the heterogeneous treatment effect literature constructs subgroups mainly based on some mean heterogeneity criterion. Arguably closest to us is the work of \cite{powers2018some} which analyzes a tree algorithm whose criterion instead uses a standardized heterogeneity of the form $(\tau(S_1)-\tau(S_2))^2/(\sigma(S_1)^2 + \sigma(S_2)^2)$ \citep[already proposed in][]{athey2016recursive}, where $S_1, S_2$ is a candidate split, $\tau(S_i)$ is the mean effect on each subgroup and $\sigma(S_i)$ the standard deviation. However, this criterion tries to identify the modes of heterogeneity and does not target the selection of subgroups directly.
	
	Our work also relates to recent work on false discovery rate control in multiple and adaptive hypothesis testing \citep{ramdas2019unified,lei2017star}. Unlike our setting, the FDR literature typically considers the statistically more daunting task of testing on the same data that was used to construct the adaptive hypothesis. Here we consider a sample-splitting setting, where testing will be constructed on a separate data set and we specifically focus on the out-of-sample power of the test. Arguably closest to our work from the FDR literature is the concurrent work of \cite{duan2021interactive}, which instantiates the FDR line of work on the heterogeneous treatment effect problem and presents an interactive algorithm that controls the false discovery rate of the constructed subgroup to be treated.
	The work of \cite{Armstrong2015-gc} considers inference on optimal treatment assignments, where the goal is to obtain a subgroup for which all conditional average treatment effects are positive with high confidence.
	Relative to this work, we instead provide a subgroup that aims to maximize evidence for a positive average subgroup effect on hold-out data, thus sidestepping the multiple-testing challenge in \cite{Armstrong2015-gc}.
	
	Our setup and motivation are very closely related to concurrent and independent work in \cite{Talisa2021-ud} that also considers the objective of finding subgroups with statistically significant treatment effects. Relative to their approach, we construct recursive-partitioning and model-based solutions where model selection itself is directly optimized for maximizing power.
	Relatedly, \cite{McFowland2018-wo} considers the problem of identifying most-affected subgroups, which it operationalizes as a problem of detecting anomalous patterns.
	In a similar spirit, \cite{Ladhania2020-ln} considers the selection of subgroups with consistently large treatment effects, which are then tested on separate data.
	
	Recent work on providing robust versions of off-policy learning is similar in spirit to our motivation. 
	\cite{Dwivedi2020} aims to identify subgroups with heterogeneous treatment effects that are stable (i.e. do not vary a lot) and are well-calibrated out of sample. Finally, concurrent work in \cite{leqi2021median} provides a way of estimating assignment policies that maximize the median policy value, instead of the expected policy value, which typically is more robust to outlier observations. Both of these works could lead to better out-of-sample power of statistical tests since they lead to more robust subgroup heterogeneity identification.
	
	\section{Setup and Goal}
	\label{sect:setup}
	
	We aim to find a subgroup of observations for which we can provide evidence of a positive treatment effect based on data from a randomized experiment.
	We learn this subgroup from a training sample of $n$ iid draws $(Y_i,W_i,X_i)$ from a population distribution of random variables $(Y,W,X)$, where 
	the outcome $Y$ is real-valued,
	treatment $W \in \{0,1\}$ is assigned randomly with probability $\E[W] = p \in (0,1)$,
	and there are additional covariates $X \in \X$ available that are not affected by the treatment.
	We assume that $Y = Y^{(W)}$ with potential outcomes $(Y^{(1)},Y^{(0)})$ that represent the outcomes an observation would have in treatment ($W=1$) and control ($W=0$), respectively. We then define the treatment effect on an observation as $Y^{(1)} - Y^{(0)}$.
	Randomization of $W$ allows us to identify from the data not only the average treatment effect $\tau = \E[Y^{(1)} -  Y^{(0)}]$, but also conditional average treatment effects $\tau(x) = \E[Y^{(1)} - Y^{(0)}|X=x]$.
	
	We formalize the selection of a subgroup by an assignment policy $a: \X \rightarrow \{0,1\}$  that maps a covariate $x \in \X$ to the indicator $a(x) \in \{0,1\}$ of whether an observation with covariates $x$ is in the chosen subgroup.
	For a subgroup of the population defined in this way,
	we can define the total effect of assigning observations in this subgroup to treatment as
	\[
	u(a) = \E[a(X) \: Y^{(1)} + (1-a(X)) \: Y^{(0)}] - \E[Y^{(0)}] = \E[a(X) \: \tau(X)].
	\]
	In this article, we are concerned with finding such subgroups for which we can provide evidence that $u(a) > 0$, i.e. that the group-wise average treatment effect $\E[\tau(X_i)|a(X_i) {=} 1] = u(a) / \P(a(X_i) = 1)$ is positive.
	We could also interpret this goal within the context of policy learning as follows: we seek a policy $a$ that assigns a subgroup of observations to treatment, where we have evidence that if we deploy this assignment policy on future observations from the same population, this leads to positive lift $u(a)$ relative to assigning everybody to control.
	
	Existing methods in the policy-learning literature are typically focused on finding a subgroup assignment $a$ that maximizes the expected lift $u(a)$. Note that a subgroup assignment that maximizes $u(a)$ is $\tilde{a}(x) = \Ind\left(\tau(x) > 0\right)$, suggesting two related standard strategies. The first strategy is to estimate heterogeneous treatment effects $\tau(x)$ directly by $\hat{\tau}(x)$ and then assign based on $\hat{a}(x) = \Ind\left(\hat{\tau}(x) > 0\right)$. The second strategy is to perform cost-sensitive classification so as to maximize the empirical objective $\frac{1}{n} \sum_{i=1}^n a(X_i) \: \hat{\tau}_i$. This latter approach would be based on an unbiased estimate $\hat{\tau}_i$ of $Y_i^{(1)} - Y_i^{(0)}$ for each observation $i$, such as the estimate $\tilde{Y}_i$ we consider in \autoref{subsect:estimation} below.
	
	We depart from existing methods by instead fundamentally incorporating hypothesis testing into our target criterion. 
	In practice, after estimating an assignment $\hat{a}$, we often have to validate that the chosen subgroup is indeed is associated with a positive effect $u(\hat{a})> 0$ by providing evidence in the form of a statistical hypothesis test.
	We therefore assume that we also have access to an additional hold-out dataset with $N$ draws $(Y_j,W_j,X_j)$ from the same distribution as the experiment, on which we perform a test of the form
	\begin{align*}
		\widehat{T}(\hat{a}) = \Ind\left(\frac{\hat{u}(\hat{a})}{\sqrt{\widehat{\Var}(\hat{u}(\hat{a}))}} \geq z_{1-\alpha}\right),
	\end{align*}
	where $\hat{u}(\hat{a})$ is an estimate of $u(\hat{a})$, $\widehat{\Var}(\hat{u}(\hat{a}))$ is an estimate of the measurement noise $\Var(\hat{u}(\hat{a})|\hat{a})$ from the hold-out data, and $z_{1-\alpha}$ is the critical value of a one-sided Normal test with significance level $\alpha$ (that is, the $1-\alpha$-quantile of a standard Normal).
	The test is justified under standard regularity conditions and for appropriate estimators $\hat{u}(a),\widehat{\Var}(\hat{u}(a))$ by
	\begin{align*}
		\frac{\hat{u}(\hat{a})}{\sqrt{\widehat{\Var}(\hat{u}(\hat{a}))}}
		\cd \N(0,1)
	\end{align*}
	under the null hypothesis $u(\hat{a}) = 0$, where we assume throughout that $\hat{a}$ assigns a sufficiently large part of the sample to treatment such that the limit remains well-behaved.
	
	Our goal is to maximize the probability of passing the test,
	so we want to choose a subgroup assignment $\hat{a}$ with maximal expected probability
	\[
	U(\hat{a})=\E[\widehat{T}(\hat{a})|\hat{a}]
	=
	\P\left(\frac{\hat{u}(\hat{a})}{\sqrt{\widehat{\Var}(\hat{u}(\hat{a}))}} \geq z_{1-\alpha}\middle|\hat{a}\right),
	\]
	where expectation and probability are over the hold-out sample.
	For an assignment policy that chooses nobody for the subgroup, we can in practice assume that the test statistic is standard Normal (that is, 
	$p$-values are uniformly distributed), yielding baseline utility $U(\hat{a}) = \alpha$.
	By splitting the sample, optimizing for such a goal does not distort inference as long as the hold-out data is only used for the test.

	\section{Population Objective and Optimal Subgroups}
	\label{sect:objective}
	
	In the previous section, we motivated the objective function
	\begin{align*}
		U(a)&=\E[\widehat{T}(a)]
		=
		\P\left(\frac{\hat{u}(a)}{\sqrt{\widehat{\Var}(\hat{u}(a))}} \geq z_{1-\alpha}\right)
	\end{align*}
	for the expected lift $u(a) = \E[a(X) \: \tau(X)]$ from the subgroup chosen by $a$.
	In this section, we consider a specific estimator for estimating the lift $u(a)$ and constructing a test. We then solve for the theoretically optimal (but generally infeasible) oracle maximizers of our testing objective when we apply a large-sample approximation for the training sample.
	
	\subsection{Treatment-effect and variance estimation}
	\label{subsect:estimation}
	
	In order to derive results on maximizing the power of the hold-out test, we now consider a specific class of estimators $\hat{u}(a)$ for the lift $u(a)$ and their variance $\Var(\hat{u}(a)|a)$ in the hold-out sample $(Y_j,W_j,X_j)_{j=1}^N$.
	An unbiased estimator for $u(a)$ based on inverse propensity score weighting is
	\begin{align*}
		\hat{u}_0(a) &= \frac{1}{N} \sum_{j=1}^N a(X_j) \: \left(\frac{W_j}{p} - \frac{1-W_j}{1-p}\right) Y_j.
	\end{align*}
	Even though it is unbiased, this estimator can have high variance, which we can typically reduce by appropriate residualization.
	To this end, note that the estimator
	\begin{align*}
		\hat{u}(a) &= \frac{1}{N} \sum_{j=1}^N a(X_j) \: \tilde{Y}_j,
		&
		\tilde{Y}_j
		&=
		\left(\frac{W_j}{p} - \frac{1-W_j}{1-p}\right) (Y_j - \hat{c}(X_j))
	\end{align*}
	based on the centered outcome $Y - \hat{c}(X)$ is still unbiased for $u(a)$, given any fixed centering function $\hat{c}:\X \rightarrow \R$ that does not use any data from the hold-out set. The variance of $\hat{u}(a)$ can then be estimated by $\widehat{\Var}(\hat{u}(\hat{a})) = \frac{1}{N (N-1)} \sum_{j=1}^N (\hat{a}(X_j) \:\tilde{Y}_j - \hat{u}(\hat{a}))^2$.
	
	While our theory remains applicable for any choice of centering function $\hat{c}$ that does not depend on hold-out data, we consider the specific choice that estimates
	\[
	c^*(x) = (1-p) \E[Y| X=x, W=1] + p \E[Y| X=x, W=0].		
	\]
	Choosing $c^*$ minimizes the variance of $u(\hat{a})$ and leads to the standard semi-parametric efficient estimator \citep[see e.g.][]{Wu2017}.
	Furthermore, \citep{spiessoptimal} shows that this optimal centering function can be written as the minimizer of the weighted least squares regression problem%
	\footnote{The first-order condition for each $x$ is: $\frac{1}{p} \E[Y| X=x, W=1] + \frac{1}{1-p} \E[Y| X=x, W=0] = \left(\frac{1}{p} + \frac{1}{1-p}\right) c(x)$.
		Re-arranging yields $c(x) = (1-p) \E[Y| X=x, W=1] + p \E[Y| X=x, W=0]$ as desired. \cite{Wu2017} proposes a leave-one-out estimator of average treatment effects in experiments based on residualization by this score. \cite{spiessoptimal} argues that this specific score solves the weighted prediction problem in \eqref{eqn:weightedprediction}. \cite{opper2021improving} also considers an implementation for average effects in experiments using cross-fitted centering.
	}
	\begin{equation}
		\label{eqn:weightedprediction}
		\E\left[ \left(\frac{W}{p^2} + \frac{1-W}{(1-p)^2}\right) (Y - c(X))^2\right]
	\end{equation}
	which can subsequently be estimated by arbitrary non-parametric machine-learning methods in a data-adaptive manner.
	Across our implementations, we estimate $c^*$ in this way from the training data, yielding some estimate $\hat{c}$. We then perform tests based on the centered outcome $\tilde{Y} = Y - \hat{c}(X)$.
	
	\subsection{Oracle subgroups and population objective}
	\label{subsect:oracle}
	
	We now solve for optimal subgroup assignments given the above treatment-effect estimators. Expanding on the notation from \autoref{sect:setup}, we write
	\begin{align*}
		\tau(x) &= \E[Y^{(1)} - Y^{(0)}|X=x] = \E\left[\left(\frac{W}{p} - \frac{1-W}{1-p}\right) (Y-c(X)) \middle|X=x\right] = \E[\tilde{Y}|X=x],
		\\
		\sigma^2_c(x) &= \Var\left(\left(\frac{W}{p} - \frac{1-W}{1-p}\right) (Y-c(X)) \middle|X=x\right) = \Var(\tilde{Y}|X=x),
	\end{align*}
	for the conditional average treatment effect and the variance in its estimation at covariate value $x \in \X$,
	where $Y-c(X)$ is the centered outcome for some centering function $c: \X \rightarrow \R$.
	In the case of the optimal centering function $c^*$, we obtain the minimal variance $\sigma^2_{*}(x) = \frac{\Var(Y^{(1)}|X{=}x)}{p} + \frac{\Var(Y^{(0)}|X{=}x)}{1-p}$, where we write $\sigma^2_{*}(x) = \sigma^2_{c^*}(x)$.
	We focus on the case where treatment effects $\tau(x)$ are small and of the order of magnitude $\frac{1}{\sqrt{N}}$, represented with the notation
	$
	\tau(x) = \frac{\mu(x)}{\sqrt{N}}.
	$
	This assumption allows us to model non-trivial testing in the limit, and expresses the use case where data is so noisy or effects so small that testing is meaningful.

	In the local-to-zero limiting regime and for the estimators outlined above,
	for any fixed rule $a$ with $\E[a(X)] > 0$ and any centering function $\hat{c}$ we have that
	\begin{equation}
		\frac{\hat{u}(a)}{\sqrt{\widehat{\Var}(\hat{u}(a))}}
		\cd \N\left( \frac{\E[a(X) \: \mu(X)]}{\sqrt{\E[a^2(X) \: \sigma^2(X)]}}, 1 \right),
		\label{eqn:localapprox}
	\end{equation}
	where we write $\sigma^2 = \sigma^2_{\hat{c}}$.
	Since the asymptotic power of the resulting test is thus
	\begin{align*}
		\P\left(\N( T(a), 1 ) > z_{1-\alpha}\right)
		&=
		\Phi(T(a) - z_{1-\alpha}),
		&
		T(a) &= \frac{\E[a(X) \: \mu(X)]}{\sqrt{\E[a^2(X) \: \sigma^2(X)]}}
	\end{align*}
	(with the Normal cdf $\Phi$),
	maximizing asymptotic power corresponds to maximizing
	\begin{align*}
		T(a) &= \frac{\E[a(X) \: \mu(X)]}{\sqrt{\E[a^2(X) \: \sigma^2(X)]}}
		=
		\sqrt{N} \frac{\E[a(X) \: \tau(X)]}{\sqrt{\E[a^2(X) \: \sigma^2(X)]}}.
	\end{align*}
	
	\begin{proposition}\label{prop:oracle}
		Asymptotically as $N\rightarrow 0$ and under standard regularity assumptions,
		the policy that maximizes the testing objective
		solves
		\begin{equation}
			\label{eqn:populationobj}
			a_* = \argmax_{a} U(a) = \argmax_{a} \frac{\E[a(X) \: \tau(X)]}{\sqrt{\E[a^2(X) \: \sigma^2(X)]}}
		\end{equation}
		when there are positive treatment effects, and otherwise $a_* \equiv 0$.
		Note that the optimizer does not depend on the hold-out size $N$, or the size $\alpha$ of the test.
	\end{proposition}
	
	\begin{proof}
		See \autoref{sect:proof_prop_oracle}.
	\end{proof}
	
	The result holds more generally for continuous relaxations $a: \X \rightarrow [0,\infty)$ of the assignment problem, to which we return in \autoref{sect:extensions}.
	In the binary case we study here, $a^2(x) = a(x)$, so the criterion becomes $\E[a(X) \: \tau(X)] / \sqrt{\E[a(X) \: \sigma^2(X)]}$.
	In additional results in \autoref{sect:theory}, we show that there are cases in which optimizing for the empirical equivalent of \eqref{eqn:populationobj} can lead to strictly more powerful tests than standard approaches from policy assignment that aim to maximize the expected lift $u(a)$.

	\section{Implementation Using Significance Trees}
	\label{sect:significance_trees}
	
	In \autoref{sect:objective}, we characterized an oracle solution to the goal of finding an assignment that optimizes for passing a hypothesis test on hold-out data in terms of its population objective.
	In principle, one could estimate such an assignment using any estimation method for treatment effects $\tau(x)$ and variances $\sigma^2(x)$ on the training data. In this and the following section, we study how to optimize the sample analog of the criterion in \autoref{prop:oracle} in the training data.
	Specifically, our goal is now to find an assignment $\hat{a}$ based on $n$ observations $(Y_i,W_i,X_i)_{i=1}^n$ from the training sample that approximately maximizes the population objective \eqref{eqn:populationobj}.
	
	A popular approach in data science is recursive partitioning via binary decision trees \citep{athey2016recursive}. Given the success of greedy tree-based heuristics in the statistical estimation literature, we focus on such heuristic solutions to this maximization problem. In particular, we define a greedy tree construction algorithm that at every step investigates the expansion of each leaf into two child nodes by splitting on a particular feature and at a particular threshold. Subsequently, we investigate the combination of the two possible treatment configurations of the two candidate child nodes and evaluate the empirical analog
	\begin{align}
		\label{eqn:empiricaltstat}
		\widehat{T}(a) &= \frac{\frac{1}{n} \sum_{i=1}^n a(X_i) \: \tilde{Y}_i}{ \sqrt{\frac{1}{n} \sum_{i=1}^n a^2(X_i) \: \tilde{Y}_i^2 - \left(\frac{1}{n} \sum_{i=1}^n a(X_i) \: \tilde{Y}_i\right)^2}}
		&
		&\text{for}
		&
		\tilde{Y}_i
		&=
		\left(\frac{W_i}{p} - \frac{1-W_i}{1-p}\right) (Y_i - \hat{c}(X_i))
	\end{align}
	of the population objective from \eqref{eqn:populationobj} for each candidate assignment $a$, choosing the candidate split that maximizes the statistic.
	This criterion corresponds to the training-sample equivalent of the hold-out test statistic.
	We note that unlike typical tree criteria, our $t$-statistic criterion cannot be solely computed based on the observations in a given node, since the standard deviation in the denominator is a global property. Thus at every candidate split we need to use the whole assignment policy to evaluate the performance of a new split. The details of our heuristic are presented in \autoref{alg:main}.
	
	\begin{algorithm}[htpb]
		\SetKw{Continue}{continue}
		\textbf{Input:} $n$ training observations $(X_i, Y_i - \hat{c}(X_i), W_i)_{i=1}^{n}$ of $J$-dimensional feature vector $X_i\in \R^J$, centered outcome $Y_i - \hat{c}(X_i) \in \R$, and treatment assignment $W_i\in \{0, 1\}$
		
		\textbf{Input:} minimum score increase $\epsilon_{\min}$, maximum depth $d_{\max}$, minimum number of leaf observations $n_{\min}$
		
		\medskip
		
		Set $\tilde{Y}_i = \frac{W_i - p}{p (1-p)} (Y_i - \hat{c}(X_i))$ for each $i \in \{1,\ldots, n\}$\;
		Calculate initial $t$-statistic $T_0 = \sfrac{\frac{1}{n}\sum_{i=1}^n 
			\tilde{Y}_i}{\sqrt{\frac{1}{n}\sum_{i=1}^n 
				\tilde{Y}^2_i - (\frac{1}{n}\sum_{i=1}^n 
				\tilde{Y}_i)^2}}$\;
		Add root node $(id = 0, N = \{1,\ldots,n\}, d = 0, w = \Ind(T_0 > 0))$ to queue\;
		Initialize $T_* = T_0 \: \Ind(T_0 > 0)$, $n_{\textnormal{nodes}}=1$, $a_i = a_0$ for each $i \in \{1,\ldots, n\}$\;
		
		\While{queue is not empty}{
			Pop a node $(id, N, d, w)$ from the head of the queue\;
			If $d \geq d_{\max}$: \Continue\;
			Set $\textit{split}=\emptyset$\;
			\For{$j \in \{1,\ldots, J\}$}{
				Draw random set of thresholds $\{\theta_1,\ldots, \theta_{\mtry}\}$ uniformly without replacement from $\{X_{i,j}\}_{i\in N}$\;
				\For{$\theta \in \{\theta_1,\ldots, \theta_{\mtry}\}$}{
					Set $N_{\ell}=\{i\in N; X_{i,j} < \theta\}$ and $N_r = \{i\in N; X_{i,j}\geq \theta\}$\;
					If $|N_\ell|\leq n_{\min}$ and $|N_r| \leq n_{\min}$: \Continue\;
					\For{$c\in \{\ell, r\}$}{
						Initialize candidate assignment policy $\tilde{a} = a$\;
						Update candidate policy $\tilde{a}_{i} = \Ind(i \in N_c)$ for each $i \in N$\;
						Set $\tilde{T} = \sfrac{\frac{1}{n}\sum_{i=1}^n 
							\tilde{a}_i \tilde{Y}_i}{\sqrt{\frac{1}{n}\sum_{i=1}^n 
								\tilde{a}_i \tilde{Y}^2_i - (\frac{1}{n}\sum_{i=1}^n 
								\tilde{a}_i \tilde{Y}_i)^2}}$\;
						\If{$\tilde{T} \geq T_* + \epsilon_{\min}$}{
							Set $T_* = \tilde{T}$ and $a = \tilde{a}$\;
							Set $\textit{split}=\{(n_{\textnormal{nodes}}, N_\ell, d+1, \Ind(\ell=c)), (n_{\textnormal{nodes}} + 1, N_r, d+1, \Ind(r=c))\}$\;
							Set $\textit{feature}(id)=j$ and $\textit{threshold}(id)=\theta$\;
							Set $\textit{left}(id) = n_{\textnormal{nodes}}$ and $\textit{right}(id)=n_{\textnormal{nodes}}+1$\;
						}
					}
				}
			}
			
			\uIf{$\textit{split}=\{N_\ell, N_r\}$}{
				Append $N_\ell$ and $N_r$ at the end of the queue and set $n_{\textnormal{nodes}} = n_{\textnormal{nodes}} + 2$\;
			}
			\Else{ 
				Set $\textit{assign}(id) = w$\;
			}
		}
		
		\medskip
		
		\textbf{Output:} An assignment policy $\hat{a}: \X\to \{0,1\}$ defined by the constructed binary tree as expressed by the $\textit{split}, \textit{feature}, \textit{left}, \textit{right}, \textit{assign}$ mappings
		
		\caption{Significance Tree}\label{alg:main}
	\end{algorithm}
	
	\FloatBarrier
	
	\section{Approximate Solutions Based on Submodular Minimization}
	\label{sect:submodular}
	
	In this section, we return to the objective formalized in \autoref{sect:objective}, and outline an implementation based on submodular optimization.
	In order to do so, we first derive an alterative representation of an empirical analog to the objective from \autoref{prop:oracle}. We then provide submodular optimization solutions separately for finite- and infinite-valued covariates. Throughout, our results remain applicable at the given sample size $n$ of the training sample, and not only asymptotically for the local-to-zero regime of \autoref{subsect:oracle}.
	
	\subsection{Empirical objective}
	\label{subsect:empirical_and_population_objectives}
	
	Again considering centered outcomes $Y_i - \hat{c}(X_i)$ with centering function $c$ (which can itself be estimated using cross-fitting on the training sample), 
	we now solve for an assignment by maximizing the empirical criterion
	\begin{align}
		\label{eqn:empiricalobj}
		\widehat{V}(a) &= \frac{\frac{1}{n} \sum_{i=1}^n a(X_i) \: \tilde{Y}_i}{\sqrt{\frac{1}{n} \sum_{i=1}^n a^2(X_i) \: \tilde{Y}_i^2}},
		&
		\tilde{Y}_i &= \left(\frac{W_i}{p} - \frac{1-W_i}{1-p}\right) (Y_i - \hat{c}(X_i))
	\end{align}
	in the training sample.
	In order to see the connection of the empirical objective from this section to the population objective from \autoref{subsect:oracle}, note that maximizing \eqref{eqn:empiricalobj} is equivalent to maximizing
	\begin{align*}
		\left(\frac{1}{\widehat{V}^{-2}(a) - 1}\right)^{\sfrac{1}{2}}
		&=
		\left(\frac{\left(\frac{1}{n} \sum_{i=1}^n a(X_i) \: \tilde{Y}_i\right)^2}{\frac{1}{n} \sum_{i=1}^n (a(X_i) \: \tilde{Y}_i)^2 - \left(\frac{1}{n} \sum_{i=1}^n a(X_i) \: \tilde{Y}_i\right)^2}\right)^{\sfrac{1}{2}}
	\end{align*}
	among all assignments with $\sum_{i=1}^n a(X_i) \: \tilde{Y}_i > 0$ (assuming that there is at least one positive $\tilde{Y}_i$), where maximizing the last expression is then also equivalent to maximizing $\widehat{T}(a)$ from \eqref{eqn:empiricaltstat} in the training sample.
	
	To understand the properties of maximizing the empirical objective \eqref{eqn:empiricalobj}, it will be helpful to define the population analog
	\[
	V(a) = \frac{\E[a(X) \tau(X)]}{\sqrt{\E[a(X) \: \varsigma^2(X)]}}
	\]
	where $\varsigma^2_c(x) = \E[\tilde{Y}^2|X=x]$ for $\tilde{Y} =\left(\frac{W}{p} - \frac{1-W}{1-p}\right) (Y - c(X))$ and we write $\varsigma^2(x) = \varsigma_{\hat{c}}^2(x)$.
	To connect this objective to its empirical analog, the following properties will later be helpful.
	
	\begin{lemma}\label{lem:basic-props}
		Under standard regularity conditions, we have
		\begin{align*}
			\tau(x)&= \E[\tilde{Y}_i| X_i=x] = \E[Y_i^{(1)} - Y_i^{(0)}| X_i=x]\\
			\varsigma^2_c(x) &= \E[\tilde{Y}_i^2 | X_i=x] = \sigma(x)^2 + \tau(x)^2,\\
			&= \frac{\Var(Y^{(1)}| X=x)}{p} + \frac{\Var(Y^{(0)}| X=x)}{1-p} + \E^2[Y^{(1)} - Y^{(0)}| X=x] + \frac{(c(x) - c^*(x))^2}{p(1-p)}\\
			& =\varsigma^2_{*}(x) + \frac{(c(x) - c^*(x))^2}{p(1-p)} \text{ for } \varsigma^2_{*}(x) = \varsigma^2_{c^*}(x).
		\end{align*}
		Moreover, maximizing $V(a)$ is equivalent to maximizing $\frac{\E[a(X) \tau(X)]}{\sqrt{\E[a(X) \sigma^2(X)] + \Var(a(X) \tau(X))}}$.
	\end{lemma}
	
	\begin{proof}
		See \autoref{sect:proof_lem_basic-props}.
	\end{proof}
	
	Note here that the objective $\frac{\E[a(X) \tau(X)]}{\sqrt{\E[a(X) \sigma^2(X)] + \Var(a(X) \tau(X))}}$ for binary $a(x)$ only differs from the population objective in \autoref{prop:oracle} by the $\Var(a(X) \tau(X))$ term in the denominator, which is of lower order in the local-to-zero asymptotic approximation from \autoref{subsect:oracle}.

	\subsection{A submodular minimization solution for finite-valued covariates}
	\label{subsect:appxopt}
	
	We now consider the problem of minimizing the empirical objective from \autoref{eqn:empiricalobj} when the covariates only take on a finite number of values.
	Observe that for a finite set $\X$, the empirical optimization problem of maximizing $\widehat{V}(a)$ is equivalent to finding a subset $S\subseteq\X$ that maximizes $\frac{\sum_{x\in S} w(x)}{\sqrt{\sum_{x\in S} v(x)}}$ for $v(x)= \frac{1}{n} \sum_{i \in \{1,\ldots,n\}: X_i=x} \tilde{Y}_i^2 \geq 0$ and $w(x)=\frac{1}{n} \sum_{i\in \{1,\ldots,n\}: X_i=x} \tilde{Y}_i$ (where we assume that the value if the denominator vanishes). Without loss of generality, when solving the empirical maximization problem, we can discard features $x\in \X$ for which $w(x) \leq 0$ (which includes all those with $v(x) = 0$). Thus when solving the optimization problem, it suffices to consider the case when $v(x), w(x)>0$, for all $x \in \X$. 
	
	The problem of maximizing $\frac{\sum_{x\in S} w(x)}{\sqrt{\sum_{x\in S} v(x)}}$ over $S\subseteq \X$ with $w(x), v(x) >0 $ for all $x \in \X$ is equivalent to minimizing the ratio of a monotone submodular function over a monotone modular function, noting that we converted our original maximization problem to a minimization problem by taking the reciprocal of the objective function. Corollary 3.3 of \cite{bai2016algorithms} shows that the latter can be solved up to a $(1+\epsilon)$-approximation with $O\left(\log(1/\epsilon)\right)$ calls to a submodular minimization process which can be solved in polynomial time \citep{schrijver2000combinatorial}; this process minimizes $\sqrt{\sum_{x\in S} v(x)} - \lambda \sum_{x\in S} w(x)$, which is also submodular, for some appropriately chosen $\lambda$ in each call. We note however that the result in \cite{bai2016algorithms} requires solving a constrained submodular minimization, subject to the constraint of a non-empty solution. This constraint is non-trivial and not the standard submodular minimization problem solved by existing solvers (covering constraints can render submodular minimization NP-hard). In \autoref{alg:submod-finite} and \autoref{prop:finite-x} below, we show how to modify the result and the algorithm in \cite{bai2016algorithms} to only require an unconstrained submodular minimization oracle.
	
	\begin{algorithm}
		\SetKw{Continue}{continue}
		\textbf{Input:} A finite set of $k$ elements $Q$\;
		\textbf{Input:} Two positive value functions $v: Q\to \R_+$ and $w:Q \to \R_+$\;
		\textbf{Input:} Approximation target $\epsilon>0$\;
		
		\medskip
		
		For any $S\subseteq Q$, let $v(S)= \sum_{x\in S}v(x)$ and $w(S)=\sum_{x\in S} w(x)$\;
		Set $\lambda_{\max} = (1 + \epsilon)\, \frac{\sqrt{v(Q)}}{w(Q)}$ and $\lambda_{\min}= \frac{\sqrt{\min_{x\in Q} v(x)}}{w(Q)}$\;
		\While{$\lambda_{\max} \geq (1+\epsilon) \lambda_{\min}$}{
			Set $\bar{\lambda}=(\lambda_{\max} + \lambda_{\min})/2$ and solve unconstrained submodular minimization problem
			\[
			\hat{S} = \argmin_{S\subseteq \{1,\ldots,m\}} \sqrt{v(S)} - \bar{\lambda} w(S)
			\]
			{\bf If} $\hat{S}=\emptyset$ or $\frac{\sqrt{v(S)}}{w(S)} \geq \bar{\lambda}$ set $\lambda_{\min} = \bar{\lambda}$ {\bf else} set $\lambda_{\max} =\bar{\lambda}$\;
		}
		
		\medskip
		
		\textbf{Output:} Return set $\hat{S} = \argmin_{S\subseteq Q} \sqrt{\sum_{x \in S} v(x)} - \lambda_{\max} \sum_{x \in S} w(x)$
		\caption{Finding Significant Subgroups via Submodular Minimization for Finite $\X$}\label{alg:submod-finite}
	\end{algorithm}
	
	\begin{proposition}\label{prop:finite-x}
		Let $\hat{S}$ be the output of \autoref{alg:submod-finite} with inputs $v(x)= \frac{1}{n} \sum_{i \in \{1,\ldots,n\}: X_i=x} \tilde{Y}_i^2\geq 0$ and $w(x)=\frac{1}{n} \sum_{i\in \{1,\ldots,n\}: X_i=x} \tilde{Y}_i$, and $Q = \X_+=\{x\in \X; w(x)>0\}$. Assume that $\X_+ \neq \emptyset$ and let $\hat{a}$ be the policy implied by $\hat{S}$, i.e. $\hat{a}(x) = \Ind(x\in \hat{S})$. Then $\hat{S}$ and $\hat{a}$ satisfy that
		\begin{align*}
			\frac{\sum_{x \in \hat{S}} w(x)}{\sqrt{\sum_{x \in \hat{S}} v(x)}} &\geq \frac{1}{1+\epsilon} \max_{\emptyset \neq S \subseteq \X_+} \frac{\sum_{x \in S} w(x)}{\sqrt{\sum_{x \in S} v(x)}} & &\Longleftrightarrow & \widehat{V}(\hat{a}) &\geq \frac{1}{1+\epsilon} \max_{a: \X\to \{0,1\}} \widehat{V}(a).
		\end{align*}
		Moreover, the algorithm will terminate after $O\left(\log(1/\epsilon) + \log(v(Q)/\min_{x\in Q}v(x))\right)$ iterations. 
	\end{proposition}
	
	\begin{proof}
		See \autoref{sect:prop_finite-x}.
	\end{proof}
	
	\autoref{prop:finite-x} provides an approximation guarantee on the empirical criterion. However, for finite $\X$ of size $k$, a simple uniform deviation bound implies that the empirical objective converges to the population objective for any assignment $a: \X\to \{0,1\}$, i.e. w.p. $1-\delta$,
	\[
	\max\left(
	\sup_{a} \left| \frac{1}{n} \sum_{i=1}^n a(X_i)\ \tilde{Y}_i - \E[a(X) \tau(X)]\right|,
	\sup_{a} \left| \frac{1}{n} \sum_{i=1}^n a(X_i)\ \tilde{Y}_i^2 - \E[a(X)\varsigma^2=(X_i)]\right|
	\right)
	\leq \epsilon(n, \delta)
	\]
	for any fixed centering function $\hat{c}$ and for $\epsilon(n, \delta)= O\left(\sqrt{\frac{k\log(1/\delta)}{n}}\right)$. Moreover, by \autoref{lem:basic-props},
	\[
	\varsigma^2(x) - \varsigma^2_{*}(x) = \frac{(\hat{c}(x) - c^*(x))^2}{p(1-p)}.
	\]
	In fact, we can invoke tighter uniform deviation bounds to obtain the following proposition.
	
	\begin{proposition}\label{prop:finite-x-pop}
		In addition to the assumptions of \autoref{prop:finite-x}, assume that the centering function $\hat{c}$  is trained on a separate sample, that $k = |\X| < \infty$, that $\varsigma^2_{*}(x)\geq \underline{\varsigma}^2$ and $\P(X=x)>\underline{\pi}$ for all $x\in \X$, and that $|\tilde{Y}|\leq H$ a.s.\ for some constant $H$.
		Let $a_*$ be any policy (e.g. the policy maximizing the population objective $V$) and
		\begin{align*}
			\kappa(n,\eta,\delta) = \sqrt{\frac{2\,k\, \log(2/\delta)}{n}} + \frac{7\,H\,k\,\log(2/\delta)}{3(n-1)} + \frac{1}{2(1+\eta)^{3/2} \underline{\pi}^{3/2}\, \underline{\varsigma}^3} \left(\frac{\E[(\hat{c}(X) - c^*(X))^2]}{p(1-p)}+ \frac{2 H\, k\log(2/\delta)}{\eta\,n}\right).
		\end{align*}
		Then $\hat{a}$ satisfies w.p. $1-\delta$:
		\begin{align*}
			\frac{\E[\hat{a}(X) \tilde{Y}]}{\sqrt{\E[\hat{a}(X)\varsigma^2_{*}(X)]}} \geq \sqrt{\frac{1-\eta}{1+\eta}} \frac{1}{1+\epsilon} \frac{\E[a_*(X) \tilde{Y}]}{\sqrt{\E[a_*(X) \varsigma^2_{*}(X)]}} - 2\kappa(n,\eta,\delta)
		\end{align*}
		Moreover, the algorithm will terminate after $O\left(\log(1/\epsilon) + \log\left(\sfrac{\sum_{x\in\X^*}v(x)}{\min_{x\in \X^*}v(x)}\right)\right)$ iterations. 
	\end{proposition}
	
	\begin{proof}
		See \autoref{sect:prop_finite-x-pop}.
	\end{proof}
	
	Thus, for finite $\X$, we can approximate the optimal assignment arbitrarily well with respect to our objective function. Moreover, the latter function is a special case of a submodular function that can be solved much more efficiently than generic submodular minimization \citep{Stobbe2010}.
	
	This reduction also draws strong connections between our approach and the variance penalization literature in policy learning. In particular, \cite{swaminathan2015counterfactual} proposes that one maximizes an objective of the form (after bringing it to our notation)
	\[
	\sum_{x\in \X} a(x) w(x) - \lambda_n \left(\sum_{x\in \X} a(x) v(x) - \left(\sum_{x\in \X} a(x) w(x)\right)^2\right)
	\]
	with a vanishing $\lambda_n$ at a roughly $O(n^{-1/2})$ rate when $a$ belongs to a VC class. We see that unlike variance penalization we balance the standard deviation with the mean policy outcome and we do not necessarily have a vanishing penalty $\lambda$. Even closer to our objective is the second moment penalization in \cite{foster2019orthogonal}, which recommends maximizing
	\[
	\sum_{x\in \X} a(x) w(x) - \lambda \sum_{x\in \X} a(x) v(x).
	\]
	Again this objective has the same discrepancy with our objective in that we balance the root of the second moment and we do not necessarily have a vanishing $\lambda$. Moreover, in our objective the $\lambda$ is auto-tuned so as to minimize the empirical $t$-statistic.
	
	\subsection{Submodular minimization for infinite or exponentially-sized covariate spaces}
	
	Above, we considered maximizing the empirical objective \eqref{eqn:empiricalobj} for finite $\X$. 
	However, we in many applications $\X$ may not be finite, but either uncountable or exponential in some implicit representation.
	For instance, if $x$ is represented by $J$ binary features, then $|\X|=2^J$.
	Moreover, typically, $x$ will be represented by $J$ continuous features.
	In this setting we do not only need to find an appropriate subset for a set of finite pre-defined subgroups, but we also need to construct how these subgroups are defined. In other words, we need to partition the $\X$ space in a finite set of partitions and then find the optimal assignment with respect to this partition. However, the choice of partitions also needs to happen in a manner that will lead to a large $t$-statistic.
	
	As a first step towards maximizing our target objective for complex $\X$, we observe that it generally suffices to consider assignments that treat observations with covariates $x$ equivalently if neither their mean treatment effect $\tau(x)$ nor $\varsigma^2(x)$ changes.
	
	\begin{lemma}\label{lem:partition-exact}
		There exists an optimal policy such that all elements in $\X$ with both $\tau(x)=\tau$ and $\varsigma^2(x)=\varsigma^2$, for some $\tau$ and $\varsigma^2 \geq 0$, are either all assigned or not assigned to the subgroup.
	\end{lemma}
	
	\begin{proof}
		See \autoref{sect:proof_lem_partition-exact}.
	\end{proof}
	
	Specifically, if we can find a partition $\X = P_1 \cup \ldots \cup P_k$ such that for all $\ell \in \{1,\ldots,k\}$ and for all $x,x'\in P_i$ we have that $\tau(x)=\tau_\ell = \tau(x')$ and $\varsigma^2(x)=\varsigma^2_\ell=\varsigma^2(x')$, then it suffices to optimize over assignments $a(x) = r(\pi(x))$, where $\pi(x)$ denotes the index of the partition that $x$ lies in.
	Equivalently, it suffices to identify the subset $S \subseteq \{1,\ldots,k\}$ of partitions to treat, with value $\frac{\sum_{\ell \in S} p_\ell \tau_\ell}{\sum_{\ell \in S} p_\ell \varsigma_\ell^2}$ for $p_\ell = \P(X\in P_\ell)$.
	Subsequently, in that case, we can apply a reduction to submodular minimization for finite elements $\X$, but now with the partitions $\{P_1, \ldots, P_k\}$ taking the role of the finitely many elements of $\X$. Hence, a $(1+\epsilon)$-approximate optimal policy can be found with $O\left(\log(1/\epsilon)\right)$ oracle calls to a submodular minimization oracle for decomposable submodular functions, as in \autoref{prop:finite-x}.
	
	We now expand this idea of selecting groups of covariates with the same $\tau(x)$ and $\varsigma^2(x)$ to selecting groups of units that are similar in terms of their $\tau(x)$ and $\varsigma^2(x)$.
	Unlike a policy that defines similarity based on the full covariate vector $x$, such an assignment only relies on similarity in the low-dimensional space of $\tau(x)$ and $\varsigma^2(x)$.
	
	\begin{lemma}\label{lem:partition-approx}
		Let $\X = P_1 \cup \ldots \cup P_k$ be a partition of the space $\X$ such that for all $\ell \in \{1,\ldots,k\}$ and for all $x,x'\in P_i$ we have that
		\begin{align*}
			\sqrt{p_\ell}\, |\tau(x) - \tau(x')|&\leq \epsilon, & 
			\sqrt{p_\ell}\, |\varsigma^2(x) - \varsigma^2(x')|&\leq \epsilon,
		\end{align*}
		where $p_\ell = \P(X\in P_\ell)$.
		For all $\ell \in \{1,\ldots,k\}$ let $\tau_\ell=\E[\tau(X)| X\in P_\ell]$ and $\varsigma_\ell^2=\E[ \varsigma^2(X)| X\in P_\ell]$.
		Suppose that $a_*$ is an optimal policy, that $\min_{\ell\in \{1,\ldots,k\}} \varsigma_\ell^2 \geq \underline{\varsigma}^2$, and that $\max_{\ell\in \{1,\ldots,k\}} \sfrac{\tau_\ell}{\varsigma_\ell^2} \leq \bar{\kappa}$.
		Then there exists an assignment policy $a_{\epsilon}$ that for every $\ell \in \{1,\ldots,k\}$ assigns the same treatment to all $x\in P_\ell$ and that satisfies $\widehat{V}(a_{\epsilon}) \geq \widehat{V}(a_*) - \frac{2\,k}{\underline{\varsigma}} \left(1 + \bar{\kappa}\right)\epsilon$.
	\end{lemma}
	
	\begin{proof}
		See \autoref{sect:proof_lem_partition-approx}.
	\end{proof}
	
	Under the assumptions of the proposition, the ability to compute optimal assignments as in \autoref{subsect:appxopt} thus extends to the case where $\tau(x)$ and $\varsigma^(x)$ are only approximately constant within sets $P_\ell$.
	Specifically, \autoref{lem:partition-approx} implies that we can again apply a reduction to submodular minimization for finite elements $\X$, and translate a feasible approximate optimal policy for that case to one that still provides a good solution for maximizing $\widehat{V}$ in the case of more complex $\X$.
	
	As a concrete application of our framework, we now focus on the case when $\X=\{0,1\}^J$. Here, we assume that both $\tau(x)$ and $\varsigma^2_{*}(x)$ are $r$-sparse functions, in the sense that they can be written as $\tau(x) = \tilde{\tau}(x_R)$ and $\varsigma^2_{*}(x) = \tilde{\varsigma}^2(x_R)$ for some subset $R\subseteq \{1,\ldots,J\}$ (where $x_R = (x_i)_{i\in R}$), with $|R|=r$. Furthermore, for simplicity of exposition we assume that $\tau(x), \varsigma^2_{*}(x) \in [-\sfrac{1}{2},+\sfrac{1}{2}]$ and that $\tilde{Y}_i \in [-1, +1]$. Our results can trivially be extended to any finite bound $H$ on the random variables at an extra multiplicative cost of $H$ in our bounds.
	
	Importantly, we assume that the variables in $R$ are $\beta$-strictly relevant, in the sense that the variance reduction incurred by conditioning on one of the variables in $R$ is larger than conditioning on any variable in $\{1,\ldots,J\} \setminus R$, by an additive factor $\beta$ . This condition is formally expressed in the following definition.
	
	\begin{definition}
		For any function $g: \{0,1\}\to [-\sfrac{1}{2},+\sfrac{1}{2}]$, let $W(S;g) = \E[\E[g(X)| X_S]^2]$. We say that the function $g$ satisfies the $(\beta,r)$-strong sparsity condition if there exists a set $R$ of size $|R|=r$, such that, for all $i \in R$ and $j \in \{1,\ldots,J\} \setminus R$ and $S \subseteq \{1,\ldots,J\} \setminus \{i\}$
		\[
		W(\{i\}\cup S;g) - W(S;g) \geq W(\{j\} \cup S; g) - W(S;g) + \beta.
		\]
	\end{definition}
	
	We now apply the above sparsity condition to the functions $\tau(x)$ and $\varsigma^2_{*}(x)$.
	We have the following corollary of the results in \cite{syrgkanis2020estimation} (in particular Theorem~C.10.2).
	
	\begin{lemma}\label{lem:subgroup-id}
		Suppose that both $\tau(x)$ and $\varsigma^2_{*}(x)$ satisfy the $(\beta, r)$-strong sparsity condition.
		Suppose that for any $x_R\in \{0, 1\}^r$, $\P(X_R=x_R)\geq \zeta / 2^r$ for some $\zeta > 0$ and $n=\Omega\left( \frac{2^r\log(J/\delta)}{\beta^2} + \frac{2^r\log(1/\delta)}{\zeta}\right)$.
		Moreover, suppose that the centering function $\hat{c}$ is estimated based on a separate sample that is large enough for $\hat{c}$ to satisfy $\E[(\hat{c}(X) - c^*(X))^2] \leq \frac{\beta\,p\,(1-p)}{2}$. Then a greedy level-split regression tree trained on labels $\tilde{Y}_i = (Y_i - \hat{c}(X_i))\left(\frac{W_i}{p} - \frac{1-W_i}{1-p}\right)$ based on the Breiman criterion and with depth $d\geq r$ satisfies w.p. $1-\delta$ that every $x, x'$ that fall in the same leaf of the tree fulfil $\tau(x)=\tau(x')$. Similarly, a greedy level-split regression tree trained on labels $\tilde{Y}_i^2$ based on the Breiman criterion and with depth $d\geq r$ satisfies w.p. $1-\delta$ that every $x, x'$ that fall in the same leaf of the tree fulfil $\varsigma^2_{*}(x)=\varsigma^2_{*}(x')$.
	\end{lemma}
	
	\begin{proof}
		See \autoref{sect:proof_lem_subgroup-id}.
	\end{proof}
	
	Thus by taking the intersection of the partitions defined by two greedy decision trees with depth $d=r$, we guarantee the property required by \autoref{lem:partition-exact}. This partition can also be thought of as the leaves of a decision tree with depth $d=2\, r$.
	
	Subsequently, for any policy value defined on the leaves of a binary tree of depth $d$, adaptively chosen based on the data, we want to understand the concentration of the numerator and denominator of our objective $\widehat{V}$. If we let $A$ denote the space of binary policies defined on the leaves of an adaptively built binary decision tree of depth $d$, then we need to control the uniform deviations, that is, we want w.p.\ $1-\delta$,
	\[
	\max\left(
	\sup_{a \in A} \left| \frac{1}{n} \sum_{i=1}^n a(X_i)\ \tilde{Y}_i - \E[a(X) \tau(X)]\right| \leq \epsilon(n, \delta),
	\sup_{a \in A} \left| \frac{1}{n} \sum_{i=1}^n a(X_i)\ \tilde{Y}_i^2 - \E[a(X)\varsigma^2_c(X_i)]\right|\right) \leq \epsilon(n, \delta).
	\]
	Assuming that $\tilde{Y}_i$ is bounded in $[-1, +1]$, then by standard results in statistical learning theory, both of the above deviations are satisfied with $\epsilon(n,\delta) = O\left(\sqrt{\frac{\textnormal{VC}(A)}{n}} + \sqrt{\frac{\log(1/\delta)}{n}}\right)$ \citep{shalev2014understanding}, where $\textnormal{VC}(A)$ is the VC-dimension of policy space $A$. By existing results in statistical learning theory, the VC dimension of a binary classification tree of depth $d$ with $J$ features is $O\left(d\, 2^d \log(J)\right)$ (see \cite{Mansour97pessimisticdecision} for the first claim of this fact and e.g. Section A.2 of \cite{syrgkanis2020estimation} for a proof, or Lemma~4 of \cite{zhou2018offline}, or \cite{Leboeuf2020} for a more extensive treatment). This yields the following lemma.
	
	\begin{lemma}\label{lem:vc-bound}
		Let $A$ denote the space of binary policies defined on the leaves of an adaptively built binary decision tree of depth $d$. Let $\hat{c}$ be a fixed centering function. Let $v=4d2^d + 2^{d+1}\log(2J)$. Assume that $|\tilde{Y}|\leq H$ a.s.. Then w.p. $1-\delta$, for all $a\in A$
		\begin{align*}
			\left|\frac{1}{n} \sum_{i=1}^n a(X_i)\ \tilde{Y}_i - \E[a(X) \tau(X)]\right| \leq \sqrt{\frac{\E_n[a(X)\varsigma^2(X)]\, 2\, v \log(2\,e\,n/\delta)}{n}} + H \frac{7\, v\log(2\,e\, n/\delta)}{3(n-1)},\\
			\left|\frac{1}{n} \sum_{i=1}^n a(X_i)\ \tilde{Y}_i^2 - \E[a(X) \varsigma^2(X)]\right| \leq H\sqrt{\frac{\E[a(X)\tilde{Y}^2]\, 2\,v\, \log(e\,n/\delta)}{n}} + H\frac{v\,\log(2\,e/\delta)}{3n}.
		\end{align*}
	\end{lemma}
	
	\begin{proof}
		See \autoref{sect:proof_lem_vc-bound}.
	\end{proof}
	
	We can now combine these results into a final proposition about the approximation provided by a binary decision tree combined with submodular optimization on sparse data.
	
	\begin{proposition}\label{prop:main-model}
		Suppose that both $\tau(x)$ and $\varsigma^2_{*}(x)$ satisfy the $(\beta, r)$-strong sparsity condition. Suppose that for any $x_R\in \{0, 1\}^r$, $\P(X_R=x_R)\geq \zeta / 2^r$ for some $\zeta > 0$ and $n=\Omega\left( \frac{2^r\log(J/\delta)}{\beta^2} + \frac{2^r\log(1/\delta)}{\zeta}\right)$. Let $\hat{a}$ be the output of \autoref{alg:model-based} with the following inputs:
		$n$ observations $(X_i, Y_i - \hat{c}(X_i), W_i)_{i=1}^n$ with a centering function $\hat{c}$ trained on a separate sample, sparsity bound $r$, and target approximation level $\epsilon$. Moreover, suppose that the centering function $\hat{c}$ satisfies $\E[(\hat{c}(X) - c^*(X))^2] \leq \frac{\beta\,p\,(1-p)}{2}$. Finally, suppose that $\varsigma^2_{*}(x)\geq \underline{\varsigma}^2$ for all $x\in \X$ and that a.s.\ $|\tilde{Y}|\leq H$. Let $v=4 r 2^r + 2^{r+1} \log(2J)$ and
		\[
		\kappa(n,\eta,\delta) = \sqrt{\frac{2\,v\, \log(2\ e\,n/\delta)}{n}} + \frac{7\,H\,v\,\log(2e\,n/\delta)}{3(n-1)} + \frac{2^{3r/2}}{2(1+\eta)^{3/2} \zeta\, \underline{\varsigma}^3} \left(\frac{\E[(\hat{c}(X) - c^*(X))^2]}{p(1-p)}+ \frac{2 H\, v\log(2e\,n/\delta)}{\eta\,n}\right).
		\]
		Let $a_*$ be any policy, such as the one maximizing $V$. Then $\hat{a}$ satisfies w.p. $1-\delta$:
		\[
			\frac{\E[\hat{a}(X) \tilde{Y}]}{\sqrt{\E[\hat{a}(X)\varsigma^2_{*}(X)]}} \geq \sqrt{\frac{1-\eta}{1+\eta}} \frac{1}{1+\epsilon} \frac{\E[a_*(X) \tilde{Y}]}{\sqrt{\E[a_*(X) \varsigma^2_{*}(X)]}} - 2\kappa(n,\eta,\delta)
		\]
		Moreover, the algorithm runs in time polynomial in $\{2^r, \log(1/\epsilon), p, n\}$.
	\end{proposition}
	
	\begin{proof}
		See \autoref{sect:proof_prop_main-model}.
	\end{proof}
	
	Finally, in the latter proposition we note that we can remove the $\log(n)$ factor in the leading term of the error, if the regression trees for learning $\tau(x)$ and $\varsigma^2(x)$ are also trained on a separate sample. This will lead to a leading error term of $\sqrt{2^{r+2}\log(2/\delta)/n}$, which could lead to an improvement for large $n$. 
	
	\begin{algorithm}
		\SetKw{Continue}{continue}
		\textbf{Input:} $n$ observations $\left(X_i, Y_i - \hat{c}(X_i), W_i\right)_{i=1}^{n}$ of features $X_i\in \R^J$, centered outcome $Y_i - \hat{c}(X_i) \in \R$, and treatment $W_i\in \{0, 1\}$\;
		\textbf{Input:} Sparsity bound $r$\;
		\textbf{Input:} Approximation target $\epsilon>0$\;
		
		\medskip
		
		Set $\tilde{Y}_i = \frac{W_i - p}{p (1-p)} (Y_i - \hat{c}(X_i))$ for each $i \in \{1,\ldots, n\}$\;
		Estimate a conditional average treatment effect $\hat{\tau}(x)$ by running a greedy level-splits regression tree construction algorithm with Breiman's criterion \citep[Algorithm~1 of][]{syrgkanis2020estimation} with labels $\tilde{Y}_i$, features $X_i$, and depth $d=r$\;
		Estimate a conditional second moment $\hat{\varsigma}^2(x)$ by running a greedy level-splits regression tree construction algorithm with Breiman's criterion \citep[Algorithm~1 of][]{syrgkanis2020estimation} with labels $\tilde{Y}_i$, features $X_i$, and depth $d=r$\;
		Let $\{P_1, \ldots, P_k\}$ with $k=2^{2r}$ be the partition of $\X$ defined by the intersection of the leaves of the two trees, i.e. $x,x'\in P_\ell$ if and only if $x, x'$ fall in the same leaf on both trees\;
		Apply \autoref{alg:submod-finite} with $Q=\{P_1, \ldots, P_k\}$, target approximation level $\epsilon$, and value functions $v(P_\ell)=\frac{1}{n}\sum_{i \in \{1,\ldots,n\}; X_i\in P_\ell} \tilde{Y}_i^2$ and $w(P_\ell)=\frac{1}{n}\sum_{i\in \{1,\ldots,n\}; X_i\in P_\ell} \tilde{Y}_i$ for all $\ell \in \{1,\ldots, k\}$\;
		Let $S_*$ be the output of the algorithm\;
		
		\medskip
		
		\textbf{Output:} Assignment policy $\hat{a}: \X\to \{0,1\}$ given by $\hat{a}(x) = \Ind(x\in \cup_{\ell \in S_*} P_\ell)$
		
		\caption{Finding Significant Subgroups via Sparse Non-Parametric Model Estimation}\label{alg:model-based}
	\end{algorithm}
	
	\section{Simulation Study}
	\label{sect:simulation}
	
	In this section, we take the empirical implementations outlined in \autoref{sect:significance_trees} and \autoref{sect:submodular}, and evaluate experimentally how they compare to alternative methods in the policy learning and heterogeneous treatment effect literature. We consider two experimental designs. In the first setting, we consider a stylized data-generating process that highlights vividly the regime where our algorithm substantially outperforms alternative approaches. Subsequently, we consider a semi-synthetic experiment where we use the features and treatment assignments from a real-world dataset of an HIV trial \citep{duflo2019hiv}. This was a four-armed randomized trial among youth in western Kenya, assessing the effect of various interventions on herpes type 2 (HSV-2) prevalence. We find that in both settings our significance tree achieves strictly higher or comparable out-of-sample $p$-values, as compared to the treatment subgroups identified by the alternative methods. This holds across multiple signal-to-noise regimes, with the difference being very stark in some conditions.
	
	\subsection{Methods and benchmarks}
	
	We compare the performance of different methods by using them to estimate a subgroup assignment on the training data, and then evaluating this assignment on hold-out data from the same distribution.
	The methods we compare are:
	
	\begin{enumerate}
		\item \textbf{all}: Choose all units for the test.
		
		\item \textbf{cate}: First, we estimate the conditional average treatment effect $\tau(x)$ by $\hat{\tau}(x)$ based on the $R$-learner method \citep[double machine learning for heterogeneous effects,][]{nie2017quasi}, which is one of the most popular methods for heterogeneous treatment effect estimation and has several favorable formal statistical properties \citep{foster2019orthogonal,chernozhukov2018double}. In particular, we use random-forest regression and classification to estimate the residual outcome and residual treatment required in the $R$-learner algorithm. Moreover, we use a single decision tree as a final stage treatment effect model, to make it comparable to the model that is constructed by our policy tree. Subsequently, we select a subset to test by $a(x)=\Ind(\hat{\tau}(x) \geq 0)$.
		
		\item \textbf{cate\_const}: Same as \textbf{cate}, but now instead of estimating a propensity model in the $R$-learner algorithm, we use a constant propensity, since we are in a randomized trial. This could potentially reduce variance injected by propensity estimation.
		
		\item \textbf{evidence}: Our significance tree algorithm but without centering in the training sample, i.e. $\hat{c}(x) \equiv 0$ in \autoref{alg:main}.
		
		\item \textbf{evidenceres}: Our proposed significance tree solution from \autoref{alg:main} with optimal centering as in \autoref{subsect:estimation}.
		
		\item \textbf{policy\_submod}:
		The \textbf{evidenceres} method is used to define the subgroups as the leaves of the tree. Subsequently, submodular minimization (\autoref{alg:submod-finite}) is used to find the truly optimal assignment on the leaves.
		
		\item \textbf{submod}: Our proposed model-based solution outlined in \autoref{alg:model-based}.
		
		\item \textbf{classifier}, \textbf{classifierres}: A very popular policy learning method, where the doubly robust policy value is optimized over the space of tree policies \citep{dudik2011doubly,athey2021policy}. This problem can be framed as a weighted classification problem, where the sign of the doubly robust values for each observation $\tilde{Y}_i =  \left(\frac{W_i}{p} - \frac{1-W_i}{1-p}\right) (Y_i - \hat{c}(X_i))$ can be used as labels $\hat{Y}_i = \sign(\tilde{Y}_i)$, and the absolute value $|\tilde{Y}_i|$ of these values as  weight. We use a decision tree classifier to solve this classification problem. The variant \textbf{classifier} does not perform optimal centering and simply uses $\hat{c}(x)=\frac{1}{n}\sum_i Y_i$ in the training sample. The variant \textbf{classifierres} performs optimal centering as in \autoref{subsect:estimation}.
	\end{enumerate}
	
	All of our methods are judged by calculating the out-of-sample $p$-value of the chosen subgroup assignment $\hat{a}$. In other words, we use the optimally centered policy value $\widehat{V}(\hat{a})$ on the hold-out and transform it into a $p$-value. The centering function $\hat{c}$ for the hold-out is estimated on the training sample using a weighted random forest with loss function as in \eqref{eqn:weightedprediction}. We always estimate $\hat{c}$ this way across all methods, regardless of how the methods use $\hat{c}$ in their training procedure.
	
	\subsection{Simple synthetic setting}
	\label{sect:simple-synth}
	
	We first consider a very simple setting designed to showcase the advantage of our method over alternative approaches.
	We consider the case of a single feature $X$ that is uniformly distributed on $]0, 3]$. The feature space is partitioned in three intervals $]0,1]$, $]1,2]$, $]2,3]$, with parameters varying by interval. In each interval $t$, the value of the outcome can be described by
	\begin{align*}
		Y &= W\cdot \tau_{\lceil X \rceil} + \beta_{\lceil X \rceil} + \epsilon, & \epsilon &\sim N(0, \sigma_{\lceil X \rceil}^2)
	\end{align*}
	where $\lceil X \rceil \in \{1,2,3\}$ denotes the smallest integer not smaller than $X$.
	Here, we use as parameters for the three regions $\tau=(0, 2, 1)$, $\beta=(0, 10, 100)$, and $\sigma=(5, 10, 1)$.
	
	This simple setting has the following characteristics.
	In the first region, the treatment effect is null, and observations from this region only add unnecessary noise to the tested subgroup. In the second region, the treatment effect is very positive, but observations from that region also have high variance (which can be seen as unobserved heterogeneity). Thus even though with many draws we would have clear evidence that observations from this region contribute a positive treatment effect, in a small sample treating this region runs the risk of adding more noise than signal and getting a statistically insignificant result. The final region has a modest treatment effect, but is much less noisy. Hence, including data from this final region should improve the power of a test.
	Moreover, we observe that these three regions operate in baseline outcome regimes that differ by orders of magnitude. Thus without correct centering of the outcomes, noise in the outcome can hurt the power of any of the methods.
	
\begin{figure}[htpb]
		\centering
		\includegraphics[width=\textwidth]{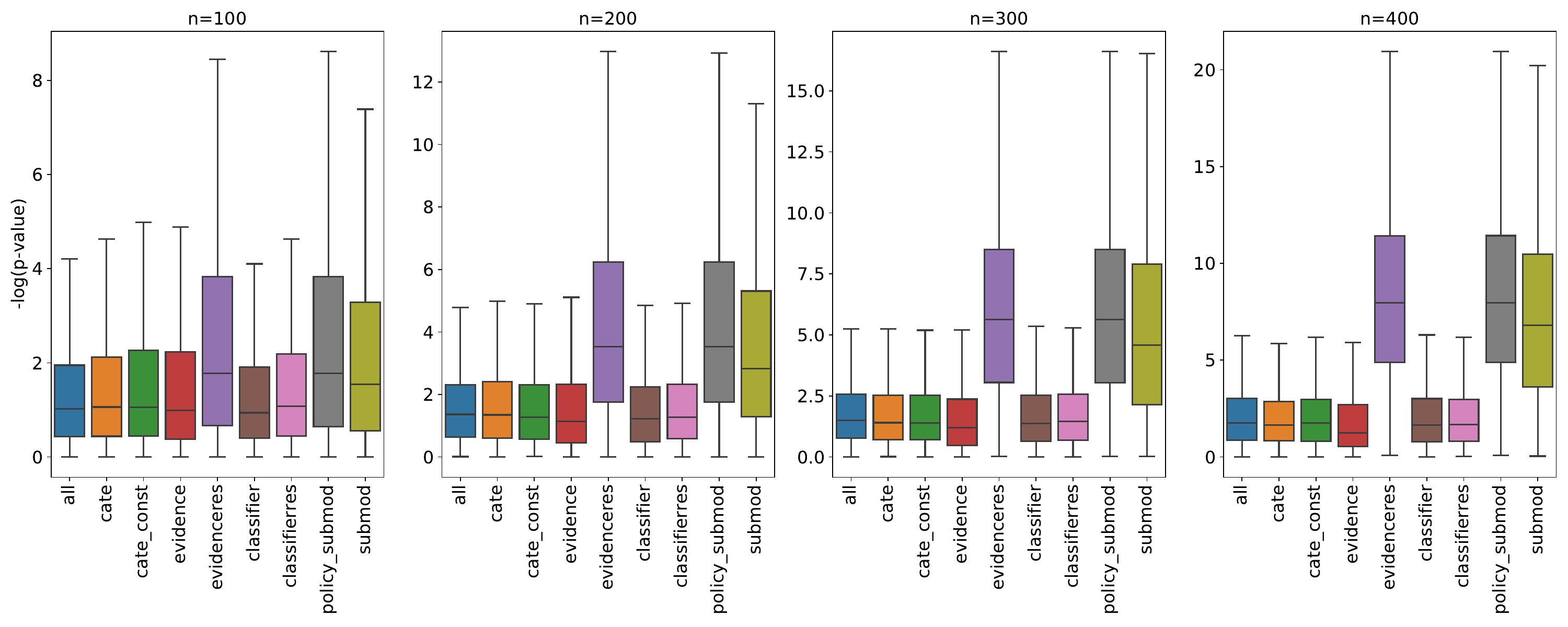}
		\includegraphics[width=\textwidth]{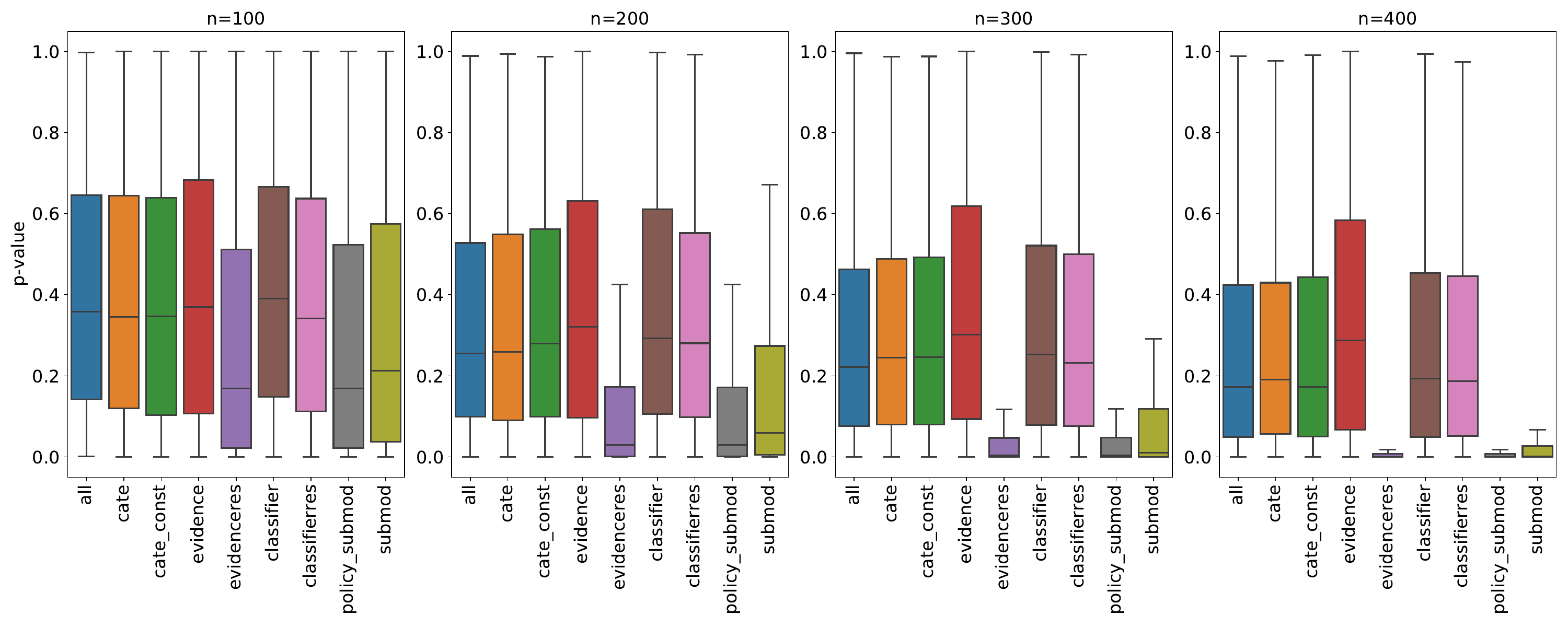}
		\includegraphics[width=\textwidth]{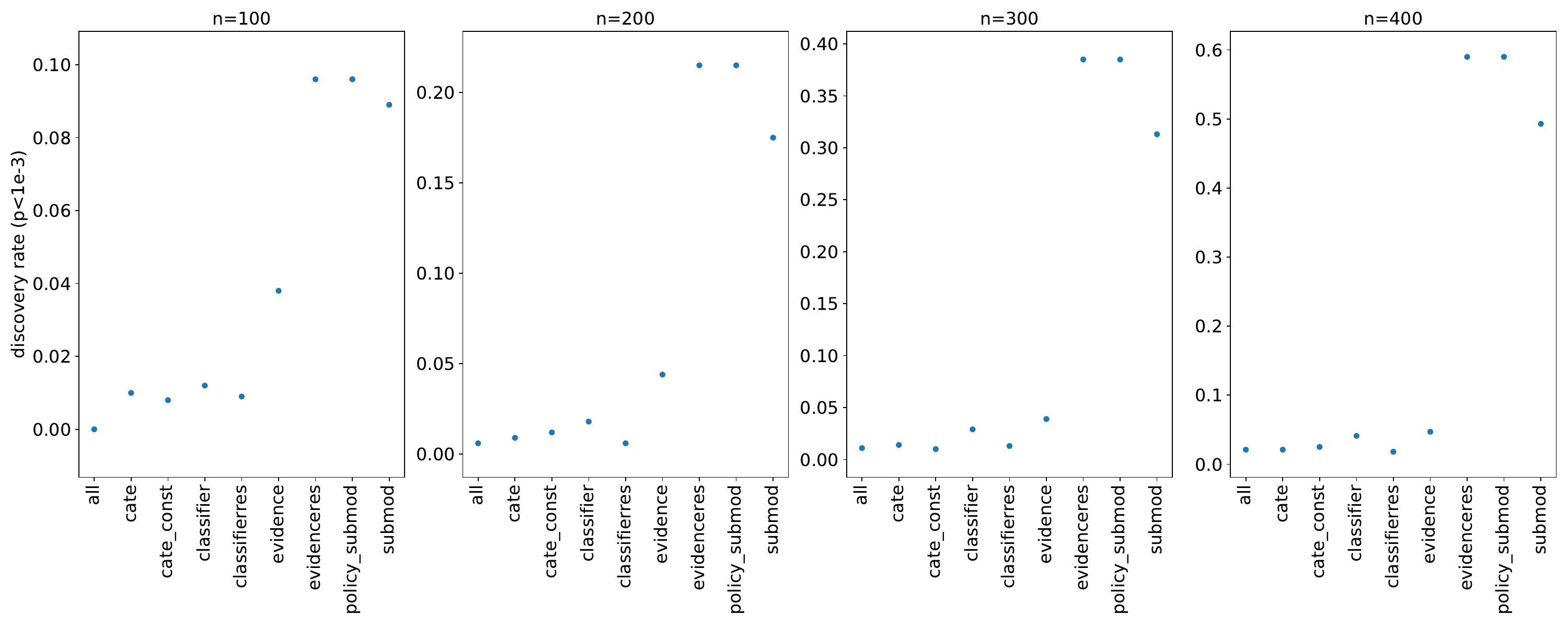}
		\caption{Out-of-sample $p$-value and negative log of $p$-value distributions of each method on $1000$ experiments. The last figure depicts the fraction of $1000$ experiments for which the out-of-sample $p$-value $\leq 1e-3$, which we declare as a discovery.}
		\label{fig:synthp}
	\end{figure}

	In this setting, a successful method should not include the first region, only include the second region when the sample size is large enough, and definitely include the third region in the tested subset.
	Furthermore, centering the outcome should only improve power.
	Among the methods we analyze, only those optimized for the power of the test fulfill these requirements, while standard tools from policy assignment tend to include regions with excess noise.
	In our experiments, we therefore find that the significance tree algorithm with optimal centering \textbf{evidenceres}, which is built to trade off signal and noise effectively, significantly out-performs all other benchmarks for a large range of sample sizes. The results are presented in \autoref{fig:synthp}.
	
	\subsection{Semi-synthetic experiments based on HIV randomized trial}
	\label{subsect:semisynth-binary}
	
	We next present results from a semi-synthetic simulation based on a randomized trial on HIV prevention from \cite{duflo2019hiv}. In the trial, youths in Kenya were randomized into one of four arms. We use one of the three treatment arms as our treatment along with the control arm, and select 2456 units that do not contain missing values in the treatment or outcome. For each unit, we have 44 observable characteristics, which we collect in a vector $X$. The outcome of interest is the prevalence of the herpes simplex virus 2 (HSV-2) two years after the treatment (binary outcome $Y \in \{0, 1\}$).
	Throughout, our simulation is informative about the performance of our method on data similar to that from \cite{duflo2019hiv}, but it does not (and is not designed to) provide any new insight into that study.
	
\begin{figure}
		\centering
		\includegraphics[width=\textwidth]{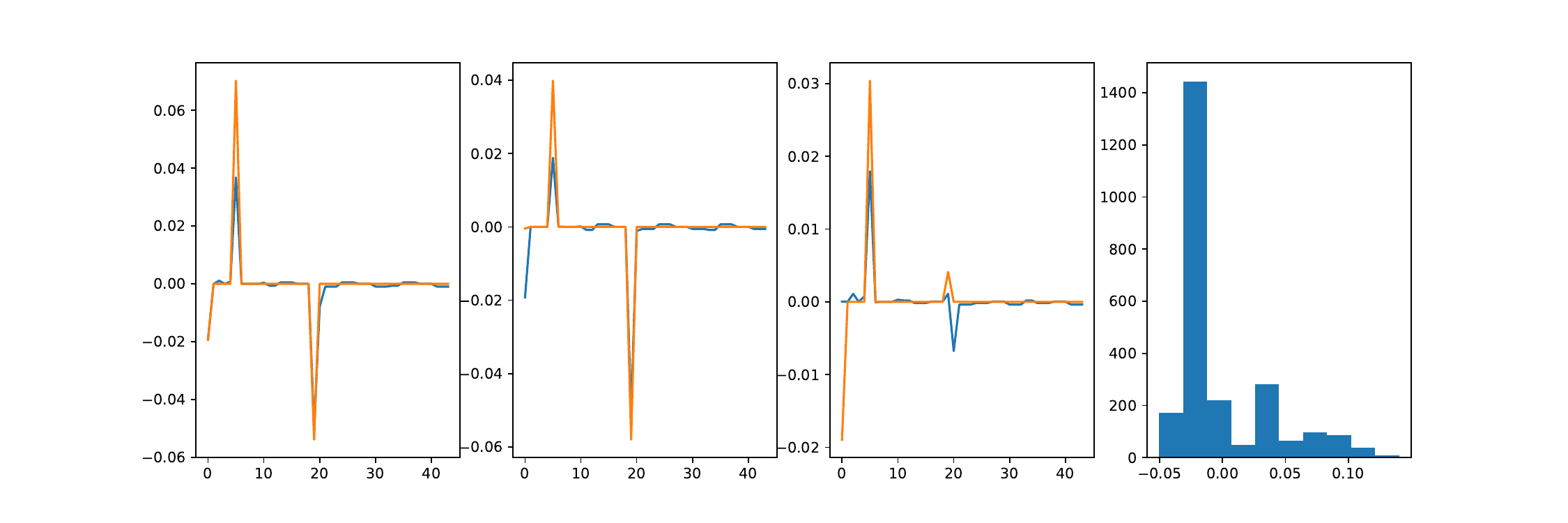}
		\caption{Learned outcome models from the raw data. The first (leftmost) plot is $\theta_1$, second plot is $\theta_0$, third plot is $\theta_1-\theta_0$ and fourth plot is the histogram of the scores $(\theta_1 - \theta_0)' (X_i - \overline{X})$, where $\overline{X}=\frac{1}{n} \sum_{i=1}^n X_i$. Orange lines are the de-biased coefficients after the un-penalized logistic regression. Blue lines are the coefficients from the first-stage penalized model.}
		\label{fig:models}
	\end{figure}

	We analyze the performance of our algorithm in a semi-synthetic experiment. For all 2456 units, we use the raw feature vector $X_i \in \R^{44}$ and treatment indicator $W_i \in \{0,1\}$ from the real data. We then train a $\ell_1$-penalized (sparse) logistic regression outcome model separately on the treated and untreated population in order to learn functions $f_w(x) = \P(Y=1| X=x, W=w) = \expit(\theta_w' x)$ for the logistic sigmoid function $\expit(z) = (1+\exp(-z))^{-1}$. Subsequently, we train un-penalized logistic regressions with the selected regressors. The learned coefficients are depicted in \autoref{fig:models}. We see that there are three significant factors of heterogeneity, with two factors increasing the effect and one factor decreasing the effect. Next, we generate new outcome data from several manipulations of the learned models. In particular, we generate binary outcome data based on the probabilistic models
	\begin{align*}
		\P(Y=1| X=x, W=1) &= \expit\left(\text{cov-strength}\cdot \theta_1' (X_i - \overline{X}) + \text{separation}\right),\\
		\P(Y=1| X=x, W=0) &= \expit\left(\text{cov-strength}\cdot \theta_0' (X_i - \overline{X}) - \text{separation}\right)
	\end{align*}
	where $\overline{X}=\frac{1}{n} \sum_{i=1}^n X_i$. The constant ``cov-strength'' controls the signal/heterogeneity contained in the features, while the constant ``separation'' controls the baseline effect of the treatment.
	
\begin{figure}
		\centering
		\includegraphics[width=\textwidth]{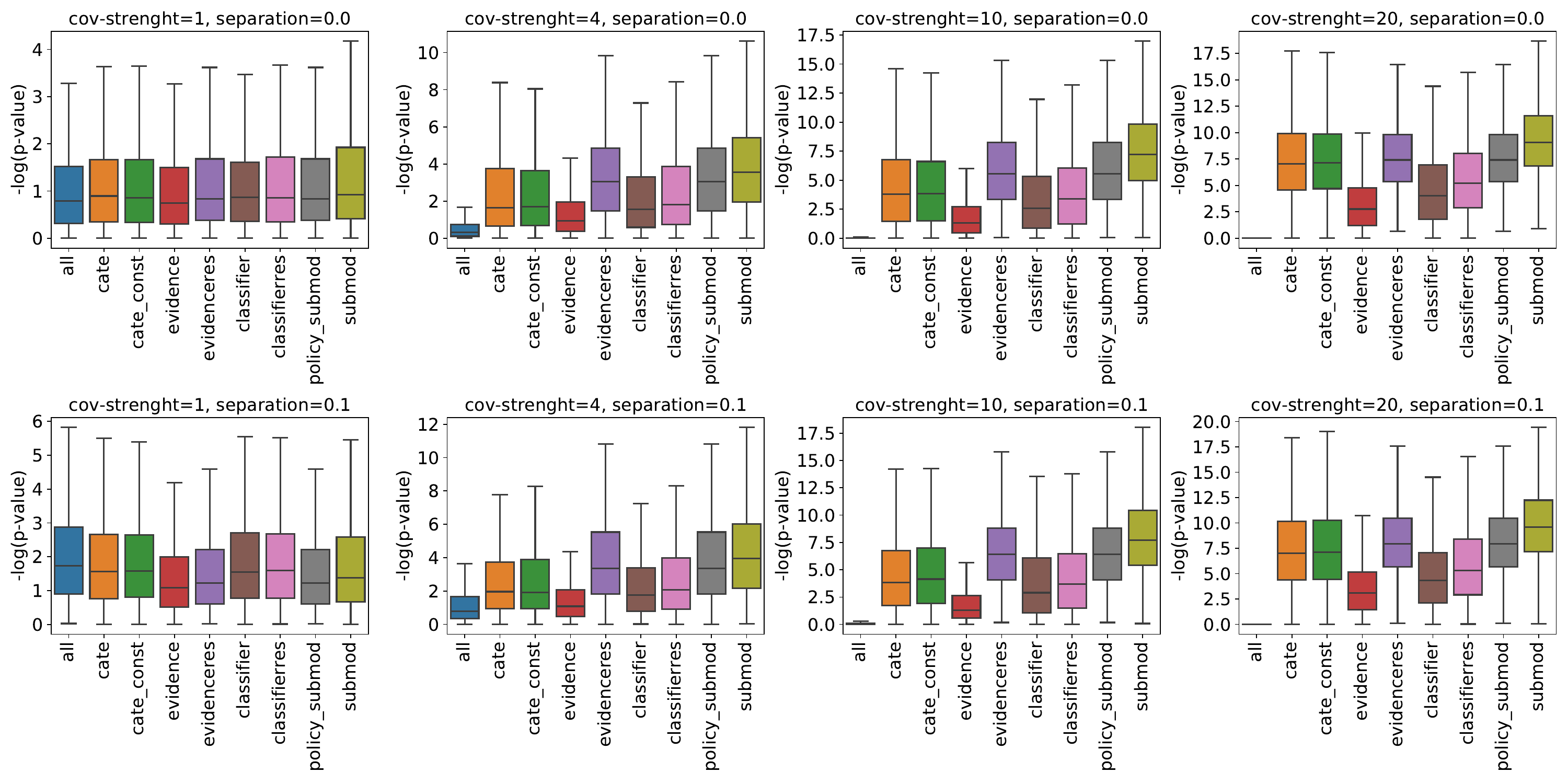}
		\caption{Out-of-sample negative log of $p$-value distributions of each method on $1000$ experiments.}
		\label{fig:semisynthlog}
	\end{figure}

\begin{figure}
		\centering
		\includegraphics[width=\textwidth]{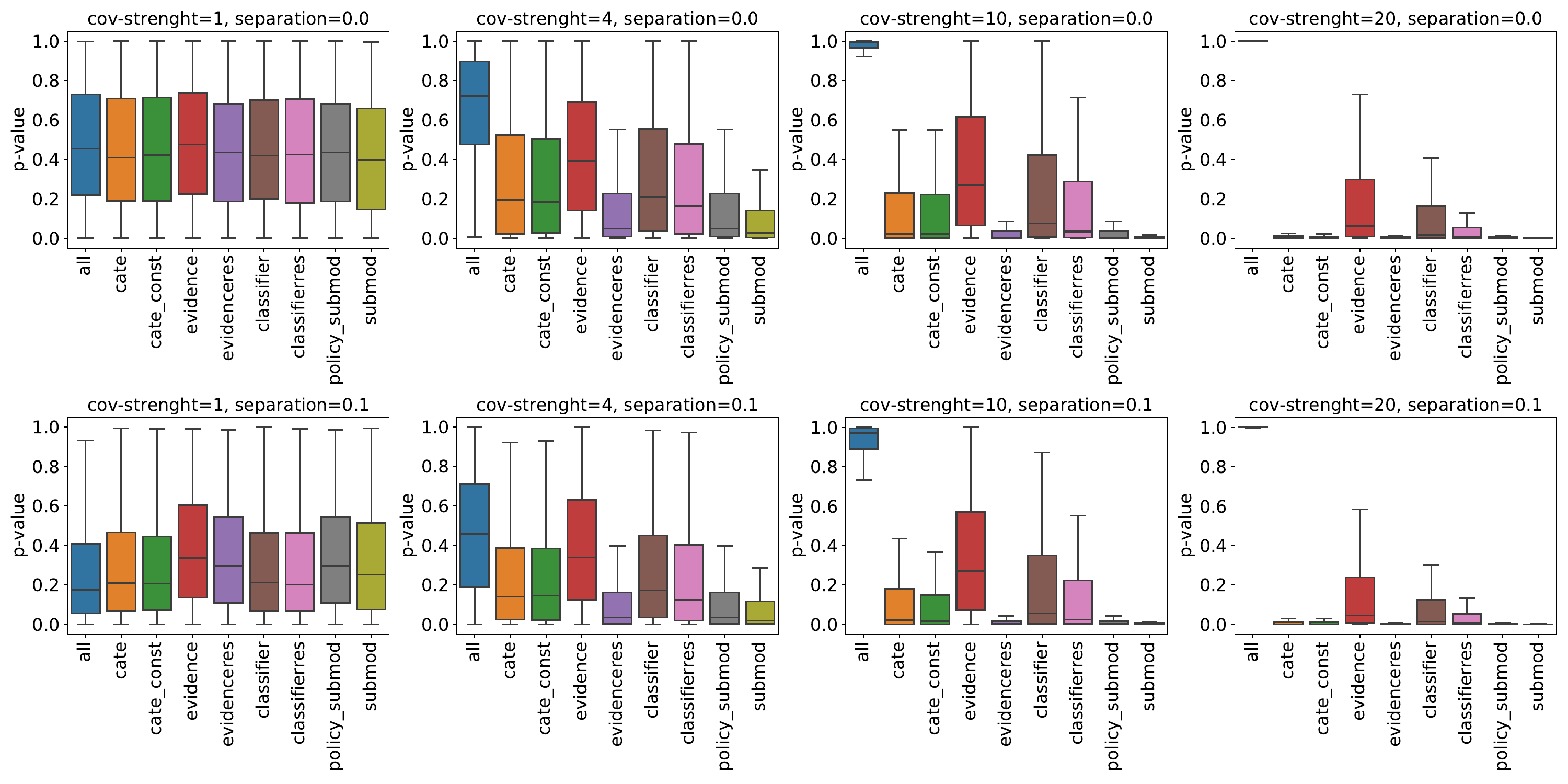}
		\caption{Out-of-sample $p$-value distributions of each method on $1000$ experiments.}
		\label{fig:semisynthp}
	\end{figure}

\begin{figure}
		\centering
		\includegraphics[width=\textwidth]{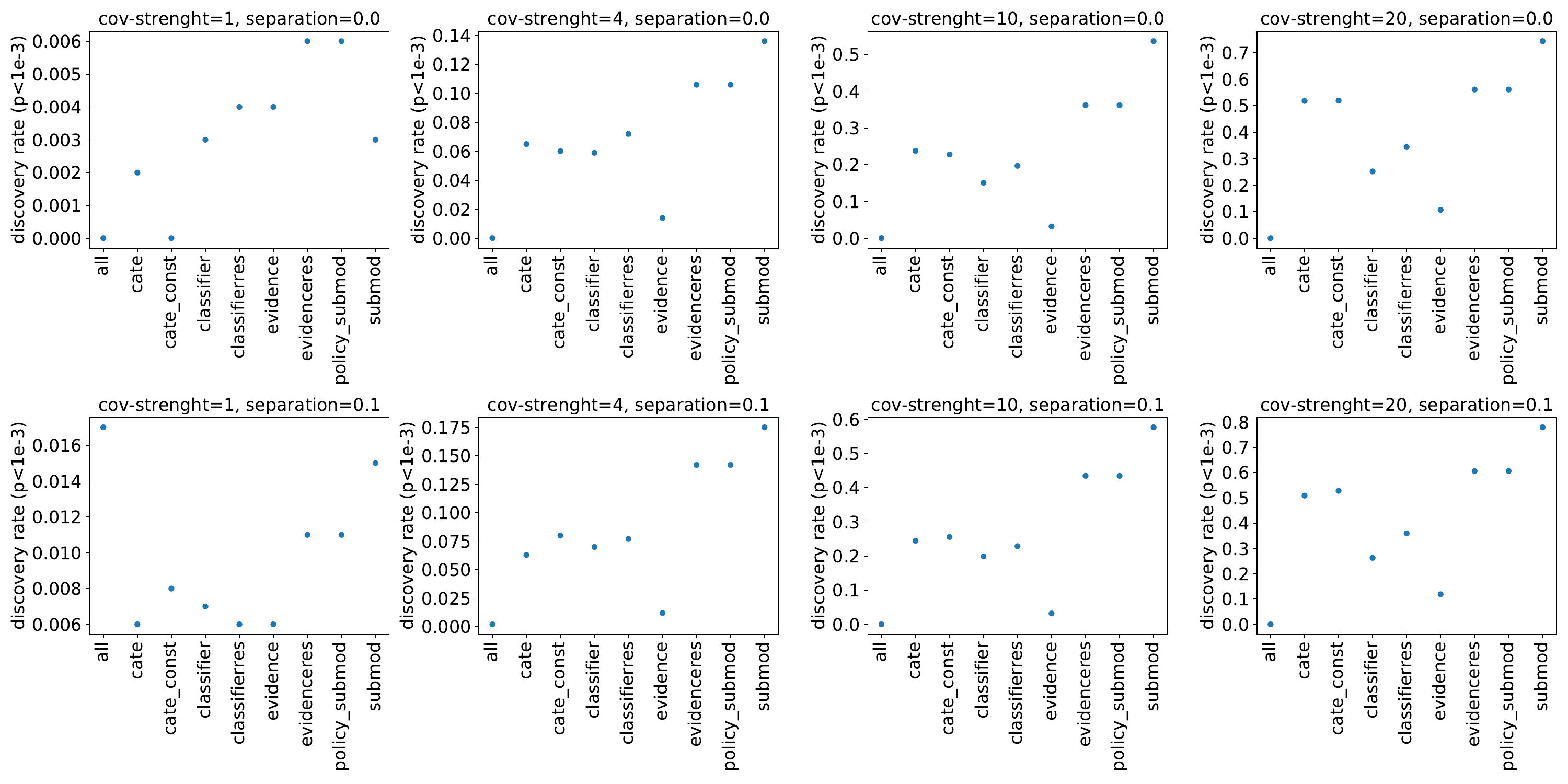}
		\caption{Fraction of $1000$ experiments where out-of-sample $p$-value $\leq 1e-3$, which we declare as a ``discovery''.}
		\label{fig:semisynthdisc}
	\end{figure}

	Our experimental results are depicted in Figures~\ref{fig:semisynthlog}, \ref{fig:semisynthp} and \ref{fig:semisynthdisc}. We find that our \textbf{evidenceres} and \textbf{submod} methods outperform the benchmarks in the medium signal regimes where the signal in the features is relatively low. Meanwhile, \textbf{evidenceres} and \textbf{submod} remain comparable with alternative methods in the two extreme regimes (very low/very high signal). Morever, the performance of both algorithms is comparable, with \textbf{evidenceres} having a slight advantage. We also see that running submodular minimization at the end of the significance tree has little to no effect. This implies not only that the greedy tree algorithm found good subgroups, but also that the assignment that it constructed in a step-by-step manner was appropriate for these subgroups.
	
	\subsection{Semi-synthetic experiments with continuous outcomes}
	
	The simulation in \autoref{subsect:semisynth-binary} is based on binary outcomes.
	We also simulate a variant of the semi-synthetic experiment with continuous outcome data. Here we use the group structure from \autoref{sect:simple-synth}, but define groups based on the real covariates from the HIV trial. Moreover, we use the treatment variable from the real data. The continuous outcome in each experiment is generated as follows:
	\begin{enumerate}
		\item Let $\tau = (-2, 2, 2)$, $\beta = (0, 10, 100)$, and $\sigma=(5, 10, 1)$. 
		
		\item For each experiment generate three indices $q=(i, j, k)$ uniformly at random from the set of feature indices $\{1,\ldots, J\}$.
		
		\item Let $\theta = (\theta_1, \theta_2, \theta_3)$, where $\theta_t$ is the median of the $q_t$ feature across all $n$ units.
		
		\item For each unit $i \in \{1,\ldots,n\}$, generate outcome data based on $Y_i = \tau(X_i)\, W_i + \beta(X_i) + \epsilon$ with $\epsilon \sim N\left(0, \sigma^2(X_i)\right)$ where
		\begin{align*}
			\tau(x) &= \sum_{t=1}^{3} \mu_t \: \Ind(x_{q_t} > \theta_t), &
			\beta(x) &= \sum_{t=1}^3 \beta_t \: \Ind(x_{q_t} > \theta_t), &
			\sigma(x) &= \sum_{t=1}^3 \rho_t \: \Ind(x_{q_t} > \theta_t).
		\end{align*}
	\end{enumerate}
	
\begin{figure}
		\centering
		\includegraphics[width=.32\textwidth]{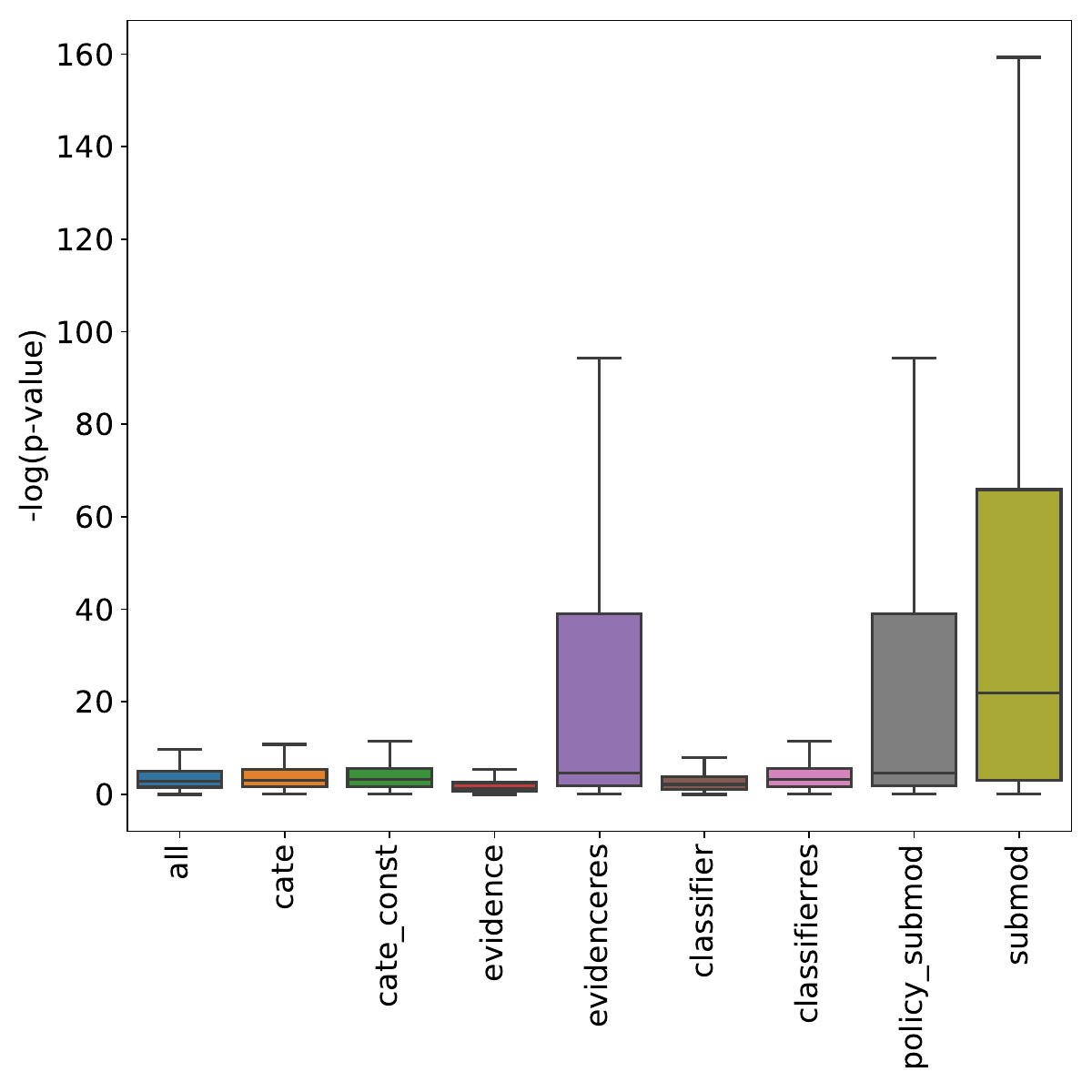}
		\includegraphics[width=.32\textwidth]{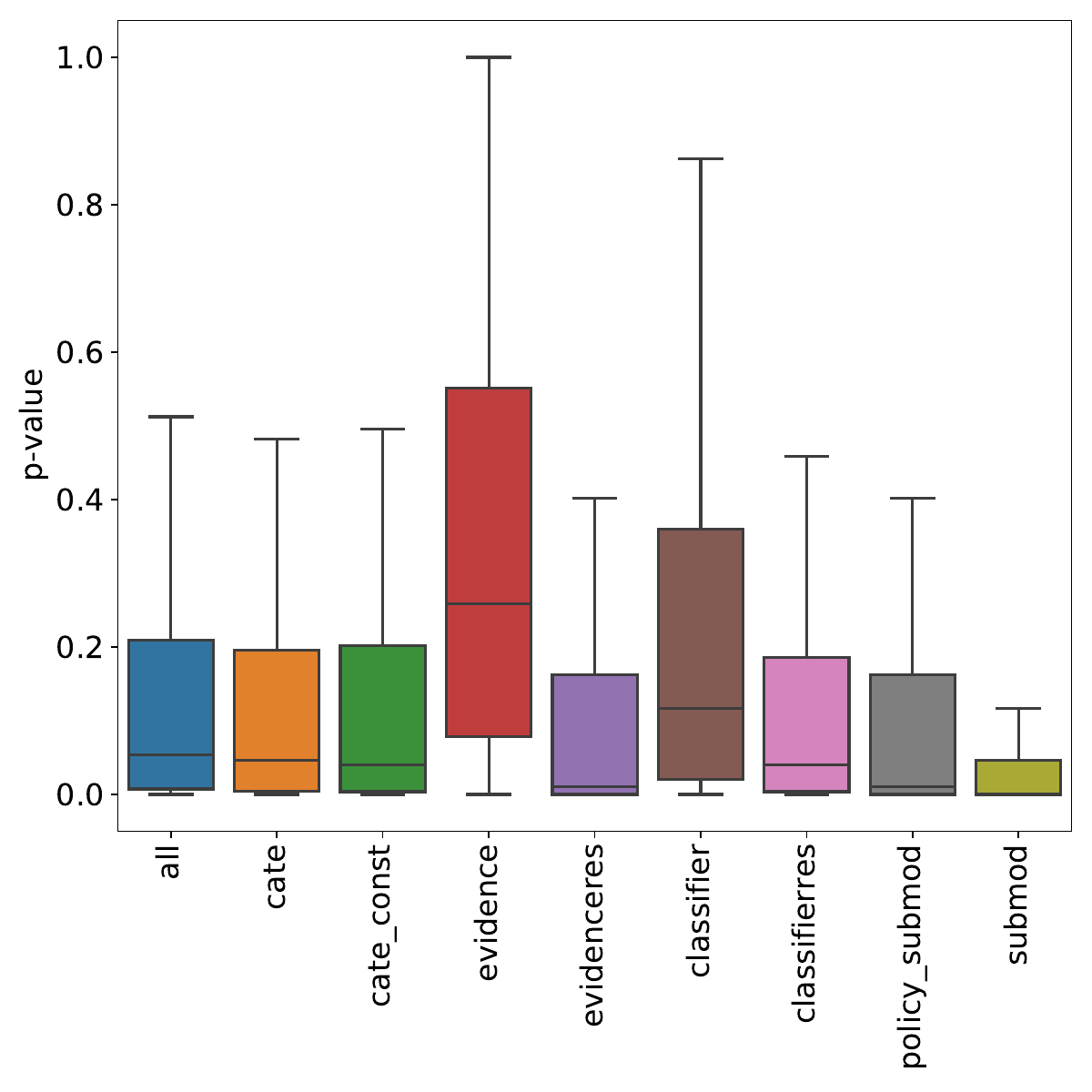}
		\includegraphics[width=.32\textwidth]{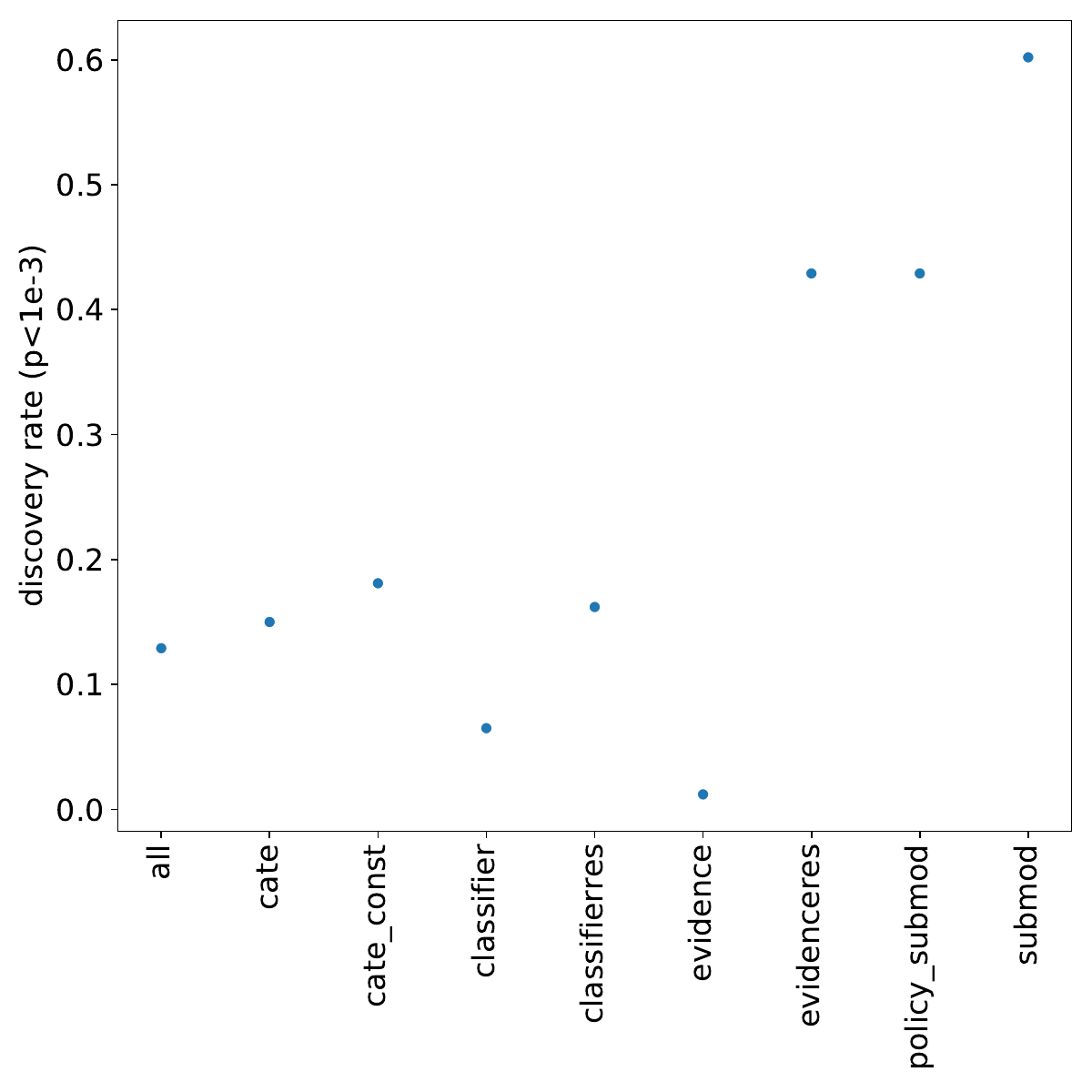}
		\caption{Out-of-sample $p$-value and negative log of $p$-value distributions of each method on $1000$ experiments. The last figure depicts the fraction of $1000$ experiments for which out-of-sample $p$-value $\leq 1e-3$, which we declare as a discovery.}
		\label{fig:semisynthcontnlog}
	\end{figure}

	Our experimental results are depicted in \autoref{fig:semisynthcontnlog}. We find that our \textbf{submod} method outperforms the benchmarks, correctly identifying not to include high-variance subgroups, even though they might have a slightly positive effect. Moreover, the \textbf{evidenceres} method comes a close second, albeit in this setting we see that \textbf{submod} is strictly better.
	
	\section{Continuous Relaxations and Significance Forests}
	\label{sect:modifications}
	
	In this section we consider two modifications that aim to improve the power of our empirical implementation.
	In \autoref{subsect:continuous_relaxation} we modify our target criterion, by generalizing our setup to permit weighted rather than binary selection into subgroups. 
	In \autoref{subsect:significance_forest} we then consider a forest-based extension of our significance-tree approach.
	We illustrate in \autoref{subsect:sim_compare_implementations} how these modifications can improve power in a simple simulation.
	
	\subsection{Continuous relaxation}
	\label{subsect:continuous_relaxation}
	
	In our baseline setup, we focused on obtaining a binary assignment $a: \X \rightarrow \{0,1\}$ that selects whether we assign an observation with covariate vector $x$ to our testing subgroup.
	We can relax this assumption and consider instead continuous weights $a: \X \rightarrow \mathbb{R}_{\geq 0}$, where our target is the expected lift $u(a) = \E[a(X) \: \tau(X)]$. Testing for $\E[a(X) \: \tau(X)] \geq 0$ can now be understood in two ways.
	
	Firstly, consider the most general case where assignment weights can take any value in  $\mathbb{R}_{\geq 0}$. This can be interpreted as a test of the null hypothesis that there is no subgroup with significant positive effects. Should we find significant continuous weights, this allows us to reject this null hypothesis. We can view the weights here as a technical tool for implementing this test. Likewise, subgroups with larger weights can be interpreted as contributing stronger evidence to our test. Note that this interpretation holds under any arbitrary re-scaling of the weights.
	
	An alternative interpretation in the context of policy learning is possible if we impose the additional requirement $a: \X \rightarrow [0,1]$. This can be achieved by appropriate scaling of the weights; for instance, we could divide all weights by the largest weight. $a$ can now be understood as a probabilistic assignment policy that assigns observations to treatment with probability $a(X)$. We can then test whether this assignment policy passes a hypothesis test on a hold-out sample.
	
	Even under this latter interpretation, note that we do not actually randomize, since the hold-out data is already given. Instead, we estimate the weighted estimand $u(\hat{a})$ without adding additional randomness, which would be statistically inefficient. Only if we were to deploy the assignment policy $\hat{a}(x)$ in the population to select who should get treated would we then randomize for an observation with covariate $x$ based on the probability $\hat{a}(x)$.

	The oracle result in \autoref{prop:oracle} remains applicable for the continuous relaxation.
	Indeed, the continuous relaxation allows for the direct solution of optimal weights in the framework of \autoref{sect:objective}:
	
	\begin{proposition}\label{prop:relaxed}
		Asymptotically as $N\rightarrow 0$,
		the probabilistic policy maximizing the power of the test on the hold-out sample is given by
		\begin{align*}
			a^*(x) &\propto \frac{\tau_+(x)}{\sigma^2(x)} 
			&
			&\text{ yielding expected power }
			&
			U(a^*)
			&=
			\Phi\left(\sqrt{N} \: \sqrt{\E \left[ \frac{\tau_+^2(X)}{\sigma^2(X)} \right] } - z_{1-\alpha}\right)
		\end{align*}
		where $\tau_+(x) = \tau(x) \Ind_{\tau(x) > 0}$.
		(Note that the absolute scale of the policy does not matter here.)
	\end{proposition}
	
	\begin{proof}
		See \autoref{sect:proof_prop_relaxed}.
	\end{proof}
	
\begin{figure}[h!]
		\centering
		\includegraphics[width=.6\textwidth]{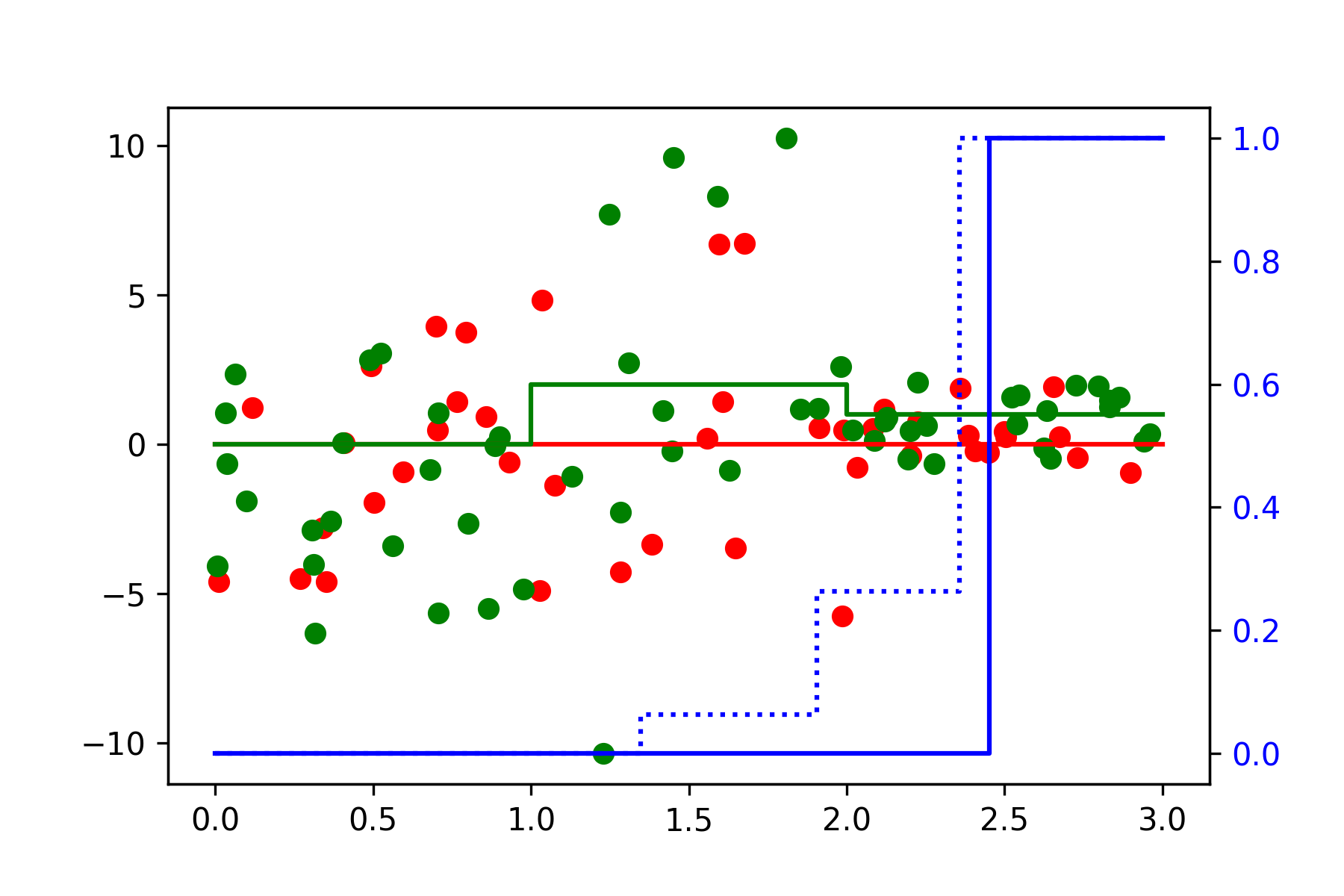}
		\caption{Example of a single tree with $n=100$ for binary (solid) and weighted (dotted) assignment, where on the blue axis is the weight for the test. Green observations are in treatment, and red observations are in control, with solid lines representing conditional means. The continuous weights are scaled to lie in $[0, 1]$, by dividing all weights by the largest such weight.}
		\label{fig:weightedexample}
	\end{figure}

	This particularly tractable formulation implies a criterion that can be used to partition the covariate space recursively using trees with binary splits as in \autoref{alg:main}. Specifically, let $\bar{\tau}_\ell$ be the average treatment effect of a leaf and let $\bar{\sigma}^2_\ell = \Var(\hat{\tau}_\ell)$ be the variance of the corresponding estimator $\hat{\tau}_\ell$.
	Then we aim to find a partition that maximizes
	$
	\sum_{\ell} \sfrac{\bar{\tau}^2_\ell}{\bar{\sigma}^2_\ell},
	$
	corresponding to weights $\bar{a}_\ell \propto \frac{\bar{\tau}_\ell}{\bar{\sigma}^2_\ell}$ on $\hat{\tau}_\ell$.
	When we also write $\hat{\sigma}^2_\ell$ for an estimator of $\bar{\sigma}^2_\ell$,
	this suggests maximizing the sample analog
	$
	\sum_{\ell} \sfrac{\hat{\tau}^2_\ell}{\hat{\sigma}^2_\ell}
	$
	over partitions of the data, which can be achieved recursively, and then setting $\hat{a}_\ell \propto \frac{\hat{\tau}_\ell}{\hat{\sigma}^2_\ell}$.
	Note here that partitions can be computed locally, in contrast to the setting with binary assignment weights.
	For illustration, \autoref{fig:weightedexample} compares a greedily fitted tree that maximizes the criterion in \autoref{prop:oracle} over binary assignments with a greedily fitted tree that recursively maximizes this weighted criterion.
	
	There may be two distinct reasons why we may prefer the continuous relaxation over binary assignment weights.
	The first one is in terms of power. Binary assignment weights are a special case of continuous assignment weights. As such, we would expect the continuous formulation to perform weakly better than the binary version with regard to our testing goal (i.e. maximizing the probability of passing a hypothesis test on hold-out data). This continuous relaxation modifies the hypothesis test to examine whether a subgroup with significant positive effects \emph{could} be found, even if we do not now find such a subgroup. By contrast, the binary assignment tests for a specific subgroup, and is thus more restrictive. The continuous formulation may lead to an improvement in power, though it has the drawback of not finding a specific subgroup. The second reason for preferring the continuous version is computational. The tree implementation for the continuous relaxation computes splits locally, whereas with binary assignments the splits have to be computed globally. Thus, the implementation of the continuous formulation is more computationally efficient.
	
	\subsection{Significance forests}
	\label{subsect:significance_forest}
	
	As decision trees are typically of limited expressiveness and high variance, we could alternatively improve performance by bagging trees into a \emph{significance forest} following the random-forest construction of \cite{breiman2001random}.
	This procedure can be used both for the binary assignment version and the continuous relaxation from \autoref{subsect:continuous_relaxation}.
	A potential implementation is as follows:
	
	\begin{enumerate}
		\item Draw $B$ bootstrap samples with replacement from the original training sample.
		\item Grow an individual tree $\hat{a}_b(x)$ on each bootstrap sample $b \in \{1, ..., B\}$ following \autoref{alg:main}. At each split-point of each tree, randomly select a subset of the total number of features to investigate as potential splitting features. (By default, we set the cardinality of the subset as equal to the square root of the total number of features, rounded down.)
		\item To return the forest's treatment assignment for a new point $x$, proceed as follows:
		\begin{enumerate}
			\item If using individual trees that output binary assignments, take the majority vote over tree-wise treatment assignments for each $x$. That is, $\hat{a}(x) = \textnormal{median} \{\hat{a}_b(x); b \in \{1,\ldots,B\}\} \in \{0,1\}$. Then the forest will also output binary assignment weights.
			\item If using individual trees that output continuous assignment weights, take the average over the outputs of the individual trees. Then the forest will also output continuous assignment weights.%
			\footnote{One could also imagine an implementation where individual trees output binary assignments, but we \emph{average} the tree outputs to obtain the forest's (continuous) assignment weights. However, we argue that this implementation is less convincing than either (a) or (b) above, as it essentially combines the drawbacks of both while sacrificing their respective advantages. This implementation would aggregate binary rather than continuous trees, leading to reduced computational efficiency compared to (b). Meanwhile, this implementation would output continuous assignment weights; relative to (a), this forgoes the interpretability of binary assignments in terms of finding a specific subgroup.}
		\end{enumerate}
	\end{enumerate}
	
	\subsection{Simulation illustration of comparative performance}
	\label{subsect:sim_compare_implementations}
	
	In this section, we compare the performance of the continuous relaxation and forest variants to the baseline implementation from \autoref{sect:significance_trees}.
	To provide a comprehensive comparison, we include nine types of subgroup-assignment algorithms in this simulation experiment:
	
	\begin{enumerate}
		\item \textbf{treat\_all}: A benchmark case where we choose every unit for testing.
		\item \textbf{binary\_tree}: A single significance tree using binary assignment weights, without centering in the outcomes.
		\item \textbf{binary\_forest}: A significance forest using binary assignment weights, without centering in the outcomes.
		\item \textbf{binary\_tree\_res}: A single significance tree using binary assignment weights, \emph{with} centered outcomes.
		\item \textbf{binary\_forest\_res}: A significance forest using binary assignment weights, \emph{with} centered outcomes.
		\item \textbf{cont\_tree}: A single significance tree using continuous assignment weights, without centering in the outcomes.
		\item \textbf{cont\_forest}: A significance forest using continuous assignment weights, without centering in the outcomes
		\item \textbf{cont\_tree\_res}: A single significance tree using continuous assignment weights, \emph{with} centered outcomes.
		\item \textbf{cont\_forest\_res}: A significance forest using continuous assignment weights, \emph{with} centered outcomes.
	\end{enumerate}
	
	Unlike the comparison in \autoref{sect:simulation}, we use the same centering strategy in training and hold-out for each of the methods. This means that those methods that do not employ centering also do not use centering for the test itself, thus treating training and hold-out consistently and highlighting the importance of residualization.
	
	For this comparison, we employ a specific data-generating process that highlights the differences between these different implementations.
	Specifically, we assume that $X = (X_1, X_2)$ is uniformly distributed on $[-2, +2]^2$, and that the outcome is generated as
	\begin{align*}
		Y &= a + b W + c \; \frac{X_{1} + X_{2}}{2} + d \: X_{1} \: W + (X_{2} + 3) \; \epsilon,
		&
		\epsilon &\sim \N(0, 1).
	\end{align*}
	We set $a = 1, b = 0$, and $c= 0.2$. $d$ then controls the strength of the treatment effect and its heterogeneity. We run our simulations on four different settings for $d$:
	\begin{multicols}{2}
		\begin{enumerate}
			\item Large treatment: $d = 1.5$
			\item Medium treatment: $d = 1$
			\item Small treatment: $d = 0.5$
			\item Null treatment: $d = 0$
		\end{enumerate}
	\end{multicols}
	As $X_1 \in [-2, +2]$, there are several regions where the treatment effect is negative, and indeed the average treatment effect is zero. This captures in a stylized sense our motivating example of an RCT where overall effects are small in a noisy environment. The extent of this noise is governed by $X_2$, with larger values of $X_2$ corresponding to noisier regions.
	
	For each treatment setting we run 1000 iterations of our simulation. Within each iteration, we draw separate training and test datasets based on the above data-generating process. Each such dataset is of size 400. We fit all the assignment policies on the training dataset, and then evaluate their respective performances by calculating their corresponding $p$-values on the test dataset. 
	
	For subgroup-assignment algorithms with centered outcomes, we compute the optimal centering function with a random forest regressor. We train one version of the centering function on the whole training dataset, which is then used to center outcomes in the hold-out test dataset. We also train another version of the centering function using cross-fitting in the training dataset: this is used to center outcomes \textit{within} the training dataset when constructing our subgroup-assignment algorithms.
	
	When using single significance trees, we control complexity by setting the minimum leaf size and maximum tree depth to 10. When aggregating these into significance forests, we permitted greater complexity in the individual trees: we set each tree to now have a minimum leaf size of 1 per arm, and did not impose a maximum tree depth.
	
\begin{figure}[htpb]
		\centering
		\includegraphics[width=\textwidth]{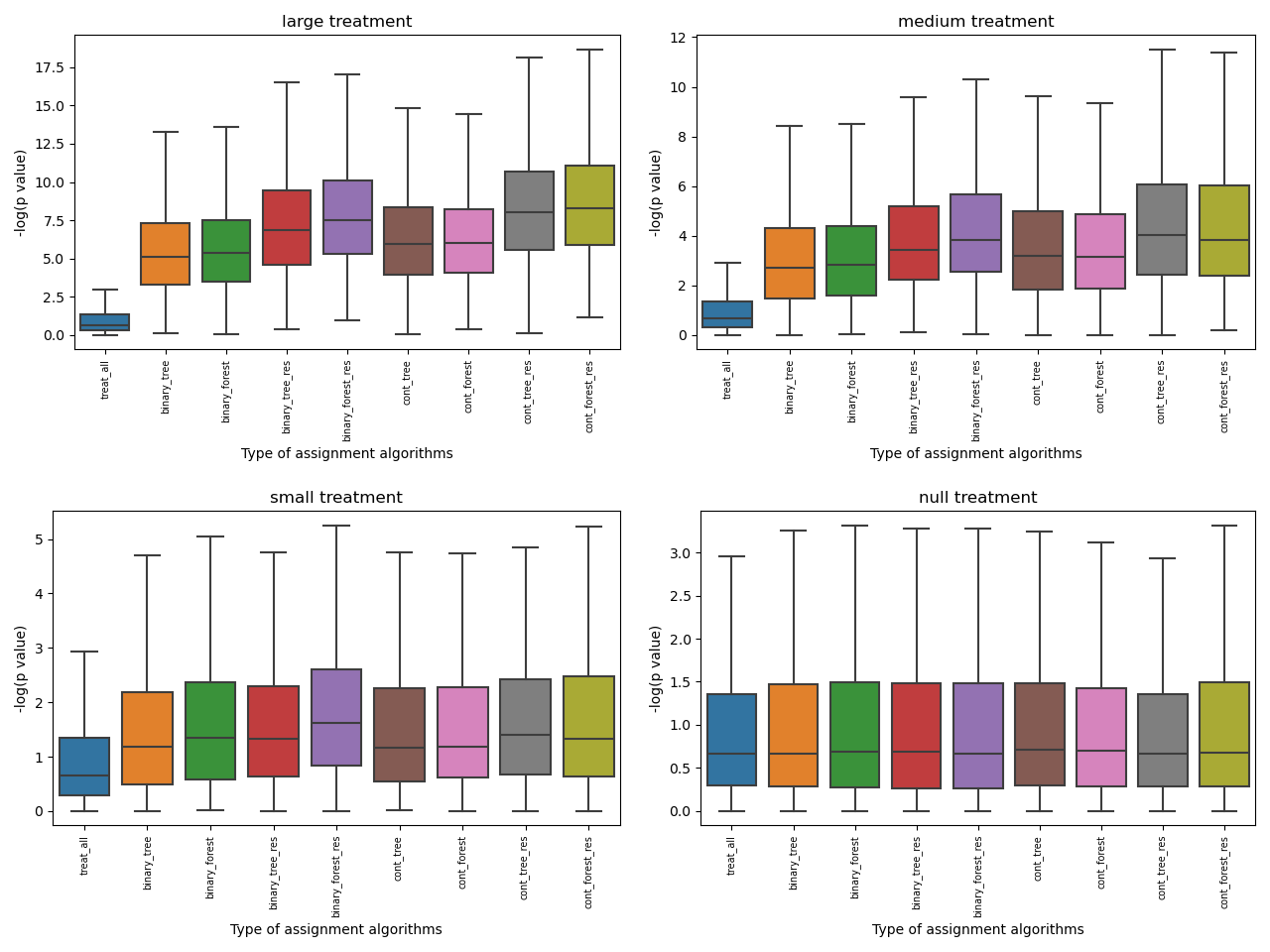}
		\caption{Out-of-sample negative log of $p$-value distributions of each assignment policy on $1000$ experiments.}
		\label{fig:synth_implement_neglogp}
	\end{figure}

\begin{figure}[htpb]
		\centering
		\includegraphics[width=\textwidth]{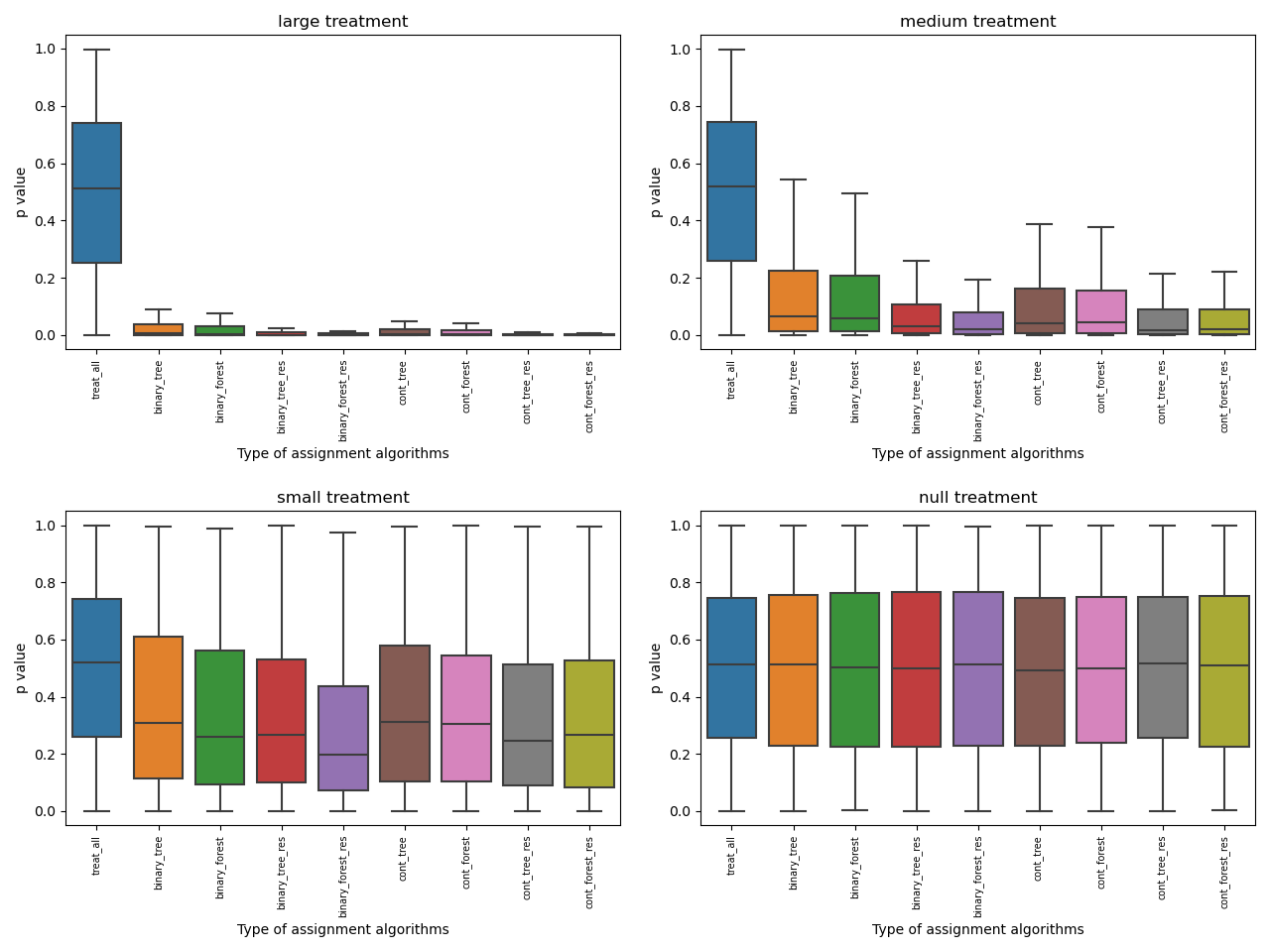}
		\caption{Out-of-sample $p$-value distributions of each assignment policy on $1000$ experiments.}
		\label{fig:synth_implement_p}
	\end{figure}

	Our results are presented in \autoref{fig:synth_implement_neglogp} and \autoref{fig:synth_implement_p}. As expected, we see that assigning everyone to treatment does relatively poorly, but our proposed algorithms deliver substantial improvements. Our results also demonstrate the performance gains in this setting from using residualized outcomes; using significance forests rather than single trees; and using continuous rather than binary assignment weights.
	
\begin{figure}[htpb]
		\centering
		\includegraphics[width=\textwidth]{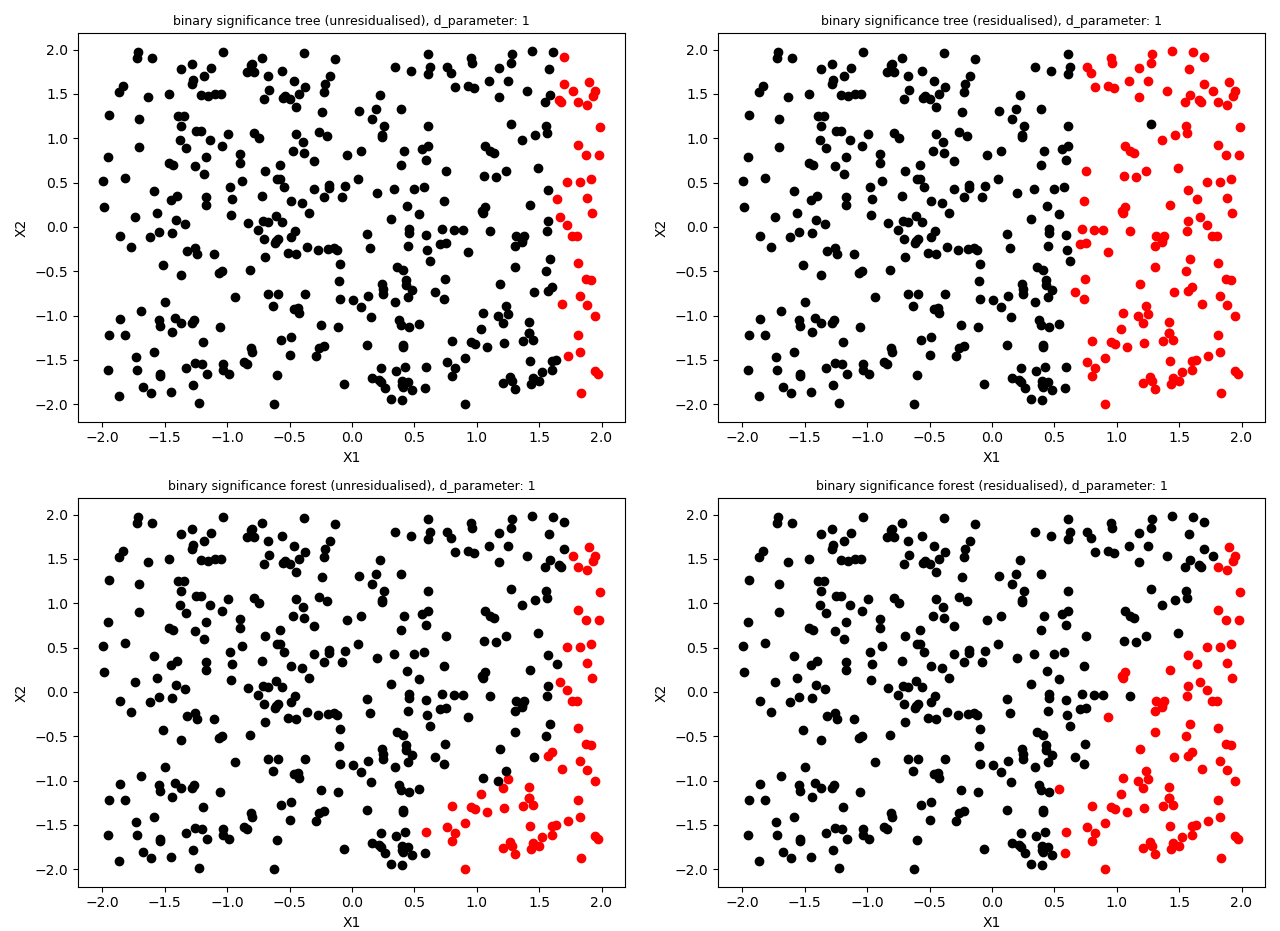}
		\caption{Binary assignment algorithms on the hold-out dataset. Black points represent units assigned to control, and red points represent units assigned to treatment.}
		\label{fig:synth_implement_binary_assign}
	\end{figure}

\begin{figure}[htpb]
		\centering
		\includegraphics[width=\textwidth]{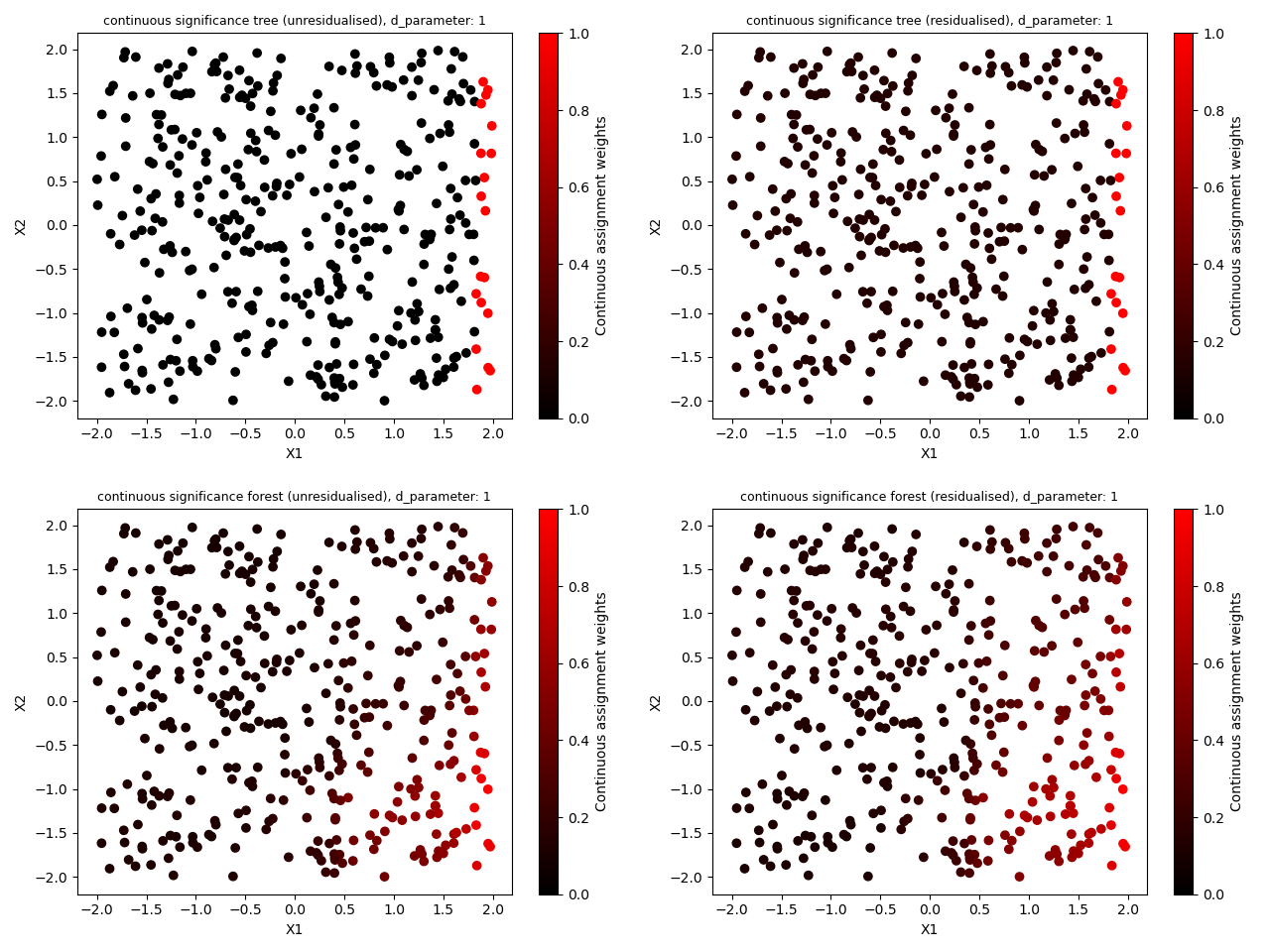}
		\caption{Continuous assignment algorithms on the hold-out dataset. Note that each set of assignment weights were scaled to lie in $[0, 1]$, by dividing those weights by the largest such weight.}
		\label{fig:synth_implement_cont_assign}
	\end{figure}

	We now further investigate the difference between these methods. To represent our subgroup-assignment algorithms graphically, we focus on a single iteration of the simulations. That is, we take one draw of a training and test dataset, with each dataset of sample size 400. We then fit all the algorithms to the training dataset. Finally, we plot the corresponding assignment weights on the test dataset. We use the medium treatment regime with $d=1$. 
	The resulting binary assignment algorithms are shown in \autoref{fig:synth_implement_binary_assign}, and continuous assignment algorithms are shown in \autoref{fig:synth_implement_cont_assign}.
	Across methods, the algorithms tend to select areas where $X_1$ is large but where $X_2$ is not too large. This is as we would expect, since this region should correspond to a good ratio of treatment effect to noise. While simple trees tend to only differentiate along the $X_1$ axis (along which the treatment effect varies), the forest-based implementations can capture variation along $X_2$ (noise variation) as well, effectively trading off both.
	
	\FloatBarrier

	\section{Extensions}
	\label{sect:extensions}
	
	To wrap up, we discuss natural extensions of our framework to different objectives and settings.
	
	\subsection{Finding significant heterogeneity} \label{sect:finding_heterogeneity}
	
	In our baseline setup, we focus on finding evidence of subgroups with positive treatment effects. We can use a similar approach to finding evidence of \textit{heterogeneous} treatment effects.
	This extension is similar to how \cite{Crump2008-cr} and \cite{Armstrong2015-gc} extend tests for non-constant and positive effects to tests for heterogeneous effects.
	Specifically, we now consider an assignment
	\begin{align*}
		a: \X \rightarrow \{-1,0,+1\}
	\end{align*}
	where units are assigned to a high-effect ($+1$), neutral ($0$), and low-effect ($-1$) group.
	We then evaluate whether the average treatment effects in the high group are above that of the low group by a hypothesis test.
	Towards this end,
	we consider the estimands
	\begin{align*}
		\Delta_1(a) &= \frac{\E[a_+(X) \tau(X)]}{\E[a_+(X)]} - \frac{\E[a_-(X) \tau(X)]}{\E[a_-(X)]} = \E[\tau(X)|a(X)=+1] - \E[\tau(X)|a(X)=-1],
		\\
		\Delta_2(a) &= \E[a(X) (\tau(X) {-} \E[\tau(X)]))] = \Cov(\tau(X),a(X))
	\end{align*}
	where $a_- = (- a)_+ \geq 0$ denotes the negative part.
	Both $\Delta_1(a) >0$ and $\Delta_2(a) > 0$ provide evidence of treatment-effect heterogeneity.
	We focus here on collecting evidence that $\Delta_2(a) > 0$, which itself is maximized for $a^*(x) = \text{sign}(\tau(X) {-} \E[\tau(X)]))$.
	This test is similar to the regression-based calibration test employed by \cite{chernozhukov2018generic} to evaluate treatment-effect heterogeneity.
	
	A test for $\Delta_2(a) > 0$ is based on
	$
	\widehat{Z}(a) = \frac{\hat{\Delta}_2(a)}{\sqrt{\widehat{\Var}(\hat{\Delta}_2(a))}}
	$
	where we set
	\begin{align*}
		\hat{\Delta}_2(a) &= \frac{1}{N} \sum_{j=1}^N \left(a(X_j) - \frac{1}{N} \sum_{k=1}^N a(X_k)\right) \left( \tilde{Y}_j - \frac{1}{N} \sum_{k=1}^N \tilde{Y}_k\right)
		\\
		\widehat{\Var}(\hat{\Delta}_2(a)) &=
		\frac{1}{N (N-1)} 
		\sum_{j=1}^N
		\left(
		\left(a(X_j) - \frac{1}{N} \sum_{k=1}^N a(X_k)\right) \left( \tilde{Y}_j - \frac{1}{N} \sum_{k=1}^N \tilde{Y}_k \right) - \hat{\Delta}_2(a)
		\right)^2.
	\end{align*}
	We now consider a local-to-zero approximation for which
	$
	\tau(x) = \bar{\tau} + \frac{\mu(x)}{\sqrt{N}}.
	$
	Under regularity conditions,
	\begin{align*}
		\widehat{Z}(a) \cd \N\left(
		\frac{\Cov(a(X),\mu(X))}{\sqrt{\E[(a(X) - \E[a(X)])^2\sigma^2(X)]}}, 1
		\right)
	\end{align*}
	and we can adapt our above approach to maximize
	\begin{align*}
		\frac{\Cov(a(X),\tau(X))}{\sqrt{\E[(a(X) - \E[a(X)])^2\sigma^2(X)]}} = \frac{\E[(a(X) - \E[a(X)])(\tau(X)-\E[\tau(X)])]}{\sqrt{\E[(a(X) - \E[a(X)])^2\sigma^2(X)]}}
	\end{align*}
	either over discrete assignments $a: \X \rightarrow \{-1,0,+1\}$ or the continuous relaxation $a: \X \rightarrow \R$.

	\subsection{Passing the test as a constraint}
	
	The testing goal ignores the actual outcome associated with our chosen subgroup. We can instead consider a second goal that also incorporates the expected outcome subject to passing the test, namely
	\begin{align*}
		U'(\hat{a})=\E[\widehat{T}(\hat{a}) \: u(\hat{a})|\hat{a}]
		=
		\P\left(\frac{\hat{u}(\hat{a})}{\sqrt{\widehat{\Var}(\hat{u}(\hat{a}))}} \geq z_{1-\alpha}\middle|\hat{a}\right) \: u(\hat{a}).
	\end{align*}
	This goal would correspond for example to a researcher wanting to find subgroups whose treatment effects are not only statistically significant, but also large in magnitude.
	It mixes the expected outcome $u(\hat{a})$ with the testing goal $\E[\widehat{T}(\hat{a})|\hat{a}]$ to trade off the value of the subgroup's outcomes with how easy it is to pass the test.
	In the approximation of \autoref{sect:objective}, the optimal subgroup assignment would then solve
	\begin{align*}
		a' &= \argmax_{a} U'(a) = \argmax_{a} \E[a(X) \: \tau(X)]	
		\:\Phi\left(\sqrt{N}	
		\frac{\E[a(X) \: \tau(X)]}{\sqrt{\E[a^2(X) \: \sigma^2(X)]}} - z_{1-\alpha}\right).
	\end{align*}
	
	\subsection{Estimation for alternative outcomes and estimated propensity score}
	
	In the main article, we considered a treatment effect that is identified in an experiment by $\tau(x) = \E\left[\tilde{Y}\middle|X=x\right] = \E\left[\frac{W - p}{p(1-p)} (Y-c(X))\middle|X=x\right]$ and can be estimated accordingly.
	We can extend the same techniques to a family of parameters $\tau(x)$ defined via the linear conditional moment equation
	\[
	\tau(x) = \E[\psi(Z; h)| X=x]
	\]
	for data $Z$, target causal parameter $\tau(x)$, and nuisance functions $h$.
	For $\tau(x)$ defined in this way, we aim to obtain an assignment that provides a significant test against the null that $u(a)=\E[a(X)  \: \tau(X)] \leq 0$. If the moment function satisfies the conditional Neyman orthogonality property \citep{chernozhukov2018double,foster2019orthogonal,oprescu2018orthogonal}, and these nuisance estimates are estimated in a cross-fitting manner and are $o(n^{1/4})$ estimable, then their error is asymptotically negligible. In this case, all our algorithms can be directly generalized by simply renaming the pseudo-outcome to
	\[
	\tilde{Y}_i = \psi(Z_i;\hat{h}).
	\]
	The estimation and optimization algorithms as well as the analysis of their finite-sample performance easily generalizes to this general pseudo-outcome setting. Moreover, Neyman orthogonality for this setting can be easily achieved via an automated-double machine learning process as outlined in the recent work of \cite{chernozhukov2018learning,chernozhukov2020adversarial}.
	
	As an example, we can generalize the case of conditional average treatment effects to cases where the propensity score is unknown and has to be estimated, in which case $h(w,x) = (c(w, x), p(x))$, with $c(w,x) = \E[Y| W=w, X=x]$, $p(x)=\E[W| X=x]$ and $\psi(Y,W,X; (c(\cdot, \cdot),p(\cdot))) = c(1,X)-c(0,X) + (Y-c(W,X)) \left(\frac{W}{p(X)} - \frac{1-W}{1-p(X)}\right)$.
	Even more straightforwardly, we can also consider the simple prediction setup where we measure the treatment effect (or outcome of interest) $Y$ directly, in which case we write $\tau(x) = \E[Y|X=x]$. Here, $\psi(y,x) = Y$.
	
	\subsection{Testing for alternative reference assignments}
	
	Our results generalize to testing effects relative to alternative null assignments other than assigning all observations to control. In the case of a reference assignment $b(X)$,
	\begin{align*}
		u(a|b) = u(a) - u(b)
		= 
		\E[(a(X) - b(X)) \: \tau(X)].
	\end{align*}
	In that case, all results go through if we relabel $Y_b^{(w)} = Y^{(|w-b(x)|)}$ for observations with $X=x$.
	
	\subsection{Cross-evaluation}
	
	So far, we have focussed on separating the training of a subgroup-assignment algorithm from its evaluation. While this may be the most applicable case in many heavily regulated settings, we may gain power through re-using training and hold-out samples by switching their roles.
	
	\section{Conclusion}
	
	In this article, we address the issue of obtaining reliable evidence about an intervention of interest, when overall estimated effects from RCT data are small or noisy.
	To do so, we propose new machine-learning tools that are optimized for finding \emph{subgroups with statistically significant treatment effects}.
	We first formalize the objective of maximizing the probability of passing a hypothesis test on additional hold-out data. We then provide empirical implementations based on decision trees and submodular minimization.
	Finally, we document the performance of these empirical implementations in a set of simulation studies, discuss modifications, and point to practically relevant extensions.
	
	\bibliography{bibliography}
	
	\newpage
	\appendix
	
	\section{Theoretical Illustration}
	\label{sect:theory}
	
	In \autoref{sect:objective} we argued in favor of choosing a subgroup that maximizes the oracle criterion
	\[\E[a(X) \: \tau(X)] / \sqrt{\E[a(X) \: \sigma^2(X)]}.\]
	In \autoref{sect:significance_trees} and \autoref{sect:submodular}, we presented strategies for approximating an optimal subgroup assignment by empirical maximization in the training data.
	Here, we demonstrate in a theoretical example how choosing a subgroup according to empirical maximization can improve power relative to a simple assignment based on estimating conditional average treatment effect alone.
	
	Specifically, for illustration, we consider a data-generating process with $k$ equally-sized covariate cells $x \in \{1,\ldots,k\}$, with all outcomes bounded.
	In the training data of size $n$,
	cell-wise average treatment effects are
	\begin{align*}
		\tau_x = \E[Y^{(1)} - Y^{(0)}|X=x] = \mu_x / \sqrt{n},
	\end{align*}
	where we again assume local-to-zero effects as in \autoref{sect:objective} to obtain non-degenerate solutions for the power of the test.
	In the training data, we obtain treatment effect estimates $\hat{\tau}_x$ with asymptotic variance $\sigma^2_x$ and variance estimates $\hat{\sigma}^2_x$ for every cell $x$, where
	\begin{align*}
		\sqrt{n/k} \: \hat{\tau}_x &\cd \N(\mu_x, \sigma^2_x),
		&
		\hat{\sigma}^2_x &\cp \sigma^2_x.
	\end{align*}
	
	We consider two ways of choosing a subset from the training data:
	\begin{enumerate}[label=(\Alph*)]
		\item 
		\label{itm:naive}
		A simple classification algorithm that targets the average outcome under assignment, choosing an assignment $\hat{a}^{\ref{itm:naive}} \in \{0,1\}^k$ that maximizes the usual assignment criterion
		\begin{align*}
			\frac{1}{k}\sum_{x=1}^k \hat{a}^{\ref{itm:naive}}_x \hat{\tau}_x
		\end{align*}
		by setting $\hat{a}^{\ref{itm:naive}}_x = \Ind\left(\hat{\tau}_x > 0\right)$;
		\item
		\label{itm:evidence}
		A significance-based algorithm that targets the sample analog of the testing criterion from in \autoref{prop:oracle}, choosing an assignment $\hat{a}^{\ref{itm:evidence}} \in \{0,1\}^k$ that maximizes
		\begin{align*}
			\frac{\frac{1}{k}\sum_{x=1}^k \hat{a}^{\ref{itm:evidence}}_x \hat{\tau}_x}{\sqrt{\frac{1}{k} \sum_{x=1}^k \hat{a}^{\ref{itm:evidence}}_x \frac{\hat{\sigma}^2_x}{n/k}}}.
		\end{align*}
	\end{enumerate}
	
	We assume that we evaluate both policies on a hold-out sample of size $N = \eta n$, where $\eta > 0$ is fixed throughout.
	We then compare the expected utility
	\begin{align*}
		\E[U(\hat{a})]
		=
		\E\left[\P\left(
		\frac{\hat{u}(\hat{a})}{\sqrt{\widehat{\Var}(\hat{u}(\hat{a}))}} \geq z_{1-\alpha}
		\middle|\hat{a}\right)\right]
	\end{align*}
	between the two approaches, where we average over the hold-out (outer expectation) and over the training data (inner probability).
	For illustration, we consider an extreme case -- that treatment effects are arbitrary, but noise levels vary so much that only one of them is worth testing in the hold-out.
	
	\begin{proposition}
		\label{prop:avoiding}
		
		Assume that treatment effects are arbitrary (and positive), with fixed $\mu^2_x > 0$.
		Then for every $\delta > 0$ there exist a constant $C$ such that
		if
		\begin{align*}
			\min_{x} \sigma^2_x &\leq \frac{1}{C},
			&
			\max_{x} \sigma^2_x &\geq C
		\end{align*}
		then
		\begin{align*}
			\E[U(\hat{a}^{\ref{itm:naive}})] &< \alpha + \frac{1-\alpha}{2} + \delta , & 1- \delta &< \E[U(\hat{a}^{\ref{itm:evidence}})]
		\end{align*}
		for $n$ sufficiently large.
		
	\end{proposition}
	
	\begin{proof}
		See \autoref{sect:proof_prop_avoiding}.
	\end{proof}
	
	This result illustrates the more general approximation results of \autoref{sect:submodular}.
	Specifically, by \autoref{prop:finite-x-pop},
	we can approximate the value of an optimal assignment up to an error that is of the same order of magnitude as the difference between different treatment assignments in this local-to-zero framework.
	Moreover, the leading term in the error in \autoref{prop:finite-x-pop} does not depend on the specific assignment or variances.
	By choosing sufficiently low variances that lead to a probability of rejection that is arbitrarily close to one for an \emph{optimal} policy, we therefore also obtain an assignment with probability of rejection that is arbitrarily close to one from the training dataset with high probability.
	The simple classification algorithm, on the other hand, will select the bucket with variance $\max_{x} \sigma^2_x \geq C$ at least half of the time in large samples, leading to an assignment with rejection probability that can be arbitrarily close to the size $\alpha$ of the test. 
	Taken together, we find some (extreme) conditions that (for $\alpha, \delta$ not too large) imply that the significance-based assignment dominates.

	\section{Proofs}
	\label{sect:proofs}
	
	\subsection{Proof of \autoref{prop:oracle}}
	\label{sect:proof_prop_oracle}
	
	\begin{proof}
		For every $a$ with $\E[a(X)] > 0$
		and with $\tilde{Y} = \left(\frac{W}{p} - \frac{1-W}{1-p}\right) (Y-c(X))$
		we have under standard regularity assumptions that
		\begin{align*}
			&\frac{\sqrt{N}}{\sqrt{\Var(a(X) \: \tilde{Y})}} (\hat{u}(a) - u(a))
			= 
			\frac{1}{\sqrt{N \: \Var(a(X) \: \tilde{Y})}} \sum_{j=1}^N (a(X_j) \: \tilde{Y}_j - \E[a(X) \: \tilde{Y}]
			\cd
			\N(0, 1),
			\\
			& \Var(a(X) \: \tilde{Y}) =
			\Var(\E[a(X) \: \tilde{Y}|X]) + \E[\Var(a(X) \: \tilde{Y}|X)]
			= \Var(a(X) \: \tau(X)) + \E[a^2(X) \: \Var(\tilde{Y}|X=x)]
			\\
			&= \frac{1}{N} \Var(a^2(X) \: \mu(X)) + \E[a^2(X) \: \Var(\tilde{Y}|X=x)] 
			\rightarrow \E[a^2(X) \: \Var(\tilde{Y}|X=x)] = \E[a^2(X) \: \sigma^2(X)]
		\end{align*}
		since $\E[\tilde{Y} | X=x] = \tau(x)$.
		Also,
		\begin{align*}
			N \widehat{\Var}(\hat{u}(a))  - \Var(a(X) \: \tilde{Y}) &= \frac{1}{N-1} \sum_{j=1}^N (a(X_j) \:\tilde{Y}_j - \hat{u}(a))^2
			- \Var(a(X) \: \tilde{Y}) \cp 0
		\end{align*}
		and thus $N \widehat{\Var}(\hat{u}(a)) \cp \E[a^2(X) \: \sigma^2(X)] > 0$.
		Since also $\sqrt{N} u(a) = \E[a(X) \: \sqrt{N}\tau(X)] = \E[a(X) \: \mu(X)]$,
		we overall have that
		\begin{align*}
			&\frac{\hat{u}(a)}{\sqrt{\widehat{\Var}(\hat{u}(a))}}
			=
			\frac{\sqrt{N} (\hat{u}(a) - u(a)) + \sqrt{N} u(a)}{\sqrt{N \widehat{\Var}(\hat{u}(a))}} \\
			&= \frac{\sqrt{N} (\hat{u}(a) - u(a)) + \E[a(X) \: \mu(X)]}{\sqrt{N \widehat{\Var}(\hat{u}(a))}}
			\cd 
			\N\left(\frac{\E[a(X) \: \mu(X)]}{\sqrt{\E[a^2(X) \: \sigma^2(X)]}}, 1\right).
			\qedhere
		\end{align*}
	\end{proof}
	
	\subsection{Proof of \autoref{lem:basic-props}}
	\label{sect:proof_lem_basic-props}
	
	\begin{proof}
		The first property is the standard derivation of the unbiasedness of the doubly robust estimator with known propensities:
		\begin{align*}
			\tau(x)&= \E[\tilde{Y}| X=x]\\
			&= \frac{\E[Y^{(1)} - c(X)| X=x] \P(W=1| X=x)}{p} - \frac{\E[Y^{(0)} - c(X)| X=x] \P(W=0| X=x)}{1-p}\\
			&= \E[Y^{(1)} - c(X)| X=x] - \E[Y^{(0)} - c(X)| X=x]\\
			&= \E[Y^{(1)} - Y^{(0)}| X=x]
		\end{align*}
		\begin{align*}
			\varsigma^2_c(x) &= \E[\tilde{Y}^2 | X=x] = \E[(Y^{(1)} - c(X))^2| X=x] \frac{1}{p} + \E[(Y^{(0)}-c(X))^2| X=x] \frac{1}{1-p}\\
			&= \frac{\Var(Y^{(1)}| X=x)}{p} + \frac{\Var(Y^{(0)}| X=x)}{1-p} + \E[Y^{(1)} - Y^{(0)}| X=x]^2\\
			~& + \frac{(c(x) - c^*(x))^2}{p(1-p)} - 2\E[Y^{(1)} - c(X))| X=x] \frac{c(x)-c^*(x)}{p}\\
			~& - 2\E[Y^{(0)}-c(X)| X=x] \frac{c(x)-c^*(x)}{1-p}\\
			&= \frac{\Var(Y^{(1)}| X=x)}{p} + \frac{\Var(Y^{(0)}| X=x)}{1-p} + \E[Y^{(1)} - Y^{(0)}| X=x]^2\\
			~& + \frac{(c(x) - c^*(x))^2}{p(1-p)} + 2 \tau(x)\, (c(x)-c^*(x))\\
			~& - 2 \tau(x) (c(x)-c^*(x))\\
			&= \frac{\Var(Y^{(1)}| X=x)}{p} + \frac{\Var(Y^{(0)}| X=x)}{1-p} + \E[Y^{(1)} - Y^{(0)}| X=x]^2 + \frac{(c(x) - c^*(x))^2}{p(1-p)}\\
			&= \sigma(x)^2 + \tau(x)^2 + \frac{(c(x) - c^*(x))^2}{p(1-p)}
		\end{align*}
		Finally, maximizing
		\[
		V(a) = \frac{\E[a(X) \tau(X)]}{\sqrt{\E[a(X) \sigma^2(X)] + \E[a(X) \tau^2(X)]}}
		\]
		is equivalent to maximizing
		\begin{align*}
			\frac{\E[a(X) \tau(X)]}{\sqrt{\E[a(X) \sigma^2(X)] + \E^2[a(X) \tau^2(X)] - \E[a(X) \tau(X)]}}&=
			\frac{\E[a(X) \tau(X)]}{\sqrt{\E[a(X) \sigma^2(X)] + \Var(a(X) \tau(X))}}.
			\qedhere
		\end{align*}
	\end{proof}
	
	\subsection{Proof of \autoref{prop:finite-x}} \label{sect:prop_finite-x}
	
	\begin{proof}
		Let $g_\lambda(S) = \sqrt{\sum_{x \in S} v(x)} - \lambda \sum_{x \in S} w(x)$. For the initial setting of $\lambda_{\max}$, observe that if we let $S_{\max} = \argmin_{S \subseteq \{1,\ldots,m\}} g_{\lambda_{\max}}(S)$, then $S_{\max}\neq \emptyset$ and $g_{\lambda_{\max}}(S_{\max})<0$, since by choosing $S=\{1,\ldots,m\}$ we achieve $g_{\lambda_{\max}}(S) = \sqrt{v(\{1,\ldots,m\})} - \lambda_{\max} w(\{1,\ldots,m\})\leq \epsilon \sqrt{v(\{1,\ldots,m\})} < 0 = g_{\lambda_{\max}}(\emptyset)$. Moreover, at any point where we decrease $\lambda_{\max}$ to $\bar{\lambda}$, we have that $\hat{S}\neq \emptyset$ and $\sqrt{\sum_{x \in \hat{S}} v(x)} < \bar{\lambda} \sum_{x \in \hat{S}} v(x)$, which implies that $g_{\bar{\lambda}}(S) < 0 = g_{\bar{\lambda}}(\emptyset)$. Thus the property that $S_{\max} \neq \emptyset$ is maintained with the new $\lambda_{\max}$. Hence, we conclude that throughout the algorithm this property is maintained.
		
		Moreover, we have that at the initial setting of $\lambda_{\min}$ that
		\begin{align*}
			&\min_{S\subseteq \{1,\ldots,m\}} g_{\lambda_{\min}}(S) = \min_{S\subseteq \{1,\ldots,m\}} \sqrt{\sum_{x \in S} v(x)} - \frac{\sqrt{\min_{x\in \X} v(x)}}{w(\{1,\ldots,m\})} \sum_{x \in S} w(x)
			\\
			&\geq \min_{\emptyset \neq S\subseteq \{1,\ldots,m\}} \sqrt{\sum_{x \in S} v(x)} - \sqrt{\min_{x\in \X} v(x)}\geq 0.
		\end{align*}
		Moreover, at any point where we change $\lambda_{\min}$ to $\bar{\lambda}$ we know that either $\hat{S}=\emptyset$, in which case $\min_{S} g_{\bar{\lambda}}(S) = 0$, or that $\sqrt{\sum_{x \in \hat{S}} v(x)} \geq \bar{\lambda} \sum_{x \in \hat{S}} w(x)$, which implies that $g_{\bar{\lambda}}(\hat{S}) = \min_{S} g_{\bar{\lambda}}(S) \geq 0$. Thus at all points during the algorithm we have the property that $\min_{S\subseteq \{1,\ldots,m\}} g_{\lambda_{\min}}(S) \geq 0$.
		
		We now look at the values of the aforementioned quantities at the termination of the algorithm. From the second property we have that for any $S\neq \emptyset$ $g_{\lambda_{\min}}(S) \geq 0$. Since $\sum_{x \in S} w(x)>0$ for any $S\neq \emptyset$, we then have that: $\min_{S\neq \emptyset} \frac{\sqrt{\sum_{x \in S} v(x)}}{\sum_{x \in S} w(x)} \geq \lambda_{\min}$. By the termination criterion of the algorithm we have that: $\lambda_{\min} \geq \frac{1}{1+\epsilon} \lambda_{\max}$. Finally, by the first property we have that $\hat{S}\neq \emptyset$ and $g_{\lambda_{\max}}(\hat{S})<0$, where $\hat{S}$ is the set returned by the algorithm. Thus we have that:
		\[
			\frac{\sqrt{\sum_{x \in \hat{S}} v(x)}}{\sum_{x \in \hat{S}} w(x)} < \lambda_{\max} \leq (1+\epsilon)\lambda_{\min} \leq (1+\epsilon) \min_{S\neq \emptyset} \frac{\sqrt{\sum_{x \in S} v(x)}}{\sum_{x \in S} w(x)}
		\]
		Equivalently, since $\sum_{x \in S} v(x)>0$, for any $S\neq \emptyset$, we also have that:
		\[
			\frac{\sum_{x \in \hat{S}} w(x)}{\sqrt{\sum_{x \in \hat{S}} v(x)}} > \frac{1}{\lambda_{\max}} \geq \frac{1}{1+\epsilon}\frac{1}{\lambda_{\min}} \geq \frac{1}{1+\epsilon} \max_{S\neq \emptyset} \frac{\sum_{x \in S} w(x)}{\sqrt{\sum_{x \in S} v(x)}}
		\]
		
		Observe that after every period, we have that $\lambda_{\max}-\lambda_{\min}$ decreases by a factor of $2$. Thus after $O\left(\log_2\left(\frac{\sqrt{v(\{1,\ldots,m\})}}{\epsilon \sqrt{\min_{x\in\X} v(x)}}\right)\right)$ iterations, we will have that $\lambda_{\max}-\lambda_{\min} \leq \epsilon \frac{\sqrt{\min_{x\in \X}v(x)}}{w(\{1,\ldots,m\})}\leq \epsilon \lambda_{\min}$. Thus $\lambda_{\max}\leq (1+\epsilon)\lambda_{\min}$ and the algorithm will terminate.
	\end{proof}
	
	\subsection{Proof of \autoref{prop:finite-x-pop}} \label{sect:prop_finite-x-pop}
	
	\begin{proof}
		Let $A$ be the space of all assignments from $\X$ to $\{0,1\}$ and observe that $|A|=2^k$. Moreover, for any random variable $X$, let $\E_n[X]=\frac{1}{n}\sum_{i=1}^n X_i$, denote the empirical average. By Bennett's inequality and empirical Bennett's inequality \citep{maurer2009empirical}, we have that w.p. $1-4\delta$, for all $a \in A$:
		\begin{align*}
			\left|\E_n[a(X)\tilde{Y}] - \E[a(X) \tilde{Y}]\right| &\leq \sqrt{\frac{\E_n[a(X)\tilde{Y}^2]\, 2\,k\, \log(2/\delta)}{n}} + H \frac{7\,k\,\log(2/\delta)}{3(n-1)}\\
			\left|\E_n[a(X)\tilde{Y}^2] - \E[a(X) \tilde{Y}^2]\right| &\leq \sqrt{\frac{\E[a(X)\tilde{Y}^4]\, 2\,k\, \log(1/\delta)}{n}} + H \frac{k\,\log(2/\delta)}{3n}\\
			&\leq H\sqrt{\frac{\E[a(X)\tilde{Y}^2]\, 2\,k\, \log(1/\delta)}{n}} + H\frac{k\,\log(2/\delta)}{3n}\\
		\end{align*}
		Thus if we let $\epsilon_1(n,\delta)= \sqrt{\frac{2\,k\, \log(2/\delta)}{n}} + H\frac{7\,k\,\log(2/\delta)}{3(n-1)}$ we have that:
		\begin{align*}
			\frac{\E_n[\hat{a}(X) \tilde{Y}]}{\sqrt{\E_n[\hat{a}(X)\tilde{Y}^2]}}&\leq \frac{\E[\hat{a}(X) \tilde{Y}]}{\sqrt{\E_n[\hat{a}(X)\tilde{Y}^2]}} + \epsilon_1(n,\delta)
		\end{align*}
		Moreover, observe that by the AM--GM inequality and the aforementioned Bennett's inequality, for any $\eta\in (0, 1)$:
		\begin{align*}
			\E_n[a(X)\tilde{Y}^2] - \E[a(X) \tilde{Y}^2]\geq - \eta \E[a(X)\tilde{Y}^2] - \frac{2 H\, k\,\log(2/\delta)}{\eta\,n} \\
			\implies \E_n[a(X)\tilde{Y}^2] \geq (1 - \eta) \E[a(X)\tilde{Y}^2] - \frac{2 H\,k\log(2/\delta)}{\eta\,n}
		\end{align*}
		If we let $\epsilon_2(n,\eta,\delta)=\frac{2H\, k\log(2/\delta)}{\eta\,n}$, then we have that:
		\begin{align*}
			\frac{\E_n[\hat{a}(X) \tilde{Y}]}{\sqrt{\E_n[\hat{a}(X)\tilde{Y}^2]}}&\leq \frac{\E[\hat{a}(X) \tilde{Y}]}{\sqrt{(1-\eta)\E[\hat{a}(X)\tilde{Y}^2] - \epsilon_2(n,\eta,\delta)}} + \epsilon_1(n,\delta) \\
			&\leq \frac{\E[\hat{a}(X) \tilde{Y}]}{\sqrt{(1-\eta)\E[\hat{a}(X)\varsigma^2_c(X)] - \epsilon_2(n,\eta,\delta)}} + \epsilon_1(n,\delta) \\
			&\leq \frac{\E[\hat{a}(X) \tilde{Y}]}{\sqrt{(1-\eta)\E[\hat{a}(X)\varsigma^2_{*}(X)] - \epsilon_2(n,\eta,\delta)}} + \epsilon_1(n,\delta) \\
			&\leq \frac{\E[\hat{a}(X) \tilde{Y}]}{\sqrt{(1-\eta)\E[\hat{a}(X)\varsigma^2_{*}(X)]}} + \frac{1}{2\,(1-\eta)^{3/2}\underline{\pi}^{3/2}\, \underline{\varsigma}^3} \epsilon_2(n,\eta,\delta)+ \epsilon_1(n,\delta)
		\end{align*}
		Similarly, we have that:
		\begin{align*}
			&\frac{\E_n[a_*(X) \tilde{Y}]}{\sqrt{\E_n[a_*(X)\tilde{Y}^2]}}\geq \frac{\E[a_*(X) \tilde{Y}]}{\sqrt{\E_n[a_*(X)\tilde{Y}^2]}} - \epsilon_1(n,\delta)\\
			\geq~& \frac{\E[a_*(X) \tilde{Y}]}{\sqrt{(1+\eta)\E[a_*(X)\tilde{Y}^2] + \epsilon_2(n, \eta, \delta)}} - \epsilon_1(n,\delta)\\
			\geq~& \frac{\E[a_*(X) \tilde{Y}]}{\sqrt{(1+\eta)\E[a_*(X)\varsigma^2(X)] + \epsilon_2(n, \eta, \delta)}} - \epsilon_1(n,\delta)\\
			\geq~& \frac{\E[a_*(X) \tilde{Y}]}{\sqrt{(1+\eta)\E[a_*(X)\varsigma^2(X)] + \epsilon_2(n, \eta, \delta)}} - \epsilon_1(n,\delta)\\
			\geq~& \frac{\E[a_*(X) \tilde{Y}]}{\sqrt{(1+\eta)\E[a_*(X)\varsigma^2_{*}(X)] + \frac{\E[(\hat{c}(X) - c^*(X))^2]}{p(1-p)}+ \epsilon_2(n, \eta, \delta)}} - \epsilon_1(n,\delta)\\
			\geq~& \frac{\E[a_*(X) \tilde{Y}]}{\sqrt{(1+\eta)\E[a_*(X)\varsigma^2_{*}(X)]}} - \frac{1}{2(1+\eta)^{3/2} \underline{\pi}^{3/2}\, \underline{\varsigma}^3} \left(\frac{\E[(\hat{c}(X) - c^*(X))^2]}{p(1-p)}+ \epsilon_2(n, \eta, \delta)\right) - \epsilon_1(n,\delta)
		\end{align*}
		Finally, since:
		\begin{align*}
			\frac{\E_n[\hat{a}(X) \tilde{Y}]}{\sqrt{\E_n[\hat{a}(X)\tilde{Y}^2]}} \geq \frac{1}{1+\epsilon} \frac{\E_n[a_*(X) \tilde{Y}]}{\sqrt{\E_n[a_*(X)\tilde{Y}^2]}}
		\end{align*}
		Thus if we denote with:
		\begin{align*}
			&\kappa(n, \eta, \delta) = \frac{1}{2(1+\eta)^{3/2} \underline{\pi}^{3/2}\, \underline{\varsigma}^3} \left(\frac{\E[(\hat{c}(X) - c^*(X))^2]}{p(1-p)}+ \epsilon_2(n, \eta, \delta)\right) - \epsilon_1(n,\delta)\\
			&= \sqrt{\frac{2\,k\, \log(2/\delta)}{n}} + \frac{7\,H\,k\,\log(2/\delta)}{3(n-1)} + \frac{1}{2(1+\eta)^{3/2} \underline{\pi}^{3/2}\, \underline{\varsigma}^3} \left(\frac{\E[(\hat{c}(X) - c^*(X))^2]}{p(1-p)}+ \frac{2 H\,k\log(2/\delta)}{\eta\,n}\right)
		\end{align*}
		We have that:
		\begin{align*}
			\frac{\E[\hat{a}(X) \tilde{Y}]}{\sqrt{\E_n[\hat{a}(X)\varsigma^2_{*}(X)]}} \geq \sqrt{\frac{1-\eta}{1+\eta}} \frac{1}{1+\epsilon} \frac{\E[a_*(X) \tilde{Y}]}{\sqrt{\E[a_*(X) \varsigma^2_{*}(X)]}} - 2\kappa(n,\eta,\delta)
		\end{align*}
		Finally, noting that $\E[a(X)\tilde{Y}]=\E[a(X)\tau(X)]$ for all $a\in A$, yields the result.
	\end{proof}
	
	\subsection{Proof of \autoref{lem:partition-exact}}
	\label{sect:proof_lem_partition-exact}
	
	\begin{proof}
		Suppose $a$ is an optimal policy and that we have two such elements $x,x' \in \X$, such that $a(x)=1$ and $a(x')=0$, while $\varsigma(x)=\varsigma(x')=\varsigma$ and $\tau(x)=\tau(x')=\tau$. First observe that for any $x$ with $\tau(x)\leq 0$, any optimal policy must assign $a(x)=0$. Thus we restrict attention to the case when $\tau>0$. Let $f(x)=\P(X=x)$. Then we can write our objective as:
		\begin{align*}
			V(a) = \frac{A + a(x) f(x) \tau}{\sqrt{B + a(x) f(x) \varsigma^2}}
		\end{align*}
		where $A, B\geq 0$, since an optimal assignment never allocates to any $x\in \X$ with $\tau(x)\leq 0$. 
		
		We first consider the case where $B > 0$, i.e. when $x$ is not the only element that is assigned the treatment. Since by removing $x$ from assignment or adding $x'$ to the assignment, we are deteriorating the objective, we must have that:
		\begin{equation}\label{eqn:opt}
			\begin{aligned}
				\frac{A + f(x) \tau}{\sqrt{B + f(x) \varsigma^2}} >~& \frac{A}{\sqrt{B}}\\
				\frac{A + f(x) \tau}{\sqrt{B + f(x) \varsigma^2}} >~& \frac{A + (f(x)+f(x'))\tau}{\sqrt{B + (f(x) + f(x'))\varsigma^2}} 
			\end{aligned}
		\end{equation}
		Consider the function:
		\begin{align*}
			F(u) = \frac{A + u \tau}{\sqrt{B + u \varsigma^2}}
		\end{align*}
		By the Equation~\eqref{eqn:opt}, we must have that $F(0) < F(f(x))$ and $F(f(x)) > F(f(x)+f(x'))$. Since for $B>0$, $F(u)$ is a smooth twice differentiable function in $[0,1]$ and we have the latter inequalities, it means that there must be a point $u_*\in (0, f(x)+f(x'))$ at which it achieves a maximum. By the first-order optimality condition at that point, we must have that:
		\begin{equation*}
			u_*\varsigma^2 \tau + 2 B \tau - A \varsigma^2 =0 \implies u_* = \frac{A\varsigma^2 - 2B\tau}{\varsigma^2\tau}
		\end{equation*}
		Since $u_*> 0$, we also have that $A\varsigma^2 - 2B\tau > 0 \implies A\varsigma^2 > 2B\tau$. Moreover, by the second-order optimality condition we must have that:
		\begin{equation*}
			u_* \varsigma^2 \tau + 4 B \tau - 3 A \varsigma^2 \geq 0 \implies B\tau - A \varsigma^2 \geq 0 \implies A\varsigma^2 \leq B\tau
		\end{equation*}
		But for $\tau\geq 0$ and $B\geq 0$, the latter two inequalities cannot hold simultaneously, leading to a contradiction. Thus either removing $x$ or adding $x'$ must be a weakly improving change, leading to an optimal policy where $x$ and $x'$ are given the same assignment. Repeatedly applying the above argument, we can arrive to an assignment policy where any pair of $x,x'$ that have $\varsigma^2(x)=\varsigma^2(x')$ and $\tau(x)=\tau(x')$ are assigned the same treatment.
		
		Now consider the edge case when $x$ is the only element assigned by the policy $a$. In this case since adding the element $x'$ to the assignment leads to a deterioration we have:
		\[
			\frac{f(x)\tau}{\sqrt{f(x)\varsigma^2}} > \frac{(f(x) + f(x'))\tau}{\sqrt{(f(x) + f(x'))\varsigma^2}} \implies \sqrt{f(x)} > \sqrt{f(x) + f(x')}
		\]
		which is a contradiction.
	\end{proof}
	
	\subsection{Proof of \autoref{lem:partition-approx}}
	\label{sect:proof_lem_partition-approx}
	
	\begin{proof}
		Consider a partition $i\in \{1,\ldots,k\}$, such that a subset $T_i\subseteq P_i$, with density $t_i$ is treated, and the complement $U_i = P_i \setminus T_i$, with density $u_i$ is not treated by $a_*$. Let $\tau_i$ be the mean value of $\tau(x)$ for all elements in $P_i$, i.e. $\tau_i = \frac{1}{p_i} \sum_{x\in P_i} f(x) \tau(x)$. Similarly, let $\varsigma_i^2$ be mean value of $\varsigma(x)^2$. By our assumption, we know that for all $x\in P_i$: $p_i|\tau(x) - \tau_i |\leq \epsilon$ and $p_i|\varsigma^2(x) - \varsigma^2_i|\leq \epsilon$. Moreover, denote with $\tau_{i,0}$ and $\varsigma_{i,0}^2$ the corresponding means on the untreated sub-population $U_i$ and $\tau_{i,1}, \varsigma_{i,1}^2$ the means on the treated sup-population $T_i$. By our assumption we also have that for all $x\in P_i$ and for any $z\in \{0,1\}$: $p_i|\tau(x) - \tau_{i,z} |\leq \epsilon$ and $p_i|\varsigma^2(x) - \varsigma^2_{i,z}|\leq \epsilon$.  
		
		Let $a_*$ be an optimal policy. We can write our objective as:
		\begin{align*}
			V(a_*) = \frac{A + t_i \tau_{i,1}}{\sqrt{B + t_i \varsigma_{i,1}^2}}
		\end{align*}
		where $A, B\geq 0$, since an optimal assignment never allocates to any $x\in \X$ with $\tau(x)\leq 0$. Moreover, $A, B$ are quantities that do not depend on how the policy behaves for any $x\in P_i$.
		
		We first consider the case where $B > 0$, i.e. when $P_i$ is not the only partition that contains elements that are assigned the treatment. Since by removing $T_i$ from assignment or adding $U_i$ to the assignment, we are deteriorating the objective by more $\delta$, we must have that:
		\begin{equation}\label{eqn:opt-approx}
			\begin{aligned}
				\frac{A + t_i \tau_{i,1}}{\sqrt{B + t_i \varsigma_{i,1}^2}} >~& \frac{A}{\sqrt{B}} + \delta\\
				\frac{A + t_i \tau_{i,1}}{\sqrt{B + t_i \varsigma_{i,1}^2}} >~& \frac{A + t_i \tau_{i,1} + u_i \tau_{i,0}}{\sqrt{B + t_i \varsigma_{i,1}^2 + u_i \varsigma_{i,0}^2}} + \delta
			\end{aligned}
		\end{equation}
		
		By the aforementioned approximation properties we have that:
		\begin{align*}
			\frac{A + t_i \tau_{i,1}}{\sqrt{B + t_i \varsigma_{i,1}^2}} &\leq \frac{A + t_i \tau_{i}}{\sqrt{B + t_i \varsigma_{i}^2}} + \frac{t_i (\tau_{i,1} - \tau_i)}{\sqrt{B + t_i \varsigma_{i}^2}} + \frac{A + t_i \tau_{i,1}}{\sqrt{B + t_i \varsigma_{i,1}^2}\sqrt{B + t_i \varsigma_{i}^2}}\left(\sqrt{B + t_i \varsigma_{i,1}^2} - \sqrt{B + t_i \varsigma_{i}^2}\right)\\
			&= \frac{A + t_i \tau_{i}}{\sqrt{B + t_i \varsigma_{i}^2}} + \frac{t_i (\tau_{i,1} - \tau_i)}{\sqrt{B + t_i \varsigma_{i}^2}} + \frac{A + t_i \tau_{i,1}}{\sqrt{B + t_i \varsigma_{i,1}^2}\sqrt{B + t_i \varsigma_{i}^2}}\frac{t_i \left(\varsigma_{i,1}^2 - \varsigma_{i}^2\right)}{\sqrt{B + t_i \varsigma_{i,1}^2} + \sqrt{B + t_i \varsigma_{i}^2}}\\
			&= \frac{A + t_i \tau_{i}}{\sqrt{B + t_i \varsigma_{i}^2}} + \frac{\sqrt{t_i} (\tau_{i,1} - \tau_i)}{\sqrt{\varsigma_{i}^2}} + \frac{\tau_{i,1}}{\varsigma_{i}^2}\frac{\sqrt{t_i} \left(\varsigma_{i,1}^2 - \varsigma_{i}^2\right)}{\sqrt{\varsigma_{i}^2}}\\
			&\leq \frac{A + t_i \tau_{i}}{\sqrt{B + t_i \varsigma_{i}^2}} + \frac{\epsilon}{\underline{\varsigma}} + \frac{\bar{\kappa}}{\underline{\varsigma}} \epsilon\\
			&\leq \frac{A + t_i \tau_{i}}{\sqrt{B + t_i \varsigma_{i}^2}} + \frac{\epsilon}{\underline{\varsigma}} \left(1 + \bar{\kappa}\right)
		\end{align*}
		Similarly, we have that:
		\begin{align*}
			\frac{A + t_i \tau_{i,1} + u_i \tau_{i,0}}{\sqrt{B + t_i \varsigma_{i,1}^2 + u_i \varsigma_{i,0}^2}} \geq~& \frac{A + p_i \tau_{i}}{\sqrt{B + p_i \varsigma_{i}^2}} - \frac{2\epsilon}{\underline{\varsigma}} \left(1 + \bar{\kappa}\right)
		\end{align*}
		Thus for $\delta \geq \frac{2\epsilon}{\underline{\varsigma}} \left(1 + \bar{\kappa}\right)$, we can combine the latter inequalities to conclude that:
		\begin{equation}
			\begin{aligned}
				\frac{A + t_i \tau_{i}}{\sqrt{B + t_i \varsigma_{i}^2}} >~& \frac{A}{\sqrt{B}}
				&
				\frac{A + t_i \tau_{i}}{\sqrt{B + t_i \varsigma_{i}^2}} >~& \frac{A + p_i \tau_{i}}{\sqrt{B + p_i \varsigma_{i}^2}}
			\end{aligned}
		\end{equation}
		Using an argument identical to that in the proof of \autoref{lem:partition-exact}, we can show that the latter leads to a contradiction. Thus either assigning all of the partition $P_i$ to the treatment or removing all of the partition $P_i$ from the treatment, can lead to a new policy that achieves as objective $V(a)$ at least an additive $\delta$ smaller than the optimal policy. 
		
		Repeating this process for every partition $i\in \{1,\ldots,k\}$, we can arrive at a new policy that achieves an objective that is at most an additive $k\delta$  smaller than that of the optimal policy and in which all partitions have a uniform assignment to either treatment or non-treatment.
	\end{proof}

	\subsection{Proof of \autoref{lem:subgroup-id}}
	\label{sect:proof_lem_subgroup-id}
	
	\begin{proof}
		The result for $\tau$ is an immediate consequence of Theorem~C.10.2 of \cite{syrgkanis2020estimation}. For the case of $\varsigma^2$ we need to make a small adjustment to account for the fact that $\hat{c}\neq c^*$. In particular, even though $\varsigma^2_{*}(x) = \varsigma^2_{c^*}(x)$ might satisfy the $(\beta, r)$-strong sparsity condition, the function $\varsigma^2(x) = \varsigma^2_{\hat{c}}(x)$ might not. Moreover, we want to guarantee that $\varsigma^2_{*}(x)=\varsigma^2_{*}(x')$ for every $x,x'$ that fall in the leaf of the constructed tree and not that $\varsigma^2(x)=\varsigma^2(x')$.
		
		However, we note that the main argument in Theorem~C.10.2 of \cite{syrgkanis2020estimation} is robust to small errors. In particular, Lemma~C.1 of \cite{syrgkanis2020estimation}, shows that the empirical splitting criterion $\widehat{\Sigma}(S)$,
		for every possible split up to depth $d$, is within an $\epsilon(n,\delta)$ of its population counterpart w.p. $1-\delta$, for $\epsilon(n,\delta) = O\left(\sqrt{\frac{2^d d\log(J\, d) + \log(1/\delta))}{n}}\right)$. Moreover, the population criterion is $\Sigma_{\hat{c}}(S)$ where we denote with
		$\Sigma_{c}(S) = \E[\E^2[\varsigma^2_c(X)| X_S]]$
		the population splitting criterion with centering function $c$.
		Since by \autoref{lem:basic-props} $\varsigma_{\hat{c}}^2(x) = \varsigma^2_{c^*}(x) + \frac{(\hat{c}(x)-c^*(x))^2}{p(1-p)}$, we thus have that:
		\begin{align*}
			\left|\Sigma_{\hat{c}}(S) - \Sigma_{c^*}(S)\right| &= \left|\E\left[\E^2[\varsigma^2(X)| X_S] - \E^2[\varsigma^2_{*}(X)| X_S] \right]\right|\\
			&\leq 2  \E\left[\left|\E[\varsigma^2(X)| X_S] - \E[\varsigma^2_{*}(X)| X_S]\right|\right]\\
			&\leq \frac{2\E[(\hat{c}(X) - c^*(X))^2]}{p(1-p)}
		\end{align*}
		Thus as long as $\epsilon(n,\delta) + \frac{2\E[(\hat{c}(X) - c^*(X))^2]}{p(1-p)} < \frac{\beta}{2}$, then we have that w.p. $1-\delta$, at every step of the greedy splitting algorithm we will only be choosing a variable in the relevant set $R$ for the function $\varsigma^2_{*}(x)$, since for any $i\in R$ and $j \in \{1,\ldots,J\} \setminus R$ and $S \subseteq \{1,\ldots,J\} \setminus \{i\}$:
		\begin{align*}
			\widehat{\Sigma}(\{i\}\cup S) \geq~& \Sigma_{\hat{c}}(\{i\}\cup S) - \epsilon(n,\delta)\\
			\geq~& \Sigma_{c^*}(\{i\}\cup S) - \epsilon(n,\delta) - \frac{2\E[(\hat{c}(X) - c^*(X))^2]}{p(1-p)}\\
			\geq~& \Sigma_{c^*}(\{j\}\cup S) + \beta - \epsilon(n,\delta) - \frac{2\E[(\hat{c}(X) - c^*(X))^2]}{p(1-p)}\\
			>~& \Sigma_{c^*}(\{j\}\cup S) + \beta/2\\
			\geq~& \Sigma_{\hat{c}}(\{j\}\cup S) + \beta/2 - \frac{2\E[(\hat{c}(X) - c^*(X))^2]}{p(1-p)}\\
			\geq~& \widehat{\Sigma}(\{j\}\cup S) + \beta/2 - \epsilon(n,\delta) - \frac{2\E[(\hat{c}(X) - c^*(X))^2]}{p(1-p)}\\
			>~& \widehat{\Sigma}(\{j\}\cup S) 
		\end{align*}
		Thus variable $i$ will be chosen over variable $j$.
	\end{proof}
	
	\subsection{Proof of \autoref{lem:vc-bound}}
	\label{sect:proof_lem_vc-bound}
	
	\begin{proof}
		By Section A.2 of \cite{syrgkanis2020estimation}, the VC dimension of $A$ is at most $v=4 d 2^d + 2^{d+1} \log(2J)$. Thus the growth number of this class by Sauer's lemma is at most $\tau(n) = (e\,n/v)^v$. Hence, the $\ell_{\infty}$ empirical covering number of $A$ on $n$ observations is at most $\tau(n)=(e\,n/v)^v$. Thus we have by the empirical and population Bennett inequality of \cite{maurer2009empirical} that w.p. $1-\delta$:
		\begin{align*}
			\left|\frac{1}{n} \sum_{i=1}^n a(X_i)\ \tilde{Y}_i - \E[a(X) \tilde{Y}]\right| &\leq \sqrt{\frac{\E_n[a(X)\varsigma_c(X)^2]\, 2\, v \log(2\,e\,n/\delta)}{n}} + H \frac{7\, v\log(2\,e\, n/\delta)}{3(n-1)}\\
			\left|\E_n[a(X)\tilde{Y}^2] - \E[a(X) \tilde{Y}^2]\right| &\leq \sqrt{\frac{\E[a(X)\tilde{Y}^4]\, 2\,v\, \log(e\,n/\delta)}{n}} + H \frac{v\,\log(2e\,n/\delta)}{3n}\\
			&\leq H\sqrt{\frac{\E[a(X)\tilde{Y}^2]\, 2\,v\, \log(e\,n/\delta)}{n}} + H\frac{v\,\log(2\,e/\delta)}{3n}
		\end{align*}
		Finally, observe that $\E[a(X)\tilde{Y}] = \E[a(X)\tau(X)]$ and $\E[a(X)\tilde{Y}^2] = \E[a(X)\varsigma^2_c(X)]$.
	\end{proof}
	
	\subsection{Proof of \autoref{prop:main-model}}
	\label{sect:proof_prop_main-model}
	
	\begin{proof}
		By \autoref{lem:subgroup-id}, we have that w.p. $1-\delta$, the partitions defined by the leaves of the two trees will contain $x$'s that all have the same value of $\tau(x)$ and $\varsigma^2_{*}(x)$. Thus the policy space that is being searched over in the last step of the algorithm contains an optimal policy $a_*$. Moreover, combining \autoref{lem:vc-bound} and \autoref{lem:basic-props} and a proof almost identical to that of \autoref{prop:finite-x-pop} (replacing $\underline{\pi}$ with $\zeta/2^r$ and $k\log(2/\delta)$ with $v\log(2\, e\,n/\delta)$, where $v=4 r 2^r + 2^{r+1} \log(2J)$), we can show that if we let:
		\begin{align*}
			\kappa(n,\eta,\delta) = & \sqrt{\frac{2\,v\, \log(2\ e\,n/\delta)}{n}} + \frac{7\,H\,v\,\log(2e\,n/\delta)}{3(n-1)}
			\\
			&+ \frac{2^{3r/2}}{2(1+\eta)^{3/2} \zeta\, \underline{\varsigma}^3} \left(\frac{\E[(\hat{c}(X) - c^*(X))^2]}{p(1-p)}+ \frac{2 H\, v\log(2e\,n/\delta)}{\eta\,n}\right)
		\end{align*}
		then $\hat{a}$ satisfies w.p. $1-\delta$:
		\begin{align*}
			\frac{\E[\hat{a}(X) \tilde{Y}]}{\sqrt{\E_n[\hat{a}(X)\varsigma^2_{*}(X)]}} &\geq \sqrt{\frac{1-\eta}{1+\eta}} \frac{1}{1+\epsilon} \frac{\E[a_*(X) \tilde{Y}]}{\sqrt{\E[a_*(X) \varsigma^2_{*}(X)]}} - 2\kappa(n,\eta,\delta)
			\qedhere
		\end{align*}
	\end{proof}
	
	\subsection{Proof of \autoref{prop:relaxed}} \label{sect:proof_prop_relaxed}
	
	\begin{proof}
		$U$ in \autoref{prop:oracle} is monotonically increasing in
		\begin{align*}
			\frac{\E[a(X) \: \tau(X)]}{\sqrt{\E[a^2(X) \: \sigma^2(X)]}}.
		\end{align*}
		By $a(x) \geq 0$ and the Cauchy--Schwartz inequality,
		\begin{align*}
			\E[a(X) \: \tau(X)]
			\leq \E[a(X) \sigma(X) \: \tau_+(X) / \sigma(X)]
			\leq \sqrt{\E[a^2(X) \sigma^2(X)]} \sqrt{\E[\tau^2_+(X) / \sigma^2(X)]}
		\end{align*}
		and hence
		\begin{align*}
			\frac{\E[a(X) \: \tau(X)]}{\sqrt{\E[a^2(X) \: \sigma^2(X)]}}
			\leq \sqrt{\E[\tau^2_+(X) / \sigma^2(X)]},
		\end{align*}
		where the upper bound does not depend on $a$ and is attained for $a(x) \propto \tau_+(X) / \sigma^2(X)$.
	\end{proof}
	
	\subsection{Proof of \autoref{prop:avoiding}}
	\label{sect:proof_prop_avoiding}
	
	\begin{proof}
		For $\alpha, \delta > 0$ let $K$ be such that $2^k \: \Phi(-K) < \delta / 3$, $L$ such that $\Phi( - L + z_{1-\alpha}) < \delta / 3$, and $\varepsilon > 0$ such that $\Phi(\varepsilon - z_{1-\alpha}) < \alpha + \delta$.
		Fix $(\mu_x)_{x=1}^k$ with $\mu_x > 0$ for all $x$,
		and let $C = \max\left\{\frac{2 K + L}{\min_x \mu_x}, \frac{2 k \max_x \mu_x}{\varepsilon} \right\}$.
		For any fixed $(\sigma^2_x)_{x=1}^k$ with $\min_x \sigma^2_x \leq \frac{1}{C}, \max_x \sigma^2_x > C$ and any fixed assignment $a \in \{0,1\}^k$,
		we have that
		\begin{align}
			\label{eqn:CLT}
			&\frac{\frac{1}{k}\sum_{x=1}^k a_x \hat{\tau}_x}{\sqrt{\frac{1}{k} \sum_{x=1}^k a_x \frac{\hat{\sigma}^2_x}{n/k}}}
			-
			\frac{\sum_{x=1}^k a_x \mu_x}{\sqrt{\sum_{x=1}^k a_x \sigma^2_x}}
			\\
			&=
			\frac{\sum_{x=1}^k a_x (\sqrt{n} \hat{\tau}_x - \mu_x)}{\sqrt{ \sum_{x=1}^k a_x \hat{\sigma}^2_x}}
			+ \frac{\sum_{x=1}^k a_x \mu_x}{\sqrt{\sum_{x=1}^k a_x \sigma^2_x} \sqrt{ \sum_{x=1}^k a_x \hat{\sigma}^2_x}}
			\underbrace{\left( \sqrt{ \sum_{x=1}^k a_x \hat{\sigma}^2_x} - \sqrt{\sum_{x=1}^k a_x \sigma^2_x} \right)}_{\cp 0}
			\cd \N(0,1)
			\nonumber
		\end{align}
		as $n \rightarrow \infty$ on the training sample.
		Since there are only $2^k$ assignments,
		\begin{align*}
			&\P\left(
			\max_{a \in \{0,1\}^k} \left|
			\frac{\frac{1}{k}\sum_{x=1}^k a_x \hat{\tau}_x}{\sqrt{\frac{1}{k} \sum_{x=1}^k a_x \frac{\hat{\sigma}^2_x}{n/k}}}
			-
			\frac{\sum_{x=1}^k a_x \mu_x}{\sqrt{\sum_{x=1}^k a_x \sigma^2_x}} \right| < K
			\right)
			\\
			&\geq
			1 - \sum_{a \in \{0,1\}^k} \P\left(
			\left|
			\frac{\frac{1}{k}\sum_{x=1}^k a_x \hat{\tau}_x}{\sqrt{\frac{1}{k} \sum_{x=1}^k a_x \frac{\hat{\sigma}^2_x}{n/k}}}
			-
			\frac{\sum_{x=1}^k a_x \mu_x}{\sqrt{\sum_{x=1}^k a_x \sigma^2_x}} \right| \geq K
			\right)
			> 1 - \frac{\delta}{2}
		\end{align*}
		for $n$ sufficiently large.
		Hence, with probability at least $1-\delta/2$ and for large $n$,
		\begin{align*}
			\frac{\sum_{x=1}^k \hat{a}^{(B)}_x \mu_x}{\sqrt{\sum_{x=1}^k \hat{a}^{(B)}_x \sigma^2_x}}
			\geq
			\max_{a \in \{0,1\}^k}
			\frac{\sum_{x=1}^k a_x \mu_x}{\sqrt{\sum_{x=1}^k a_x \sigma^2_x}}
			- 2 K
			&
			\geq \frac{\min_x \mu_x}{\min_x \sigma_x} - 2K
			\geq C \min_x \mu_x - 2 K
			\geq L.
		\end{align*}
		By \eqref{eqn:CLT} applied to the test sample, on that sample
		\begin{align*}
			\frac{\frac{1}{k}\sum_{x=1}^k \hat{a}^{(B)}_x \hat{\tau}_x}{\sqrt{\frac{1}{k} \sum_{x=1}^k \hat{a}^{(B)}_x \frac{\hat{\sigma}^2_x}{n/k}}}
			> z_{1-\alpha}
		\end{align*}
		with probability at least $1 - \delta/2$ for $N$ sufficiently high.
		Taken together, we reject with probability at least $1-\delta$ once $n,N$ are sufficiently high.
		
		On the other hand, for the simple classifier,
		$\P(\hat{a}^{(1)}_x = 1) > \frac{1}{2}$ for all $x$ and $n$ sufficiently large, including for $x$ some $x$ with $\sigma_x > C$.
		In that case,
		\begin{align*}
			\frac{\sum_{x=1}^k \hat{a}^{(A)}_x \mu_x}{\sqrt{\sum_{x=1}^k \hat{a}^{(A)}_x \sigma^2_x}}
			\leq \frac{k \max_x \mu_x}{C}
			< \varepsilon / 2
		\end{align*}
		and we therefore reject with overall probability at most
		\begin{align*}
			\frac{1}{2} + \frac{1}{2} \Phi(\varepsilon -z_{1-\alpha})
			< \frac{1 + \alpha}{2} + \delta
		\end{align*}
		as $n,N$ sufficiently large.
	\end{proof}
	
\end{document}

%% file: header.tex
\usepackage{booktabs} %
\usepackage[titlenumbered,ruled]{algorithm2e}
\usepackage{amsmath}
\usepackage{amsfonts}
\usepackage{mathtools}
\usepackage{bbm}
\usepackage{enumitem}
\usepackage{fullpage}
\usepackage{placeins}
\usepackage{setspace}
\usepackage[authoryear]{natbib}
\usepackage{amsthm}
\usepackage{xcolor}
\usepackage{xfrac}
\usepackage[backref=page,breaklinks=true,hidelinks]{hyperref}
\hypersetup{
	colorlinks,
	linkcolor={red!50!black},
	citecolor={blue!50!black},
	urlcolor={blue!80!black}
}
\renewcommand*{\backref}[1]{}
\renewcommand*{\backrefalt}[4]{%
	\ifcase #1 {\footnotesize(Not cited.)}%
	\or        {\footnotesize(Cited on page~#2.)}%
	\else      {\footnotesize(Cited on pages~#2.)}%
	\fi}
\usepackage{multicol}

\usepackage{todonotes}
\presetkeys{todonotes}{inline}{}

\makeatletter
\def\blfootnote{\xdef\@thefnmark{}\@footnotetext}
\makeatother

\SetAlFnt{\small}
\SetAlCapFnt{\small}
\SetAlCapNameFnt{\small}
\SetAlCapHSkip{0pt}
\IncMargin{-\parindent}

\renewcommand{\P}{\mathop{}\!\textnormal{P}}

\newcommand{\E}{\mathop{}\!\textnormal{E}}

\newcommand{\Cov}{\mathop{}\!\textnormal{Cov}}

\newcommand{\Var}{\mathop{}\!\textnormal{Var}}

\newcommand{\X}{\mathcal{X}}
\newcommand{\N}{\mathcal{N}}

\newcommand{\cd}{\stackrel{d}{\longrightarrow}}
\newcommand{\cp}{\stackrel{p}{\longrightarrow}}
\newcommand{\sign}{\mathop{\textnormal{sign}}}
\newcommand{\expit}{\mathop{\textnormal{expit}}}

\newcommand{\mtry}{\text{mtry}}

\newcommand{\R}{\mathbb{R}}
\DeclareMathOperator*{\argmin}{arg\,min}
\DeclareMathOperator*{\argmax}{arg\,max}
\newcommand{\Ind}{\mathbbm{1}}  %

\newtheorem{lemma}{Lemma}
\newtheorem{proposition}{Proposition}
\newtheorem{definition}{Definition}

\renewcommand{\varsigma}{\ensuremath{\nu}}

\usepackage{etoolbox}
\makeatletter
\patchcmd{\hyper@makecurrent}{%
	\ifx\Hy@param\Hy@chapterstring
	\let\Hy@param\Hy@chapapp
	\fi
}{%
	\iftoggle{inappendix}{%
		\@checkappendixparam{chapter}%
		\@checkappendixparam{section}%
		\@checkappendixparam{subsection}%
		\@checkappendixparam{subsubsection}%
		\@checkappendixparam{paragraph}%
		\@checkappendixparam{subparagraph}%
	}{}%
}{}{\errmessage{failed to patch}}

\newcommand*{\@checkappendixparam}[1]{%
	\def\@checkappendixparamtmp{#1}%
	\ifx\Hy@param\@checkappendixparamtmp
	\let\Hy@param\Hy@appendixstring
	\fi
}
\makeatletter

\newtoggle{inappendix}
\togglefalse{inappendix}

\apptocmd{\appendix}{\toggletrue{inappendix}}{}{\errmessage{failed to patch}}